\documentclass[12pt]{article}
\usepackage{graphicx}
\usepackage{natbib} 
\usepackage{url} 
\usepackage{xcolor}
\usepackage{algorithm}
\usepackage{algpseudocode}
\usepackage[scr=esstix]{mathalpha}

\usepackage{subcaption}
\usepackage{amsmath,amssymb,amstext} 
\DeclareMathOperator{\E}{\mathbb{E}}
\DeclareMathOperator{\V}{\mathbb{V}}

\DeclareMathOperator*{\argmax}{arg\,max}

\usepackage{amsfonts}       
\usepackage{mathtools}
\usepackage{placeins}
\usepackage{bbm}
\usepackage{bm}
\usepackage{booktabs}

\usepackage[pdftex,pagebackref=false, hypertexnames=false]{hyperref} 
\usepackage{float}

\newcommand{\supplementarysection}{%
  \setcounter{section}{0}
  \setcounter{figure}{0}
  \let\oldthefigure\thefigure
  \renewcommand\thesection{\Alph{section}}
\renewcommand\thesubsection{\thesection.\arabic{subsection}}
}

\usepackage{setspace}
\begin{document}

\def\spacingset#1{\renewcommand{\baselinestretch}%
{#1}\small\normalsize} \spacingset{1}
\newcommand{\bit}{\begin{itemize}}
\newcommand{\eit}{\end{itemize}}
\newcommand{\x}{\mathbf{x}}
\newcommand{\bfx}{\mathbf{x}}
\newcommand{\mb}{\mathbf{}}


{
  \title{\bf Fully Bayesian Sequential Design for Heteroscedastic Stochastic Simulations}
  \author{Yuying Huang 
    and 
    Samuel W.K. Wong \\
    Department of Statistics and Actuarial Science, University of Waterloo}
    \date{}
  \maketitle
} 

\bigskip
\begin{abstract}
We present a fully Bayesian sequential strategy for predicting the mean response surface of heteroscedastic stochastic simulation models. Leveraging dual Gaussian processes as the surrogate and a selection criterion based on expected Bayesian integrated mean-square prediction error, our approach sequentially selects informative design points while fully accounting for parameter uncertainty. Sequential importance sampling is employed to efficiently update the posterior distribution of the parameters. Our strategy is tailored for expensive simulation models, where achieving robust predictive accuracy under a limited budget is critical. Using synthetic examples, we illustrate its practical advantages compared to existing approaches, in terms of predictive accuracy, noise estimation, and uncertainty quantification. We then implement the proposed strategy on a real motivating application in seismic design of wood-frame podium buildings.
\end{abstract}
\noindent%
\vfill

\newpage
\section{Introduction}

Computer-based simulation is a commonly utilized tool for analyzing complex systems \citep{sacks1989designs,sacks1989design, santner2003design}. By constructing simulation models that mimic system behavior, practitioners can gain insights into the system's underlying mechanisms and dynamics. However, evaluating simulation models can be computationally intensive. As the motivating application for this work, a dynamic analysis for seismic design of podium buildings \citep{chen2020criterion} requires over 30 hours of CPU time on a standard computer for each evaluation. In such cases, practitioners may be limited in exploring large numbers of design points or conducting extensive searches of the input space. Fast surrogates of simulation models, known as metamodels, approximate the simulation model based on a limited set of input/output data \citep{barton1994metamodeling,forrester2008engineering, kleijnen2009kriging, kleijnen2017regression, gramacy2020surrogates}. These surrogate models produce predictions for unseen inputs, and serve as alternatives to expensive simulations. Some applications of metamodeling include vehicle side impact design \citep{hao2021novel}, Delta smelt population estimation \citep{zhang2022batch}, and freeway traffic management \citep{chen2018bayesian}, among others.

Simulation models can be broadly categorized according to whether their outputs are deterministic or stochastic. Deterministic simulations yield identical outputs for the same inputs, while stochastic simulations yield varying outputs due to intrinsic random noise \citep{sacks1989designs, sacks1989design}. Stochastic simulations are widely used to explore complex phenomena in various fields such as transportation, ecology, and epidemiology \citep[see, e.g.,][]{richardson1981stochastic, peleg2017advanced, mckinley2018approximate,chen2018bayesian,kennedy2023multilevel}. In stochastic simulations, the variance of the random noise, or noise level, controls the extent of deviations between the simulated outputs and the mean response surface. Many industrial simulations are stochastic with heteroscedastic noise, where the noise level depends on the inputs \citep{kleijnen2005robustness}. For instance, in the seismic design application \citep{chen2020criterion}, the response, inter-story drift, varies due to randomness in how building components, such as springs or shear walls, react to ground motions of simulated earthquakes. As the noise level is not directly observable, the presence of heteroscedastic noise makes the prediction of mean response surface more challenging. \cite{ankenman2010stochastic, yin2011kriging} apply stochastic kriging to model the mean response surface assuming multiple observations are available per input point, so that the noise variance can be estimated by the sample variance of the repeated observations. More recent studies have explored joint metamodeling of the underlying mean response surface and the noise variance to save computational resources \citep{robinson2010semi, lazaro2011variational, wang2016effects, binois2018practical, zhu2021emulation}.

Constructing surrogates for simulation models involves selecting and evaluating design points of interest in the input space. Sequential design can improve the efficiency of exploring the input space compared to one-shot designs, by utilizing information from previous observations to iteratively improve the quality of fit \citep{chernoff1992sequential, cohn1996active, kleijnen2009kriging, liu2018survey,gramacy2020surrogates,fuhg2021state}. This requires constructing a selection criterion to gauge the potential information gain at candidate design points. The criterion is then optimized to select the subsequent design points for evaluation, addressing objectives such as reducing prediction uncertainty or finding the maximum (or minimum) of the underlying simulation model. Various sequential strategies have been proposed for mean response surface prediction in stochastic simulations. For instance, \cite{chen2014sequential, chen2017sequential} established theoretical budget allocation rules to minimize prediction uncertainty using stochastic kriging. In contrast, \cite{binois2018practical, binois2019replication} proposed Gaussian process (GP) frameworks for heteroscedastic noise, leveraging Woodbury matrix identities \citep{harville1998matrix} for computational efficiency.

Existing sequential strategies tend to rely on plug-in estimators for model parameters and random noise, such as via maximum likelihood estimation (MLE) in a frequentist setting or \textit{maximum a posteriori} (MAP) estimation in an empirical Bayes setting. Using point estimates for parameters can simplify computations, but neglects full uncertainty quantification. In contrast, a fully Bayesian approach treats all model parameters as random variables in a joint posterior distribution, which in the context of a sequential strategy can better propagate parameter uncertainties into design point selection and surrogate model prediction. Potential advantages of a fully Bayesian approach that have been demonstrated in other settings are: improved model prediction performance \citep{helbert2009assessment, karagiannis2019bayesian}, more accurate inference \citep{park2013bayesian, lalchand2020approximate}, and better uncertainty quantification of parameters \citep{xu2014fully, yerramilli2023fully}. A common deterrent to adopting fully Bayesian approaches is the computational cost associated with posterior inference, e.g., via Markov chain Monte Carlo (MCMC) methods \citep{binois2018practical, baker2022analyzing}. However, in this paper we are motivated by applications where the simulation model is expensive in the overall scheme, such that the benefits of working with the full posterior distribution can outweigh the additional costs. In this setting, a fully Bayesian sequential strategy focusing on mean response surface prediction for stochastic simulations remains unexplored, to the best of our knowledge. For instance, while \cite{yuan2015calibration} propose several strategies with GP models under a Bayesian framework to optimize model calibration and prediction, they employ empirical Bayes point estimates, and their strategies only accommodate homogeneous noise.

As the main contribution of this paper, we thus propose a fully Bayesian sequential strategy for predicting the mean response surface of stochastic simulations with heteroscedastic noise levels. GPs serve as our surrogate model, since their conceptual simplicity and mathematical tractability in prediction make them well-suited for modeling stochastic simulations \citep{santner2003design, williams2006gaussian, gramacy2020surrogates}. As pioneered by \cite{goldberg1997regression}, and later adopted by \cite{kersting2007most, binois2019replication, binois2021hetgp}, we employ two independent GPs to model the underlying true function and log-noise variance, respectively. Building on the dual GP surrogate model, we develop a fully Bayesian sequential strategy that incorporates weakly informative priors for model parameters and utilizes a variance-based criterion to guide design point selection. Specifically, we propose a selection criterion based on the expected Bayesian integrated mean-square prediction error (BIMSPE). While inspired by the classic IMSPE \citep{sacks1989designs,sacks1989design}, our approach accounts for both GP hyperparameter and latent noise uncertainty by marginalizing over their posterior distributions. The efficacy of our proposed strategy is shown through several synthetic examples, where it outperforms alternatives in terms of predictive accuracy, noise estimation, and uncertainty quantification. We then present an illustrative real-world application of our methodology in seismic design.

The remainder of this paper is organized as follows. Section \ref{sec-Methodology} begins with a review of heteroscedastic GP regression, and then presents our proposed fully Bayesian sequential strategy and the expected BIMSPE criterion. We also detail computational techniques for expected BIMSPE calculations and efficient posterior updates, along with steps for practical implementation. Section \ref{sec-experiments} assesses the performance of our strategy, compared to existing ones, through synthetic examples. Section \ref{sec-real-life-example} showcases our strategy on the motivating real application to the seismic design of podium buildings. Section \ref{sec-conclusion} concludes the paper and offers some discussion and avenues for further research.

\section{Methodology}\label{sec-Methodology}
\subsection{Gaussian Process Regression under Heteroscedasticity}\label{sec-GPR}

Let $\mathcal{X} \subset \mathbb{R}^d$ represent the $d$-dimensional design space, and let $\mb{Y}(\bfx)$ denote the (random) output of a stochastic simulation model with heteroscedastic noise at design point $\bfx\in \mathcal{X}$. We assume the underlying mean response surface is represented by the function $f:\mathbb{R}^d\rightarrow \mathbb{R}$, such that $\E[Y(\bfx)]=f(\bfx)$ for any $\bfx\in \mathcal{X}$. Then the observation model is defined as $Y(\bfx)=f(\bfx)+\epsilon(\bfx)$, where the noise $\epsilon(\bfx)$ has mean zero and input-dependent variance $g(\bfx)$, and is independent across observations. We impose a zero-mean GP prior on $f(\bfx)$ with kernel function $k_f(\bfx, \bfx')$, i.e., $f(\bfx) \sim \mathcal{GP} \left(0, k_f(\bfx, \bfx') \right)$. Furthermore, assuming the noise is a smooth function of the input, it is natural to model the noise variance with a second independent GP \citep{goldberg1997regression}. Since the noise variance $g(\bfx)$ is strictly positive, we follow previous works \citep[e.g.,][]{kersting2007most, binois2019replication, binois2021hetgp} and assign a GP prior to the log-variance $v(\bfx) = \log g(\bfx)$, i.e., $v(\bfx)\sim \mathcal{GP}(\mu_0, k_v(\bfx, \bfx'))$. The constant mean parameter $\mu_0$ for the $v$-process then represents the average log-variance of the noise across $\mathcal{X}$. 

In practice, we consider a finite set of design points $\mathcal{D} \subset \mathcal{X}$ at which observations may be taken. The set $\mathcal{D}$ is selected by the experimenter to provide a representative discretization of the design space \citep{ankenman2010stochastic}. We assume the total number of design points, $|\mathcal{D}|$, is large enough such that exhaustive evaluation (with replicates) is infeasible. We note that the discretization of $\mathcal{X}$ into a sufficiently dense finite set $\mathcal{D}$ may be computationally prohibitive in high dimensions. However, this approach is often reasonable in engineering applications of low-to-moderate dimensions, where small variations in inputs typically yield negligible changes in the response $f(\bfx)$ below the threshold of physical significance \citep{santner2003design}. Moreover, this choice facilitates our fully Bayesian treatment of sequential design and selection of replicates, by keeping the candidate set $\mathcal{D}$ fixed throughout. 

The GP kernel function determines the covariance between two design points based on their distance. Most kernels have  a variance parameter $\sigma^2$, which controls the overall amplitude,  and a vector of lengthscales $\textbf{l}=(l_1, \dots, l_d)^T$, which controls the rate of correlation decay along each input dimension \citep{williams2006gaussian}. The widely used anisotropic squared exponential kernel sets $k_{\boldsymbol{\cdot}}(\bfx_i,\bfx_j)=\sigma^2 \text{exp} \left[-\sum_{h=1}^d {(\bfx_{i,h}-\bfx_{j,h})^2}/ ({2l_h^2}) \right]$, which we adopt for the simulation examples in Section \ref{sec-experiments}, given the smooth nature of the test functions involved. A common alternative is the Mat\'ern family of kernels, which is useful for physical processes that may not satisfy strong smoothness assumptions \citep{stein1999interpolation}; our proposed sequential strategy can be used with any choice of kernel function. In our GP priors, the kernel function $k_f(\bfx, \bfx')$ is then specified by the hyperparameters $\sigma_f$ and $\textbf{l}_f$, while $k_v(\bfx, \bfx')$ is parameterized by $\sigma_v$ and $\textbf{l}_v$. Together, the dual GPs are fully specified by the parameter set $\bm{\theta} = \{ \mu_0, \sigma_f, \sigma_v, \textbf{l}_f, \textbf{l}_v\}$ to be inferred from the observations. 

We next review some standard prediction results for GPs. Consider a set of $n$ observations $\textbf{y}=[y_1, y_2, \dots, y_n]^T$ at the design points  $\mathbf{X}=[\bfx_1, \bfx_2, \dots, \bfx_n]^T$. Further, let $\textbf{v} =[v_1, v_2, \dots, v_N]^T$ denote the latent log-noise process at the design points $\mathbf{X}$. For a test location $\bfx_*$, let $v_* = v(\bfx_*)$ be the corresponding latent log-noise. Following standard results \citep{williams1998prediction}, the predictive distribution of $Y(\bfx_*)|\mathbf{y}, \bm{\theta}, \textbf{v}, v_*$ is normal:
\begin{equation}
	\hat{Y}(\bfx_*)=\E\left(\mb{Y}(\bfx_*)|\textbf{y}, \bm{\theta}, \textbf{v}, v_* \right)=\textbf{k}_*^T\boldsymbol{K}_n^{-1}\textbf{y}    \label{pred_mean}
\end{equation}
\begin{equation}
	\V \left(\mb{Y}(\bfx_*)|\textbf{y}, \bm{\theta}, \mb{\textbf{v}}, v_* \right)=\sigma_f^2+ \exp(v_*)-\textbf{k}_*^T\boldsymbol{K}_n^{-1}\textbf{k}_*\label{pred_var}
\end{equation}
where $\boldsymbol{K}_n=\boldsymbol{K}_f+ \text{diag}\left\{\exp(\mathbf{v})\right\}$ is the $n\times n$ covariance matrix of the observations, and $\boldsymbol{K}_f$ is covariance matrix of the $f$-process with elements $[\boldsymbol{K}_f]_{ij} = k_f(\bfx_i, \bfx_j)$. Here, $\textbf{k}_*=\big [k_f(\bfx_1, \bfx_*),\dots, k_f(\bfx_n, \bfx_*) \big ]^T$ is the vector of covariances between $\bfx_*$ and $\mathbf{X}$. For notational simplicity, we suppress explicit conditioning over the locations $\bfx_*$ and $\mathbf{X}$. Since the $f$ and $v$-processes are GPs, the predictive distributions of $f(\bfx_*)$ and $v(\bfx_*)$ are normal and derived analogously to \eqref{pred_mean} and \eqref{pred_var}. For the response surface, the predictive mean $\hat{f}(\bfx_*)=\E [f(\bfx_*)|\textbf{y}, \bm{\theta}, \textbf{v}]$ is identical to $\hat{Y}(\bfx_*)$, while the predictive variance $\V [f(\bfx_*)|\textbf{y}, \bm{\theta}, \textbf{v}]=\sigma^2_f-\textbf{k}_*^T\boldsymbol{K}^{-1}_n\textbf{k}_*$ excludes the observation noise. For the log-noise, let $\boldsymbol{K}_v$ denote the covariance matrix of the $v$-process with $[\boldsymbol{K}_v]_{ij} = k_v(\bfx_i, \bfx_j)$ and $\textbf{k}_{v,*}=[k_v(\bfx_1, \bfx_*), \dots, k_v(\bfx_n, \bfx_*) \big ]^T$. Then $\E [v(\bfx_*)|\bm{\theta}, \textbf{v}]=\mu_0+\textbf{k}_{v,*}^T\boldsymbol{K}_{v}^{-1}(\mathbf{v}-\mu_0)$ and $\V [v(\bfx_*)|\bm{\theta}, \textbf{v}]= \sigma^2_g - \textbf{k}_{v,*}^T\boldsymbol{K}_{v}^{-1}\textbf{k}_{v,*}$.  

In the fully Bayesian setup, both the latent noise ($\textbf{v}$ and $v_*$) and hyperparameters $\bm{\theta}$ are treated as random variables, and therefore the posterior predictive distribution of $\mb{Y}(\bfx_*)$ is obtained by marginalizing over $\bm{\theta}, \textbf{v}$ and $v_*$ with respect to their joint posterior distribution given the observations $\textbf{y}$:
\begin{align}
	p(Y(\bfx_*)|\textbf{y})
	&=\iiint p(Y(\bfx_*)|\textbf{y}, \bm{\theta}, \textbf{v}, v_*) p(v_*, \textbf{v},\theta|\textbf{y}) dv_* d \textbf{v} d\bm{\theta}, \label{pred_eq}  
\end{align}
where the posterior can be expressed as $p(v_*, \textbf{v}, \bm{\theta}| \textbf{y})\propto p(v_*|\textbf{v}, \bm{\theta})p(\textbf{y}| \textbf{v}, \bm{\theta})p(\textbf{v}|\bm{\theta})p(\bm{\theta})$. In this factorization, $p(v_*|\textbf{v}, \bm{\theta})$ is the predictive distribution of $v_*$, $p(\textbf{y}| \textbf{v}, \bm{\theta})$ is the likelihood of the observations $\textbf{y}$, $p(\textbf{v}|\bm{\theta})$ is obtained from the GP prior for the $v$-process, and $p(\bm{\theta})$ is the prior for the GP hyperparameters. Further discussion of prior specification is deferred to Section \ref{prior_specification}.

Empirical Bayes is an alternative approach to marginalization, wherein some or all of the parameters are set to fixed values based on the data, e.g., via \textit{maximum a posteriori} (MAP) estimation. In this setting, an empirical Bayes approach might use MAP point estimates $\bm{\theta} = \hat{\bm{\theta}}$ and $\textbf{v}=\hat{\textbf{v}}$, such that \eqref{pred_eq} is approximated by the plug-in density $p(Y(\bfx_*)|\textbf{y}, \bm{\theta}=\hat{\bm{\theta}}, \textbf{v}=\hat{\textbf{v}}, v_*=\hat{v}_*)$ with $\hat{v}_* = \E [v(\bfx_*)|\hat{\bm{\theta}}, \hat{\textbf{v}}]$; this is commonly done in metamodeling applications \citep{kennedy2001bayesian,yuan2015calibration, ait2025bayesian} to simplify computations. However, point estimates may be insufficient when the likelihood is diffuse or multimodal, as they fail to capture important features of the full posterior distribution  \citep{svensson2015marginalizing}. In sequential design, ignoring the uncertainties in $\bm{\theta}$ and $\textbf{v}$ can lead to sub-optimal design selections. In contrast, our fully Bayesian approach can offer three key advantages: (i) improved predictive accuracy across the design space; (ii) more robust estimation of the latent noise surface; (iii) better uncertainty quantification, with credible intervals that more closely achieve nominal coverage. These advantages are demonstrated in our subsequent numerical illustrations.

\subsection{Sequential Design and Bayesian IMSPE}\label{sec-lookahead}

Evaluating expensive computer simulations is often constrained by limited computational budgets \citep{pellegrino2010fea, chen2020criterion}. When designing experiments for global metamodeling of expensive simulations, the goal is to make best use of the available budget to construct the surrogate model \citep{kleijnen2004application}. Unlike a fixed experimental design where all design points are chosen upfront, sequential strategies, also known as active learning, construct a surrogate model by iteratively adding observations \citep{settles2009active}. This can help optimize the use of computational resources, by exploring design points of interest or high uncertainty conditional on all previously observed data. Sequential design may also be classified according to the number of observations added at each iteration: single-selection procedures add one point per iteration, whereas batch-selection strategies add multiple points simultaneously. Batch sequential design has become increasingly popular because it can improve computational efficiency when it is possible to run several simulations in parallel \citep{chevalier2014corrected, Beek2020ScalableAB, zhang2022batch, Surer02012026}. However, points within the same batch cannot account for the observed responses of each other, as the model is generally only updated after the entire batch has been evaluated. 
Given that we consider computationally expensive simulations where every evaluation is costly, we prioritize the information gain of each observation and focus on single-selection design in this paper.

Sequential strategies are guided by a selection criterion (i.e., for selecting the next point), and thus choosing an appropriate criterion is an important consideration. Criteria for global metamodeling aim to minimize the predictive uncertainty of the surrogate model and are often based on entropy or variance \citep{liu2018survey,fuhg2021state}. Because of the uncertainty quantification properties of GP surrogates (i.e., flexibility in kernel function specification and analytical form of predictive distribution), variance-based criteria are widely used \citep{ gramacy2009adaptive, yuan2013sequential, yuan2015calibration, leatherman2018computer, binois2019replication}. We propose a lookahead sequential strategy inspired by Active Learning Cohn (ALC) \citep{cohn1996active}, which selects the next design point by evaluating the potential reduction in total predictive variance upon its hypothetical inclusion. Our strategy differs from the original ALC in two key ways: (i) we utilize dual GPs to model the mean response surface while accounting for the heteroscedasticity inherent in stochastic simulation models, and (ii) we employ a fully Bayesian framework, treating all quantities, including GP hyperparameters $\bm{\theta}$ and latent noise process $\mathbf{v}$, as random variables governed by their posterior distributions. This leads to the formulation of our Bayesian integrated mean squared prediction error (BIMSPE) metric.

The sequential strategy begins with an initial one-shot design \citep{fuhg2021state}, in which a small set of $n_0$ points  is evaluated to build an initial surrogate model. The sequential process then proceeds iteratively: after $N$ additional sequential observations, we denote the complete dataset with size $N+n_0$ as $D_N = \{(\mathbf{x}_{0:N}, \mathbf{y}_{0:N})\}$, where $D_0=(\mathbf{x}_0, \mathbf{y}_0)$ collectively represents the initial $n_0$ points. Let $\textbf{v}_{0:N}$ denote the vector of latent log-noise  variables corresponding to $\mathbf{x}_{0:N}$ under the heteroscedastic framework.

The original metric that guides selection in ALC is the integrated mean squared prediction error (IMSPE). Given the observed dataset $D_N$, hyperparameter point estimates $\hat{\bm{\theta}}$, and predicted log-variances $\hat{\textbf{v}}_{0:N}$, the MSPE for the mean response $f$ at $\x_*$ is
\begin{align*}
	\text{MSPE}(\x_*|D_N, \hat{\bm{\theta}}, \hat{\textbf{v}}_{0:N}) &=\mathbb{E} \left[ \left(f(\mathbf{\x}_*) - \hat{f}(\mathbf{\x}_*)\right)^2 \big| D_N, \hat{\bm{\theta}}, \hat{\textbf{v}}_{0:N} \right] 
	=\V \left(f(\x_*) | D_N, \hat{\bm{\theta}}, \hat{\textbf{v}}_{0:N} \right), 
\end{align*}
where the posterior mean and variance of $f(\mathbf{\x}_*)$ follow the standard GP expressions in Section \ref{sec-GPR} with $n = n_0 +N$.
Integrating over $\x_* \in \mathcal{X}$ yields the classic IMSPE:
$$\text{IMSPE}(D_N, \hat{\bm{\theta}}, \hat{\textbf{v}}_{0:N})=\int_{\mathcal{X}} \V \left(f(\x_*) | D_N, \hat{\bm{\theta}}, \hat{\textbf{v}}_{0:N} \right) d\x_*.$$
Previous works have shown that this integral has a closed-form expression under a GP prior for common kernel choices \citep{ankenman2010stochastic, leatherman2018computer, binois2019replication}. However, a fully Bayesian treatment of IMSPE becomes more complex, as it requires marginalizing over the posterior of $\bm{\theta}$ and $\textbf{v}_{0:N}$. By applying the law of total variance, we can decompose the MSPE of $f(\x_*)$ into two components:
\begin{equation}\V [f(\x_*)|D_N] = \E_{\bm{\theta}, \textbf{v}_{0:N}| D_N}  \left [\V \left(f(\x_*) | \bm{\theta}, \textbf{v}_{0:N}, D_N \right) \right] + \V_{\bm{\theta}, \textbf{v}_{0:N} | D_N} \left[ \E\left( f(\x_*) | \bm{\theta}, \textbf{v}_{0:N}, D_N \right)\right]. \label{eq:evve}
\end{equation}
The first term is the expected conditional variance of the mean response surface, while the second term accounts for the uncertainty in the posterior mean itself due to unknown $\bm{\theta}$ and $\textbf{v}_{0:N}$. Integrating this expression over $\x_*$ yields our Bayesian IMSPE (BIMSPE):
\begin{equation}
	\text{BIMSPE}(D_N)=\int_{\mathcal{X}} \V [f(\x_*)|D_N] d\x_*.   \label{IEMSPE}
\end{equation}
While IMSPE admits a closed-form expression with respect to integration over $\x_*$ and exact gradients for continuous optimization, BIMSPE does not. Its evaluation requires marginalization over the hyperparameters and latent noise variables; while the integration over these variables and $\x_*$ can theoretically be exchanged to yield a semi-analytic form, that cannot be leveraged for efficient gradient-based optimization as the posterior for $\bm{\theta}$ and $\textbf{v}_{0:N}$ remains analytically intractable. The concept of a ``Bayesian IMSPE'', that accounts for GP parameter uncertainty, was explored in the works of \cite{yuan2013sequential, leatherman2018computer}, albeit in the context of homoscedasticity or where the noise variance is assumed to be a known function of the input. In contrast, here $\V [f(\x_*)|D_N]$ also involves an integral over the posterior of the heteroscedastic noise process $\textbf{v}_{0:N}$. Note that BIMSPE simplifies to the standard (plug-in) IMSPE if the posterior of $\bm{\theta}$ and $\textbf{v}_{0:N}$ is treated as a Dirac delta distribution at the point estimates $\hat{\bm{\theta}}$ and $\hat{\textbf{v}}_{0:N}$; in this case, the second term in the law of total variance vanishes, leaving only the conditional variance term used in traditional ALC.

To bypass the difficulty of continuous optimization in the fully Bayesian setting, we adopt an empirical approximation by averaging the variances over the representative discretization $\mathcal{D}$ assumed in Section \ref{sec-GPR}, i.e., $\int_{\mathcal{X}}\V \left[f(\mathbf{\x}_*) \mid D_N \right] d\bfx_* \approx  \frac{1}{|\mathcal{D}|} \sum_{\mathbf{\x}_* \in \mathcal{D}} \V \left[f(\mathbf{\x}_*) \mid D_N \right]$. Such an approach, often known as ``empirical IMSPE'', has been widely utilized in sequential design strategies for neural networks \citep{cohn1993neural, seo2000gaussian, kleijnen2004application} and stochastic simulation \citep{ankenman2010stochastic}. The accuracy of this approximation inherently depends on the quality of the discretization. Up to a moderate input dimension (e.g., $d=5$ as considered in Section~\ref{sec-5-d}), a sufficiently dense space-filling design for $\mathcal{D}$ can be feasible for relatively smooth functions. As for the integrals over the posteriors of $\bm{\theta}$ and $\textbf{v}_{0:N}$, we employ Monte Carlo methods, which are discussed in detail in Section \ref{sec:calcBIMSPE} in the context of leveraging BIMSPE for lookahead sequential design.

\subsection{Computing the Expected BIMSPE Selection Criterion} \label{sec:calcBIMSPE}

Next, we consider how the subsequent design point $\bfx_{N+1}$ is selected for evaluation given $D_N$, using a sequential strategy based on the BIMSPE. Let $\tilde{\bfx}_{N+1}$ be a candidate point for hypothetical inclusion, with $\tilde{y}_{N+1}$ as the corresponding hypothetically observed value and $\tilde{v}_{N+1}$ as the hypothetical log noise. The total posterior predictive variance after the hypothetical inclusion of the candidate point may then be denoted as $\text{BIMSPE} \big(D_N \cup  \{(\tilde{\bfx}_{N+1}, \tilde{y}_{N+1}) \} \big)$. Importantly, we evaluate this metric using the current posterior $p(\bm{\theta}, \textbf{v}_{0:N}, \tilde{v}_{N+1}  | D_N)$, i.e., the distributions of the GP hyperparameters and latent noise are \textit{not} updated with the hypothetical point, as $(\tilde{\bfx}_{N+1}, \tilde{y}_{N+1})$ does not represent observed data. Then, our sequential strategy selects $\bfx_{N+1}$ by minimizing the expected BIMSPE selection criterion:
\begin{equation}
	\mathbf{x}_{N+1} = \underset{\tilde{\mathbf{x}}_{N+1} \in \mathcal{D}}{\arg\min} \ \mathbb{E}_{\tilde{y}_{N+1} \mid D_N} \left[ \text{BIMSPE} \big(D_N \cup \{(\tilde{\mathbf{x}}_{N+1}, \tilde{y}_{N+1})\}\big) \right]. \label{RC}
\end{equation}
The expectation over $\tilde{y}_{N+1}$ is natural in this Bayesian context, since $\tilde{y}_{N+1}$ influences the BIMSPE through the second term (variance of the posterior means) in \eqref{eq:evve}. This fully Bayesian treatment contrasts with the plug-in ``kriging believer'' \citep{ginsbourger2010kriging}, which substitutes the posterior predictive mean at $\tilde{\bfx}_{N+1}$ for the hypothetical observation.

For practical computation, we begin with the law of total variance decomposition
\begin{align}
	&\mathbb{E}_{\tilde{y}_{N+1} \mid D_N} \left[ \text{BIMSPE} \big(D_N \cup \{(\tilde{\mathbf{x}}_{N+1}, \tilde{y}_{N+1})\}\big) \right]
	\approx \E_{\tilde{y}_{N+1}|D_N} \bigg[ \frac{1}{|\mathcal{D}|} \sum_{\x_* \in \mathcal{D}}  \V \left[f(\x_*)|D_{N}, \tilde{y}_{N+1} \right]\bigg]
	\nonumber\\
	&= \frac{1}{|\mathcal{D}|} \sum_{\x_* \in \mathcal{D}} \bigg(  \underbrace{\E_{\bm{\theta}, \textbf{v}_{0:N}, \tilde{v}_{N+1}| D_N}  \left [\V \left(f(\x_*) | \bm{\theta}, \textbf{v}_{0:N}, \tilde{v}_{N+1}, D_N \right) \right]}_{\text{C1}} + \nonumber \\
	&\quad\quad\quad \underbrace{\E_{\tilde{y}_{N+1}|D_N} \left[ \V_{\bm{\theta}, \textbf{v}_{0:N}, \tilde{v}_{N+1} | D_N} \left[ \E\left( f(\x_*) | \bm{\theta}, \textbf{v}_{0:N}, \tilde{v}_{N+1}, D_N, \tilde{y}_{N+1} \right)\right]  \right]}_{\text{C2}} \bigg), 
	\label{eq:seqBIMSPE}
\end{align}
where the inner terms $\E \left(f(\x_*) \mid \cdot \right)$ and  $\V \left(f(\x_*) \mid \cdot \right)$ are the closed-form GP predictive mean and variance following the expressions in Section \ref{sec-GPR}, but now also conditioning on the hypothetical  $\tilde{y}_{N+1}$, $\tilde{\mathbf{x}}_{N+1}$, and  $\tilde{v}_{N+1}$. These can be computed efficiently starting from values based on $D_N$ via standard GP update formulas based on the Woodbury identity \citep{williams2006gaussian}. Next, by denoting $\tilde{\mu}_* = \E\left( f(\x_*) | \bm{\theta}, \textbf{v}_{0:N}, \tilde{v}_{N+1}, D_N, \tilde{y}_{N+1} \right)$, we may further write $\text{C2}= \E_{\tilde{y}_{N+1}|D_N} \left[ \E_{\bm{\theta}, \textbf{v}_{0:N}, \tilde{v}_{N+1} | D_N} \left(\tilde{\mu}_*^2 \right) - \left\{ \E_{\bm{\theta}, \textbf{v}_{0:N}, \tilde{v}_{N+1} | D_N} \left(\tilde{\mu}_*  \right) \right\}^2 \right]$. Note that the term C1 in \eqref{eq:seqBIMSPE} corresponds to the expected conditional variance of standard IMSPE, while C2 is the additional term that results from the fully Bayesian treatment.

To approximate the expectations associated with \eqref{eq:seqBIMSPE}, suppose $\{(\bm{\theta}^{(m)}, \textbf{v}^{(m)}_{0:N}) \}_{m=1}^{M}$ is a properly weighted sample \citep[e.g., see][]{liu1998sequential}, with normalized weights $\{\tilde{w}^{(m)}_N\}_{m=1}^{M}$, that represent the current posterior distribution $p(\bm{\theta}, \mb{\textbf{v}_{0:N}}|\textbf{y}_{0:N})$. If $\tilde{\bfx}_{N+1}$ is a design point that has not been previously observed, we draw a sample $\tilde{v}^{(m)}_{N+1}$ from the predictive distribution $p(\tilde{v}_{N+1} | \bm{\theta}^{(m)}, \textbf{v}^{(m)}_{0:N})$ for each $m = 1, \ldots, M$; if $\tilde{\bfx}_{N+1}$ is a replicate of a previous design point, we re-use the corresponding posterior sample from $\textbf{v}^{(m)}_{0:N}$ as $\tilde{v}^{(m)}_{N+1}$. Then we draw $K$ samples $\{\tilde{y}_{N+1}^{(k)}\}_{k=1}^K$ from the posterior predictive distribution $p(\tilde{y}_{N+1} | D_N)$, which is of the form in \eqref{pred_eq}. This is achieved by first sampling an index $m^* \in \{1, \dots, M\}$ with probability $\tilde{w}^{(m^*)}_N$, and then drawing $\tilde{y}_{N+1}^{(k)}$ from the normal distribution with mean and variance specified by \eqref{pred_mean} and \eqref{pred_var}, where the latter uses the sampled observation noise $\exp(\tilde{v}^{(m^*)}_{N+1})$. Finally, estimates of the C1 and C2 terms in \eqref{eq:seqBIMSPE} for candidate point $\tilde{\bfx}_{N+1}$ and one $\bfx_*$ are given by
\begin{equation}
	\widehat{C1}=\sum_{m=1}^{M} \tilde{w}^{(m)}_N  \V \left(f(\x_*) | \bm{\theta}^{(m)}, \textbf{v}_{0:N}^{(m)}, \tilde{v}^{(m)}_{N+1}, D_N \right), \label{C1_hat}
\end{equation}
\begin{equation}
	\widehat{C2}= \frac{1}{K}\sum_{k=1}^{K} \left[ \sum_{m=1}^{M}  \tilde{w}^{(m)}_N  \left(\tilde{\mu}^{(m,k)}_* \right)^2 -\bigg (\sum_{m=1}^{M} \tilde{w}^{(m)}_N \tilde{\mu}^{(m,k)}_* \bigg )^2\right],\label{C2_hat}
\end{equation}
where $\tilde{\mu}^{(m,k)}_* = \E\left( f(\x_*) | \bm{\theta}^{(m)}, \textbf{v}_{0:N}^{(m)}, \tilde{v}_{N+1}^{(m)}, D_N, \tilde{y}^{(k)}_{N+1} \right)$. The average of $\widehat{C1} + \widehat{C2}$ over all $\bfx_* \in \mathcal{D}$ completes our approximation of the expected BIMSPE selection criterion in \eqref{eq:seqBIMSPE}.

\subsection{Posterior Update via Sequential Importance Sampling} \label{sec-para_update}

Building upon the initial design, the sequential strategy selects and evaluates one design point every iteration. Each iteration consists of three main steps. At the $N$-th iteration, we first execute the lookahead step for acquisition, where $\mathbb{E}_{\tilde{y}_{N+1} \mid D_N} \left[ \text{BIMSPE} \big(D_N \cup \{(\tilde{\mathbf{x}}_{N+1}, \tilde{y}_{N+1})\}\big) \right]$ is estimated and minimized over $\mathcal{D}$ to select the next design point $\bfx_{N+1}$ for evaluation. Second, the simulation model is evaluated, thereby generating the observation $y_{N+1}$. The third step is the posterior update, which generates updated Monte Carlo samples from $p(\bm{\theta}, \textbf{v}_{0:N+1} | \textbf{y}_{0:N+1})$ that are used for the subsequent lookahead acquisition step. The sequential strategy terminates after $B$ iterations, where $B$ represents the total budget for the number of simulation model evaluations to perform beyond the initial design.

For the posterior update step, we avoid the prohibitive cost of running the MCMC sampler at every sequential iteration by updating our representation of the posterior distribution via sequential importance sampling (SIS). Briefly, SIS approximates expectations or integrals with respect to a target distribution via a weighted average of samples drawn from a convenient proposal distribution, and sequentially updates estimates by re-weighting posterior samples from previous iterations; see \citet{doucet2001introduction}. We can then perform expected BIMSPE calculations for subsequent iterations using existing samples with sequentially updated weights. To mitigate weight degeneracy, we monitor the effective sample size (ESS) of the importance weights \citep{kong1994sequential}, and use MCMC to obtain a fresh set of posterior samples only when the ESS falls below a pre-defined threshold.

Starting from the initial observed dataset $D_0$, we draw $M$ posterior samples $\{\bm{\theta}^{(m)}, \textbf{v}^{(m)}_0\}_{m=1}^{M}$ from $p(\bm{\theta}, \textbf{v}_0|D_0)$ via MCMC, with initial weights $\tilde{w}^{(m)}_0=1/M$. Now suppose at the $N$-th iteration we have properly weighted samples from $p(\bm{\theta}, \mb{\textbf{v}_{0:N}}|\textbf{y}_{0:N})$, namely $\{\bm{\theta}^{(m)},\textbf{v}^{(m)}_{0:N}) \}_{m=1}^{M}$ with weights $\{\tilde{w}^{(m)}_N\}_{m=1}^{M}$. In the acquisition step, for each candidate point $\tilde{\bfx}_{N+1} \in \mathcal{D}$, we obtain the draws $\{ \tilde{v}^{(m)}_{N+1} \}_{m=1}^M$ and $\{\tilde{y}_{N+1}^{(k)}\}_{k=1}^K$ as described in Section \ref{sec:calcBIMSPE}, and calculate the expected BIMSPE selection criterion using the estimates \eqref{C1_hat} and \eqref{C2_hat}. After locating the candidate point that minimizes the selection criterion as in \eqref{RC}, we evaluate the simulation model at the selected point $\bfx_{N+1}$ and observe $y_{N+1}$. Recall that the candidate set $\mathcal{D}$ is kept fixed throughout, i.e., even after a point has been selected, so this criterion naturally handles replication; $\mathbf{x}_{N+1}$ is a replicate if it corresponds to a previously observed location. Simultaneously, the draws $\{\tilde{v}_{N+1}^{(m)}\}_{m=1}^{M}$ generated during the lookahead at the chosen $\tilde{\x}_{N+1}$ are concatenated with the existing $\{\textbf{v}^{(m)}_{0:N}\}_{m=1}^{M}$ samples to form $\textbf{v}^{(m)}_{0:N+1}=(\textbf{v}^{(m)}_{0:N}, \tilde{v}^{(m)}_{N+1})$; the current weight $\tilde{w}^{(m)}_N$ is then associated with the joint $(\bm{\theta}^{(m)}, \textbf{v}^{(m)}_{0:N+1})$ sample for $m = 1, \ldots, M$.

To then perform the posterior update step via SIS, we decompose the target distribution as $p(\bm{\theta},\textbf{v}_{0:N+1} |\textbf{y}_{0:N+1}) \propto p(y_{N+1}|\textbf{v}_{0:N+1},  \bm{\theta}, \textbf{y}_{0:N})p(\textbf{v}_{0:N+1}, \bm{\theta}|\textbf{y}_{0:N})$. Here, $p(\textbf{v}_{0:N+1},  \bm{\theta}|\textbf{y}_{0:N})$ is treated as the proposal distribution and $p(y_{N+1}|\textbf{v}_{0:N+1}, \bm{\theta}, \textbf{y}_{0:N})$ is the incremental importance weight, i.e., the normal likelihood of the observation $y_{N+1}$. Then the $m$-th un-normalized weight at iteration $N+1$ is computed as $w^{(m)}_{N+1}  = \tilde{w}_{N}^{(m)}~ p(y_{N+1}|\textbf{v}^{(m)}_{0:N+1}, \bm{\theta}^{(m)}, \textbf{y}_{0:N})$, and the updated normalized weights are obtained as $\tilde{w}^{(m)}_{N+1}={w^{(m)}_{N+1}}/{\sum_{j=1}^{M}w^{(j)}_{N+1}}$. If the ESS of $\{\Tilde{w}^{(m)}_{N+1} \}_{m=1}^{M}$ falls below a specified threshold $\tau M$, we run MCMC to draw $M$ new samples from $p(\bm{\theta},\textbf{v}_{0:N+1} |\textbf{y}_{0:N+1})$ and reset the weights $\{\tilde{w}^{(m)}_{N+1}\}_{m=1}^{M}$ to $1/M$ for the acquisition step of the next iteration. While we have written $\textbf{v}_{0:N+1}$  for notational consistency, in practice we only need sample $\mathbf{v}$ over the set of \textit{unique} design locations in the dataset for computational efficiency, and these latent draws are shared among replicates at each location. Our complete proposed sequential strategy is summarized in Algorithm \ref{alg1}. 

\begin{algorithm}[!htbp]
	\caption{Fully Bayesian Sequential Design Strategy with BIMSPE and SIS }\label{algorithm}
	\begin{algorithmic}[1] 
		\State \textbf{Input:} Initial design $D_0$, discretized set of design points $\mathcal{D}$, budget $B$, number of samples $K$ and $M$, SIS threshold $\tau$
		\State Draw $M$ samples $\{(\bm{\theta}^{(m)}, \mathbf{v}_{0}^{(m)})\}_{m=1}^{M}$ from $p(\bm{\theta}, \mathbf{v}_0 \mid D_0)$ via MCMC  \Comment{Initialization}
		\State Set initial weights $\tilde{w}^{(m)}_0 = 1/M$ for $m=1, \dots, M$
		\For{$N = 0$ to $B-1$} \Comment{Sequential design iterations}
		\For{$\tilde{\mathbf{x}}_{N+1} \in \mathcal{D}$} \Comment{Lookahead acquisition step}
		\State Draw $K$ samples of $\tilde{y}_{N+1}$ from $p(\tilde{y}_{N+1} \mid D_N)$
		\If{$\tilde{\mathbf{x}}_{N+1}$ is a new design point}
		\State Draw $\tilde{v}_{N+1}^{(m)} \sim p(v_{N+1} \mid \mathbf{v}^{(m)}_{0:N}, \bm{\theta}^{(m)})$ for $m=1, \dots, M$
		\Else{ if $\tilde{\mathbf{x}}_{N+1}$ is a replicate}
		\State Set $\tilde{v}_{N+1}^{(m)}$ to the existing draw of $v$ at that design point for $m=1, \dots, M$
		\EndIf
		\State Estimate $\mathbb{E}_{\tilde{y}_{N+1} \mid D_N} \left[ \text{BIMSPE} \big(D_N \cup \{(\tilde{\mathbf{x}}_{N+1}, \tilde{y}_{N+1})\}\big) \right]$ using \eqref{C1_hat} and \eqref{C2_hat}
		\EndFor
		\State Select $\mathbf{x}_{N+1}$ using \eqref{RC} and observe $y_{N+1}$ \Comment{Evaluation step}
		\State $D_{N+1} = D_N \cup \{(\mathbf{x}_{N+1}, y_{N+1})\}$   
		\State Concatenate $\mathbf{v}_{0:N+1}^{(m)} = (\mathbf{v}^{(m)}_{0:N}, \tilde{v}_{N+1}^{(m)})$ using $\tilde{v}_{N+1}^{(m)}$ corresponding to selected $\mathbf{x}_{N+1}$ 
		\State $w_{N+1}^{(m)} = \tilde{w}_{N}^{(m)} \cdot p(y_{N+1} \mid \mathbf{v}^{(m)}_{0:N+1}, \bm{\theta}^{(m)}, D_N)$ for each $m$  \Comment{Posterior update step}
		\State Normalize $\tilde{w}_{N+1}^{(m)} = w_{N+1}^{(m)} / \sum_{j=1}^{M} w_{N+1}^{(j)}$ and compute $\text{ESS} = \left( \sum_{j=1}^{M} (\tilde{w}_{N+1}^{(j)})^2 \right)^{-1}$
		\If{$\text{ESS} < \tau M$} 
		\State Run MCMC to draw fresh $\{(\bm{\theta}^{(m)}, \mathbf{v}_{0:N+1}^{(m)})\}_{m=1}^{M}$  from $p(\bm{\theta}, \mathbf{v}_{0:N+1} \mid D_{N+1})$
		\State Reset weights $\tilde{w}^{(m)}_{N+1} = 1/M$
		\EndIf
		\EndFor
		\State \Return Final posterior samples $\{(\bm{\theta}^{(m)}, \mathbf{v}_{0:B}^{(m)})\}_{m=1}^{M}$ and weights $\{\tilde{w}^{(m)}_{B}\}_{m=1}^{M}$
	\end{algorithmic}  \label{alg1}
\end{algorithm}

To analyze the computational complexity of the $N$-th sequential iteration, suppose that the size of the initial design $n_0 \ll N$. In the limited budget setting, we may consider that the number of design space points necessary to capture all meaningful variability in the simulation model is large relative to $B$, so that $| \mathcal{D} | \gg B \ge N$. The lookahead step begins by pre-computing posterior predictive quantities, which is dominated by the $\mathcal{O}(N^3)$ Cholesky factorization and repeated for each of the $M$ posterior samples to yield a $\mathcal{O}(MN^3)$ cost. Obtaining predictions for each candidate point by reusing the Cholesky factors is $\mathcal{O}(N^2)$, yielding a $\mathcal{O}(MN^2 |\mathcal{D}|)$ cost over all $M$ samples and candidate points. For each candidate, completing the expected BIMSPE calculation then requires aggregating predictions across $|\mathcal{D}|$ points, $M$ samples, and $K$ draws of $\tilde{y}_{N+1}$ for $\mathcal{O}(MK |\mathcal{D}|^2)$ cost. The total cost of the lookahead step per iteration is thus $\mathcal{O}(MN^3 + MN^2 |\mathcal{D}| + MK |\mathcal{D}|^2)$. For the posterior update step, the computational cost of MCMC is only incurred if $\text{ESS} < \tau M$. In that case, the cost of each gradient evaluation in NUTS is $\mathcal{O}(N^3)$, as we maintain a full-sample representation (without collapsing replicates) for computing the exact joint posterior under heteroscedasticity. Crucially, the total lookahead cost is dominated by $\mathcal{O}(MN^3 + MN^2 |\mathcal{D}|)$, with $\mathcal{O}(MN^2 |\mathcal{D}|)$ being the main bottleneck when $| \mathcal{D} | \gg N$; this term is quadratic in $N$ but only linear in $|\mathcal{D}|$. This emphasizes the utility of our approach for scenarios where the cost of simulation evaluations outweighs the overhead of fully Bayesian sequential design. In such cases, we maintain the feasibility to search a large design space (e.g., $|\mathcal{D}|$ up to a few thousand points) and effectively extract information from a limited number of expensive observations.

\subsection{Implementation Details}\label{sec-Implementation Details}

\subsubsection{Initialization and Sampling Details \label{initialization & sampling}}
Recall that $\mathcal{D}$ is chosen to provide a representative discretization of the design space. By default, we may define $\mathcal{D}$ as a regularly-spaced grid in low (1-D and 2-D) dimensions, for ease of interpretability. For higher-dimensional problems ($d>2$), we may instead use a Latin Hypercube Design (LHD) for $\mathcal{D}$ to maintain efficient coverage of the design space. The initial design then chooses $n_0$ points in $\mathcal{D}$ at which to make the first observations $D_0$. A widely-used method for defining the initial design is a one-shot space-filling approach. In this work, we employ the maximin-distance LHD \citep{johnson1990minimax} within points in $\mathcal{D}$, which provides a preliminary exploration by spreading out the initial observations across the input space. To determine the size of this initial design, a common rule of thumb is $n_0=10d$, as first proposed by \cite{jones1998efficient} and later supported by \cite{loeppky2009choosing} for GP models. We adopt this rule as a minimum requirement for $n_0$.

When MCMC needs to be run for the posterior update step, we employ the No-U-Turn Sampler \citep[NUTS,][]{hoffman2014no}, which can be viewed as an extension of Hamiltonian Monte Carlo and is the default sampler in \textit{RStan}. To facilitate faster convergence, we warm start the sampler using the final draw from the previous MCMC run. We generate a single chain with a total of 1,000 MCMC iterations, discarding the first 500 as burn-in; the remaining $M = 500$ post–burn-in draws are used to estimate \eqref{C1_hat} and~\eqref{C2_hat}. For the expectation over $\tilde{y}_{N+1}$ in the lookahead step, empirical analysis suggests that a single draw $K= 1$ is sufficient and yields designs that are indistinguishable from those produced with larger values of $K$. See Supplement~B.3 for a comparison of $K=1$ and $K=20$; we find that while larger $K$ values tend to produce a smoother acquisition surface, the location of the resulting minimum remains fairly robust. Finally, a small jitter of $10^{-6}$ is added to the diagonal of the log-noise covariance matrix $\boldsymbol{K}_v$ to ensure numerical stability of the Cholesky decompositions.

As introduced in Section \ref{sec-para_update}, our sequential strategy incorporates SIS, which requires a choice of the threshold $\tau$. Smaller values of $\tau$ tend to allow a larger number of fast SIS posterior updates before a more costly MCMC run is triggered. However, an excessively low threshold for weight degeneracy can reduce the accuracy of the posterior approximation. We assessed the practical effect of different ESS thresholds $\tau M$ (with $\tau=0.2, 0.5, 1$) in Supplement A. Our results indicate that $\tau=0.2$ is sufficiently accurate, yielding predictive performances that are indistinguishable from the case where $\tau = 1$ (i.e., re-running MCMC after every sequential design iteration).

Parallel computing in the lookahead step is an essential feature of our sequential strategy. Specifically, the estimation of the expected BIMSPE (lines 6-12 of Algorithm \ref{alg1}) is independent for each candidate design point $\tilde{\bfx}_{N+1}$ and can be executed simultaneously across multiple cores. This parallelization notably reduces the overhead of fully Bayesian sequential design, making the strategy viable for values of $| \mathcal{D}|$ up to a few thousand points.

\subsubsection{Prior Distributions for Hyperparameters}\label{prior_specification}
The heteroscedastic GP model involves the hyperparameters $\boldsymbol{\theta}=\{ \mu_0, \sigma_f, \sigma_v, \textbf{l}_f, \textbf{l}_v\}$. To complete a fully Bayesian specification, priors need to be assigned for $\boldsymbol{\theta}$. We assume independent priors for the hyperparameters, namely $p(\boldsymbol{\theta})=p(\mu_0)p(\sigma_f)p(\sigma_v)p(\textbf{l}_f)p(\textbf{l}_v)$, which adopt the following forms:
\begin{align}
	&\mu_0 \sim \text{Normal}(0, s_0),~ 
	\sigma_f \sim \text{Half-Cauchy}(0, s_f),~
	\sigma_v \sim \text{Half-Normal}(0, s_v); \nonumber \\
	&l_{fh} \sim \text{Inv-Gamma}(\alpha_{fh}, \beta_{fh}),~
	l_{vh} \sim \text{Inv-Gamma}(\alpha_{vh}, \beta_{vh}), \text{ for }h=1,\dots, d.
	\label{eq:priorh}
\end{align}
In the absence of specific prior information, we recommend choosing values in \eqref{eq:priorh} such that the resulting priors are weakly informative. The hyperparameter $\mu_0$ represents the average log-noise level, so a diffuse normal prior with SD $s_0 = 5$ is a robust default that allows the model to accommodate a wide range of noise scales. For the variance parameters $\sigma^2_f$ and $\sigma^2_v$, we follow \citet{gelman2006prior} and place either Half-Normal or Half-Cauchy priors on their corresponding SDs. Since $\sigma_f$ relates to the scale of the observations, a heavier-tailed Half-Cauchy with $s_f = 5$ is used to allow flexibility in the $f$-process. In contrast, since $\sigma_v$ controls the variability of the \textit{log}-noise, a more concentrated Half-Normal prior with $s_v = 1$ can be a reasonable choice to prevent extreme heteroscedasticity. Lastly, we employ inverse-gamma priors for the GP lengthscales \citep{van2009adaptive}, which prevent the lengthscales from shrinking to zero. We set $\alpha_{fh} = \alpha_{vh} = 3$ for each $h = 1, \ldots, d$, which ensure the priors have finite variance; we set the scale parameters $\beta_{vh} = \beta_{fh}$ to be 0.6 times the range of the design space in the $h$-th dimension of $\mathcal{X}$. Together, this implies the lengthscales have a prior mode at 0.15 times the range and a prior mean at 0.3 times the range, while being sufficiently broad to be informed by the data. To ensure that our results are robust, we assess the sensitivity of the prediction performance to the prior specifications; further details are provided in Supplement B.1.

\section{Synthetic Examples and Results}\label{sec-experiments}

This section presents examples with synthetic simulation models to illustrate our proposed approach. We compare our fully Bayesian sequential strategy using BIMSPE and SIS (FB-BIM) to several established alternatives. First, we consider an empirical Bayes variant (EB-EIM) that employs MAP point estimates for GP hyperparameters and latent noise; with this setup, the expected BIMSPE criterion simplifies to the empirical (discretized) IMSPE with plug-in hyperparameters, allowing us to assess the impact of ignoring hyperparameter and noise uncertainty in model fitting and design selection. Second, we consider a variational inference variant (VI-BIM), which maintains the BIMSPE and SIS framework but replaces MCMC sampling with a VI approximation to the posterior to reduce computational cost. Third, we consider a homoscedastic noise variant (Homo-BIM), which retains the fully Bayesian BIMSPE criterion but assumes a constant noise variance across the input space. This allows us to quantify the importance of accounting for heteroscedasticity for mean surface estimation. Finally, we consider the state-of-the-art hetGP method \citep{binois2019replication} with an adaptive horizon, as implemented in the ``hetGP'' R package \citep{binois2021hetgp}; additional hetGP results with fixed horizon values $h=0,\dots,4$ are included in Supplement I, where higher horizon values place greater emphasis on replication. Full descriptions and implementation details for each of these alternative strategies are provided in Supplement F. 

We evaluate the performance of each strategy over the $B$ allotted sequential iterations following the initial design. Each experiment is repeated for 100 independent macro-replications.  To quantify the prediction accuracy of the mean response surface, we compute the root mean squared prediction error (RMSE) over all points in $\mathcal{D}$, defined as $RMSE^{(m)}_f=\sqrt{\frac{1}{| \mathcal{D} |}\sum_{\bfx_i \in \mathcal{D}} \left[\hat{f}_m(\bfx_i)-f(\bfx_i) \right ]^2}$,
where $\hat{f}_m(\bfx_i)$ is the GP-based predictive  mean for the $m$-th macro-replication ($m=1,\dots, 100$) at point $\bfx_i$, and $f(\bfx_i)$ is the true mean response. Similarly, the accuracy of the noise surface estimation is assessed using the RMSE for log-noise, defined as $RMSE^{(m)}_{v}=\sqrt{\frac{1}{| \mathcal{D} |}\sum_{\bfx_i \in \mathcal{D}} \left[ \hat{v}_m(\bfx_i)-\log \left(g(\bfx_i) \right) \right]^2}$, where $\hat{v}_m(\bfx_i)$ is the GP-based predictive mean log-noise for the $m$-th macro-replication, and $\log \left(g(\bfx_i) \right)$ is the true log-noise. To assess uncertainty quantification, we compute the empirical coverage for both $f$ and $v$. For each macro-replication $m$, the coverage is defined as 
$\text{coverage}^{(m)}=\frac{1}{| \mathcal{D} |} \sum_{\bfx_i \in \mathcal{D}} \mathbbm{1}\left( \text{true value}_i\in [L_i^{(m)}, U_i^{(m)}]\right)$, where $\mathbbm{1}\left( \cdot \right)$ is the indicator function, and $[L_i^{(m)}, U_i^{(m)}]$ is the 95\% interval at $\bfx_i$. For fully Bayesian methods (FB-BIM, Homo-BIM), these are credible intervals constructed based on empirical quantiles of MCMC draws. For VI-BIM, intervals are derived from the variational posterior, while for EB-EIM these are obtained from a Laplace approximation at the MAP estimate. Finally, for hetGP, we construct normal-based confidence intervals using the predictive means and variances provided by the package. All strategies are initialized with the same random seed in each macro-replication $m$, so their initial designs $D_0$ are identical to facilitate fair comparison. We record the four performance metrics at each iteration of each macro-replication. We conduct two-sided two-sample $t$-tests at a significance level of $\alpha = 0.05$ to determine whether the mean $RMSE_f$ and $RMSE_v$ metrics obtained by FB-BIM are significantly different from those of the alternative strategies.

Each example in this section adopts the priors described in Section \ref{prior_specification}. A further ablation study comparing expected BIMSPE to the plug-in empirical IMSPE as a selection criterion, with all other settings held fixed, is provided in Supplement C. The R code that implements all strategies for each example is provided in the supplementary .zip file.

\subsection{Illustrative One-dimensional Test Function}\label{sec-1-d}

\begin{figure}[!htbp]
	\captionsetup[subfigure]{labelformat=empty}
	\begin{subfigure}[b]{0.49\linewidth}
		\centering
		\includegraphics[width=1\textwidth, height=0.6\textwidth]{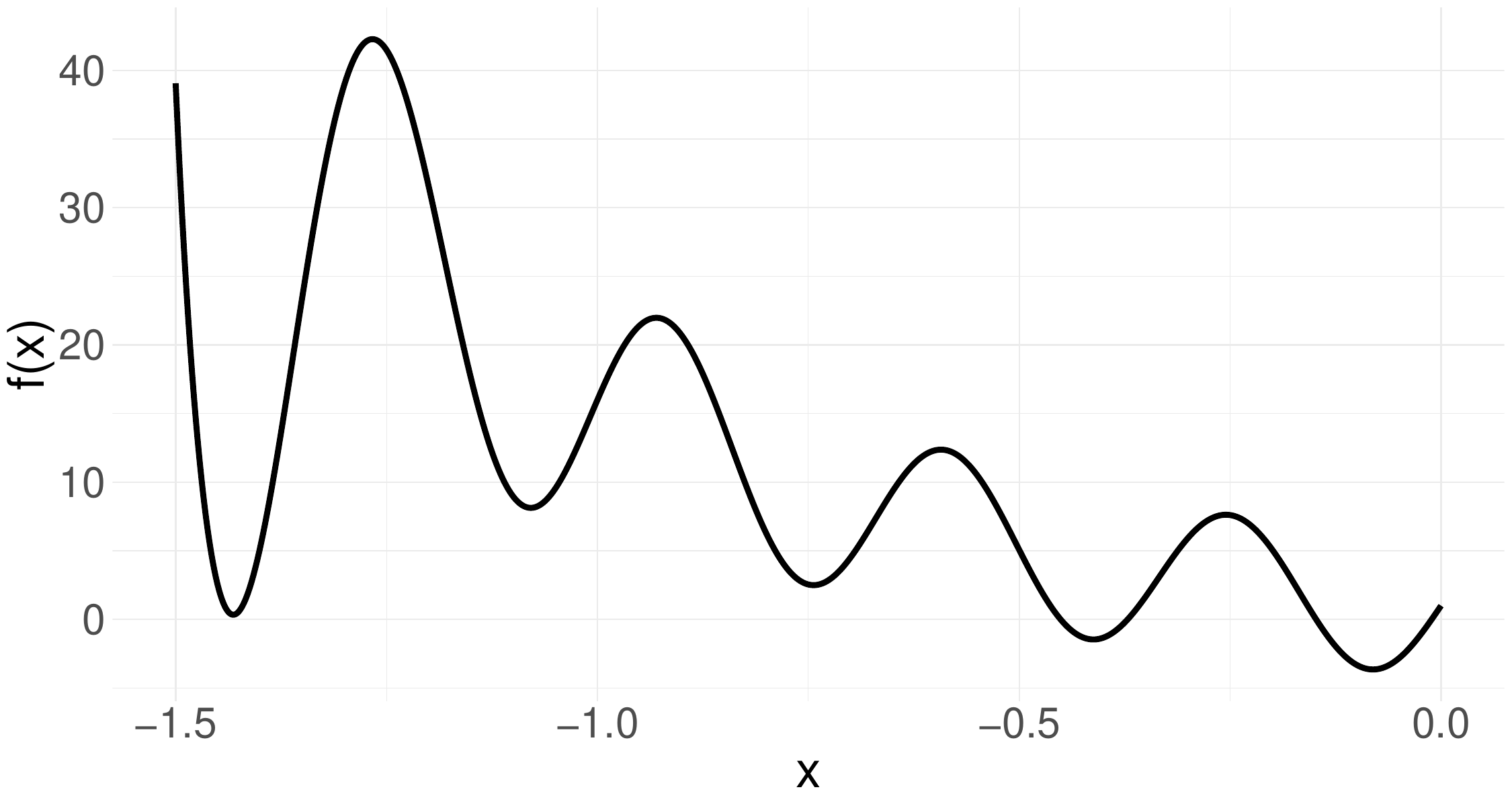}
		\caption{(a) Mean response surface $f(x)$ }
	\end{subfigure} 
	\begin{subfigure}[b]{0.49\linewidth}
		\centering
		\includegraphics[width=1\textwidth, height=0.6\textwidth]{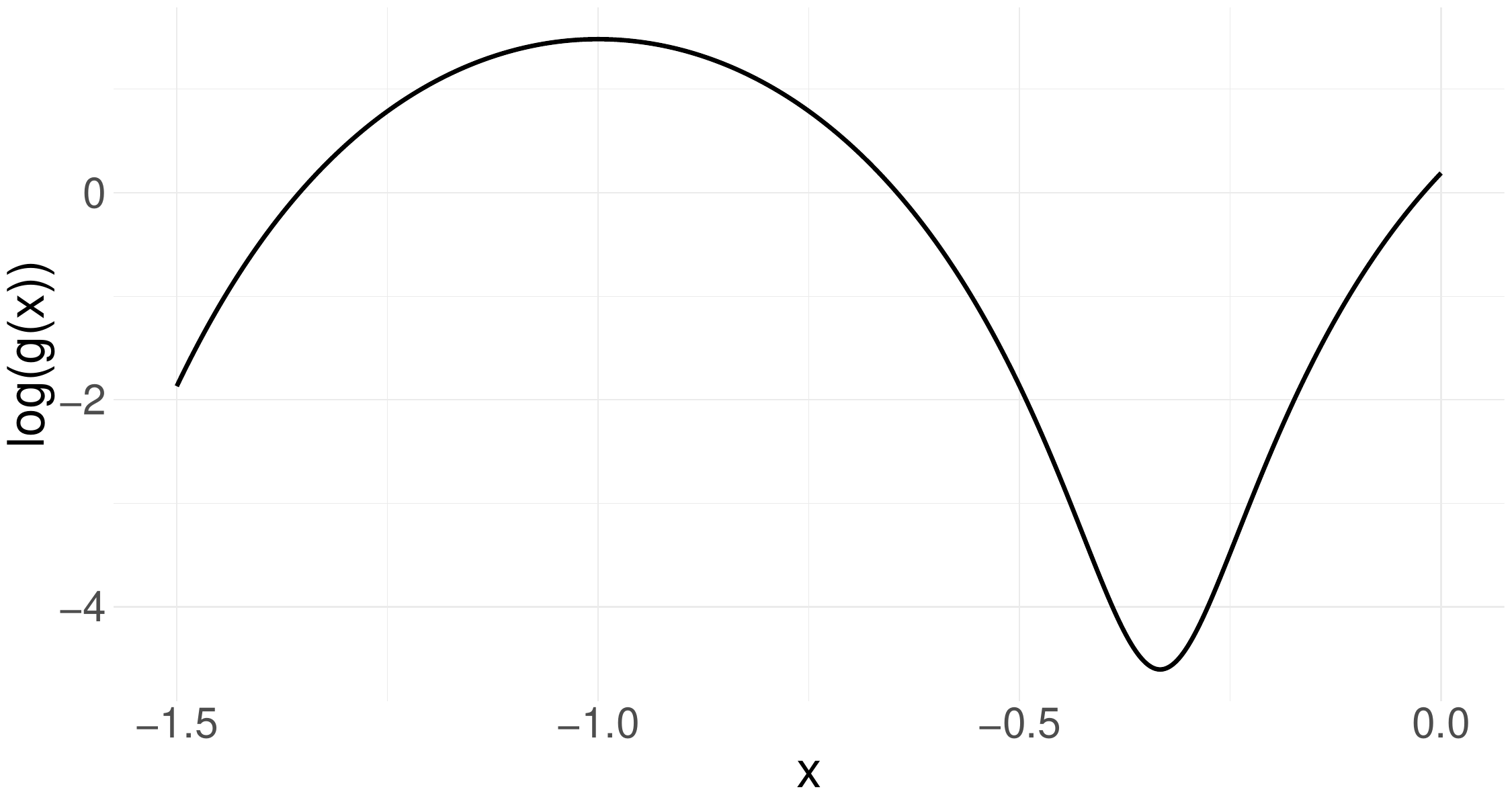}
		\caption{(b) Log noise variance $log (g(x))$}
	\end{subfigure} 
	\caption{Visualizations of the mean response surface and the log-noise variance
		surface of the 1-D example.} \label{1-d example illustration}
\end{figure}

We consider the following 1-D example inspired by the test function given by \cite{gramacy2012cases}. The true mean response surface is $f(x)= \dfrac{5\text{sin}(6\pi x)}{\text{cos}(x)}+(x-1)^4$ and the noise function is  $g(x)=\big( \text{sin}(1.5 \pi x)+1.1 \big)^2$ for $\bfx \in [-1.5, 0]$; see Figure \ref{1-d example illustration}. The design grid $\mathcal{D}$ is comprised of 151 unique points, evenly spaced within the design space with a stride of 0.01. The noise magnitude, measured as $\sqrt{g(\bfx)/R_f}$, where $R_f$ is the range of the response, spans from 1.5\% to 31\% with an average of 16\% across $\mathcal{D}$, which is considered relatively light and even. The initial design consists of $n_0=16$ uniformly spaced design points without replicates, and the total budget allocated for the sequential design is $B=100$ iterations. 

\begin{figure}[!htbp]
	\centering
	\includegraphics[width=1\linewidth]{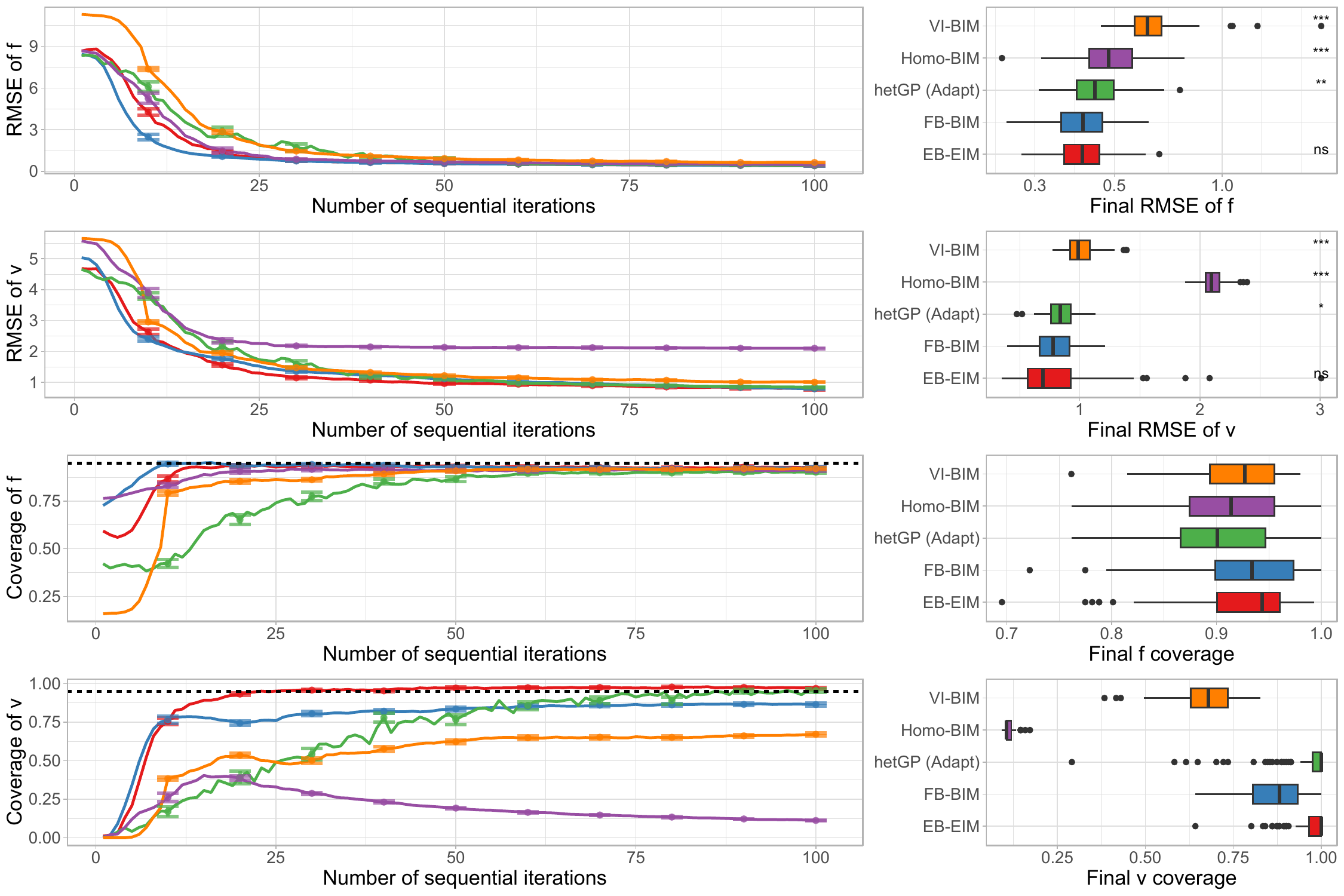}
	\caption{Line plots of prediction performance (left) and final-iteration boxplots (right) for VI-BIM, Homo-BIM, hetGP (Adapt), FB-BIM, and EB-EIM on the 1-D example, denoted by orange, purple, green, blue, and red respectively. The left column shows the trajectories of four metrics over iterations: RMSE for the mean response, RMSE for the log-noise variance, and empirical coverages for the mean response and log-noise variance. The black dashed line in the coverage plots indicates the 95\% nominal level. Error bars (mean $\pm$ standard deviation) are displayed every 10 iterations. The right column presents boxplots of the corresponding metric values at the final iteration. A two-sample t-test is conducted on the results at the final iteration to compare the mean RMSE metrics between FB-BIM  and each alternative strategy. Significance levels are indicated beside the boxplots: ``ns''(p$>$0.05), ``*''(p$<=$0.05), ``**''(p$<=$0.01), ``***''(p$<=$0.001).}
	\label{1-d RMSE boxplot comparisons}
\end{figure}

Figure \ref{1-d RMSE boxplot comparisons} shows performance results summarized over the 100 macro-replications. By recording all four metrics at every iteration, the line plots illustrate the trajectory of each metric across the sequential iterations. In this relatively simple example, FB-BIM, EB-EIM, and hetGP all perform well in terms of both $RMSE_f$ and $RMSE_v$. Mean surface estimation for Homo-BIM is slightly hindered by the misspecified homoscedastic variance, while a loss of accuracy from the variational approximation is evident for VI-BIM.

FB-BIM achieves the fastest reduction in $RMSE_f$ in the early iterations, and then becomes comparable to EB-EIM by about $N = 50$. Examining the contributions of the C1 and C2 terms in \eqref{eq:seqBIMSPE} can provide an intuitive explanation: see Figure~S10 in Supplement C which shows the evolution of C1 and C2 in the expected BIMSPE criterion for a single macro-replication. The term C2, which represents the variance of posterior means, is specific  to the fully Bayesian framework, and this source of uncertainty quantification accounts for a substantial proportion of the total expected BIMSPE (i.e., C1 + C2) during the early iterations. Hence early on, this uncertainty propagation plays a more important role in helping FB-BIM make better design selections. In this simple 1-D example, the relative contribution of C2 falls below 10\% by about $N=40$. As a result, FB-BIM and EB-EIM acquisitions would be expected to become largely similar during later iterations. At the same time, FB-BIM maintains more stable $RMSE_v$ and $v$-coverage than EB-EIM, with fewer outliers, as the quality of Laplace approximation-based credible intervals can be quite variable.

\subsection{Illustrative Two-dimensional Test Function}\label{sec-2-d}

\begin{figure}[!htbp]
	\captionsetup[subfigure]{labelformat=empty}
	\begin{subfigure}[b]{0.47\linewidth}
		\centering
		\includegraphics[width=1\textwidth, height=.6\textwidth]{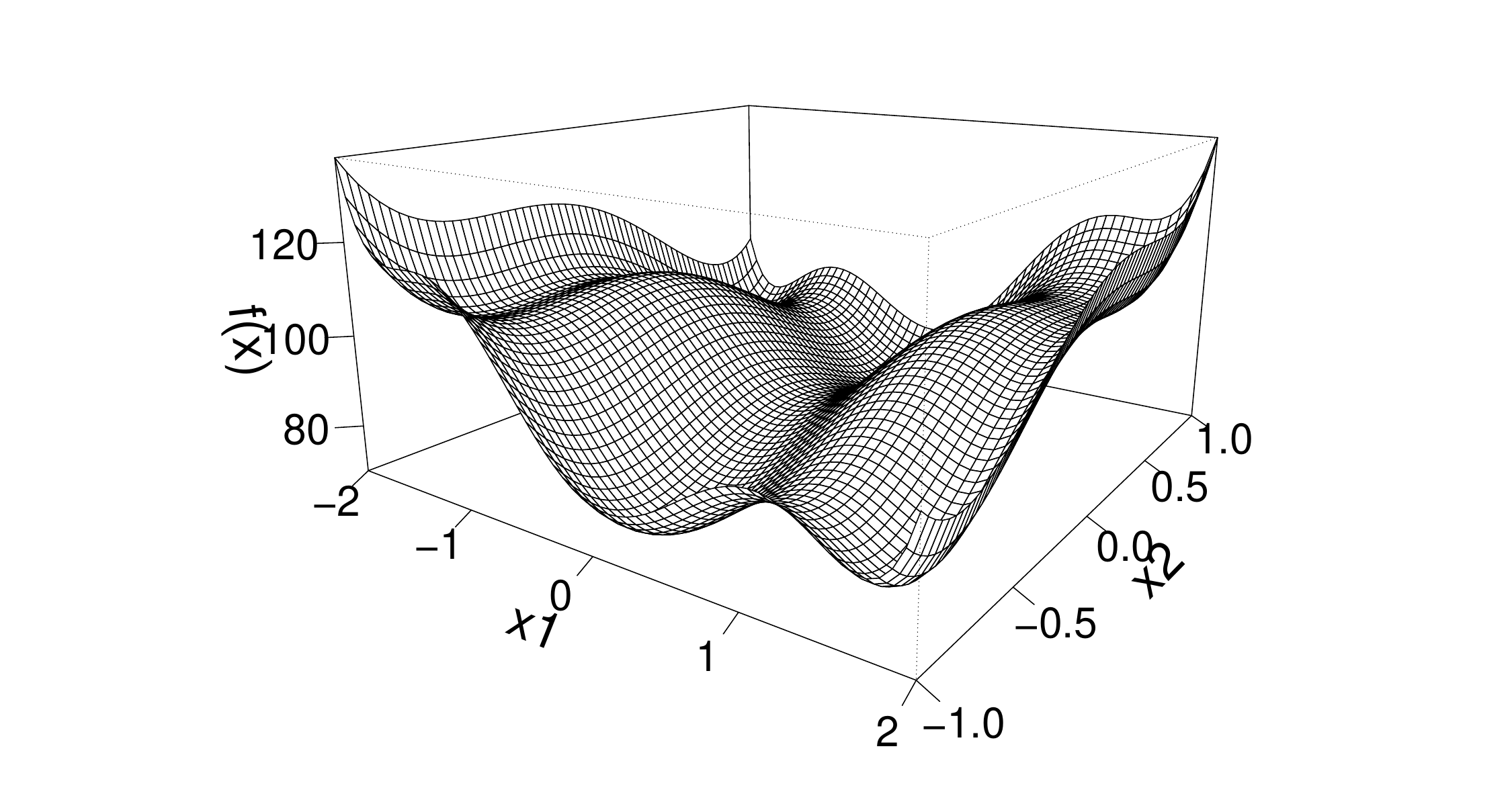}
		\caption{(a) Mean response surface f(x)}
	\end{subfigure} 
	\begin{subfigure}[b]{0.52\linewidth}
		\centering
		\includegraphics[width=1\textwidth, height=.6\textwidth]{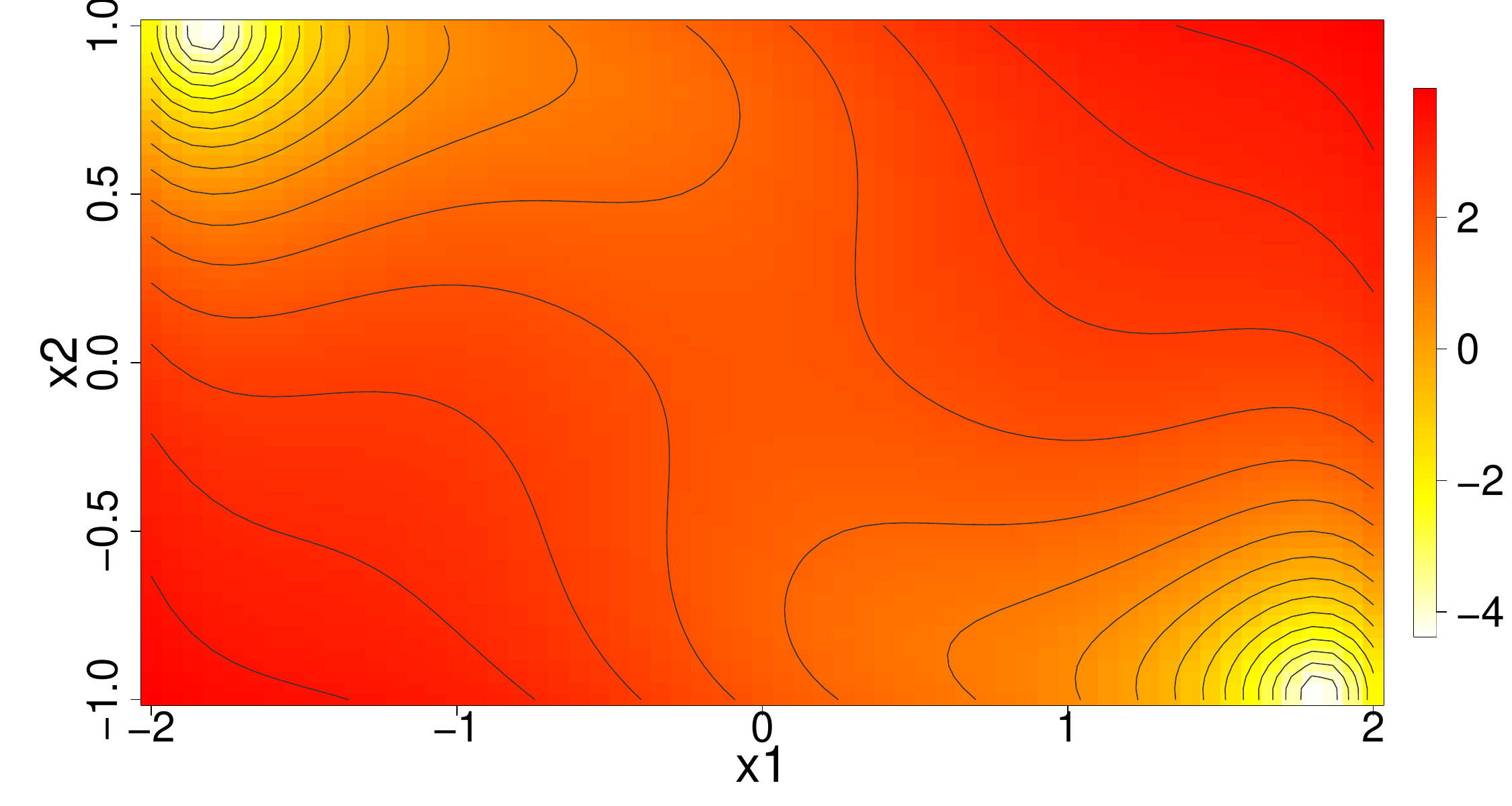}
		\caption{(b) Log noise variance $log (g(x))$}
	\end{subfigure} 
	\caption{Visualizations of the mean response surface and the log-noise variance
		surface of the 2-D example.}\label{2dfig} 
\end{figure}

Next, we consider a 2-D example where the simulation model for the mean response surface is based on a modified version of the six-hump camel function \citep{molga2005test}, defined as $f_2(x_1, x_2)=(40-21x_1^2+{10x_1^4}/3)x_1^2+10x_1x_2+(-40+40x_2^2)x_2^2+80$, together with the noise function $g_2(x_1, x_2)=\left(f_2(x_1, x_2) / \sqrt{10}+ \sqrt{10} \sin( {x_1 x_2\pi}/{6}) \right)^2$, where $\bfx=[x_1, x_2]^T$, $x_1 \in [-2, 2], \text{ and } x_2 \in [-1, 1]$; see Figure \ref{2dfig}. The chosen design grid $\mathcal{D}$ divides $x_1 \in [-2, 2]$ with a stride of 0.2, and $x_2 \in [-1, 1]$ with a stride of 0.1, resulting in a total of 441 unique design points. The noise magnitude ($\sqrt{g(\bfx)/R_f}$) varies between 1.4\% and 86\% with an average of 36\%. The overall noise level is moderate and heteroscedasticity is strong in some regions. The initial design is a LHD of 21 points from $\mathcal{D}$ with no replicates, and the total budget for the sequential design $B$ is set to 150 iterations.

\begin{figure}[!htbp]
	\centering
	\includegraphics[width=1\linewidth]{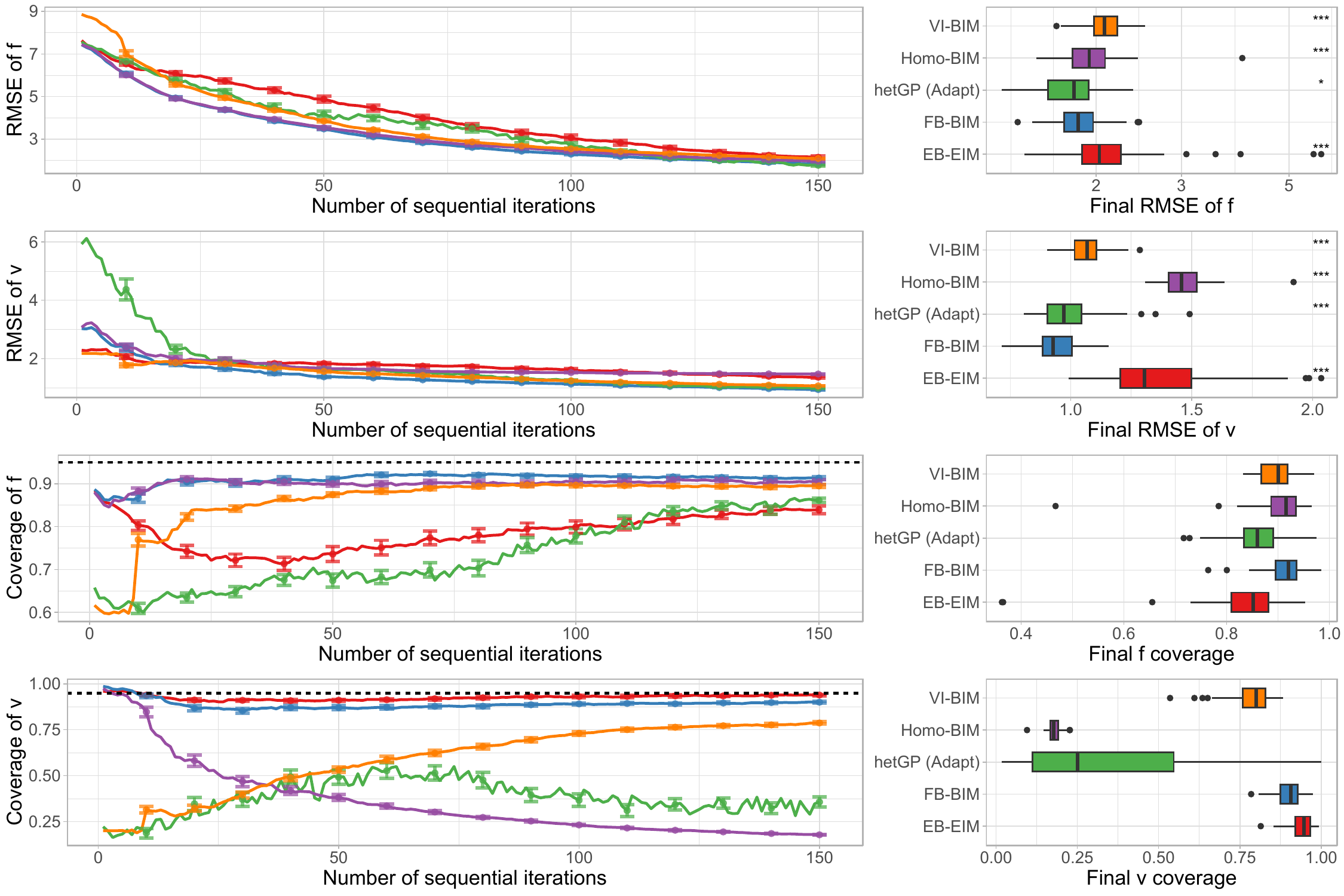}
	\caption{Line plots of prediction performance (left) and final-iteration boxplots (right) for VI-BIM, Homo-BIM, hetGP (Adapt), FB-BIM, and EB-EIM on the 2-D example, denoted by orange, purple, green, blue, and red respectively. The left column shows the trajectories of four metrics over iterations: RMSE for the mean response, RMSE for the log-noise variance, and empirical coverages for the mean response and log-noise variance. The black dashed line in the coverage plots indicates the 95\% nominal level. Error bars (mean $\pm$ standard deviation) are displayed every 10 iterations. The right column presents boxplots of the corresponding metric values at the final iteration. A two-sample t-test is conducted on the results at the final iteration to compare the mean RMSE metrics between FB-BIM and each alternative strategy. Significance levels are indicated beside the boxplots: ``ns''(p$>$0.05), ``*''(p$<=$0.05), ``**''(p$<=$0.01), ``***''(p$<=$0.001).}
	\label{2-d RMSE boxplot comparisons}
\end{figure}

Figure \ref{2-d RMSE boxplot comparisons} shows performance results summarized over the 100 macro-replications. In this case, the $RMSE_f$ of FB-BIM decreases faster than EB-EIM and maintains an advantage across all sequential iterations; Supplement Figure~S10 shows that C2 consistently contributes at least $10\%$ of the expected BIMSPE criterion. Hence, uncertainty in the hyperparameters and latent noise variables has a non-negligible effect on design selection throughout the experiment, and can explain the more efficient design-space exploration and sequential acquisition achieved by FB-BIM relative to EB-EIM.

Interestingly, hetGP has the best final $RMSE_f$ in this example, while FB-BIM has the best final $RMSE_v$. One possible explanation is that here hetGP (Adapt) places greater emphasis on exploring new design points, whereas FB-BIM tends to allocate more samples to replication. This interpretation is supported by Supplement I, where it can be seen that hetGP with fixed horizons $h=3$ and $h=4$, which favor replication by design, have notably larger $RMSE_f$ than hetGP (Adapt) for this example.

Coverage results further highlight differences among the strategies. For the mean response, FB-BIM and Homo-BIM rapidly approach the nominal 95\% level and remain stable through the sequential iterations. In contrast, hetGP (Adapt) and EB-EIM exhibit persistent under-coverage. This is likely due to these strategies' reliance on point estimates, with variance approximations which fail to fully quantify the uncertainty in the mean response.

As an extension to this 2-D example, Supplement~D presents additional experiments with known GP generating models, for examining how the FB-BIM posterior distributions converge to the true hyperparameter values under different noise levels (high, medium, and low). Therein, we also investigate the potential impacts of a misspecified GP kernel, which indicate that recovery of the mean response and log-noise  surfaces remain reasonable.

\subsection{Illustrative Five-dimensional Test Function}\label{sec-5-d}

Finally, we consider a 5-D example where the simulation model for the mean response surface is based on the Friedman function \citep{friedman1991multivariate}, defined as
$f(\bfx)=10\text{sin}(\pi x_1x_2)+ 20(x_3-0.5)^2+10x_4+5x_5$,
together with the noise function based on the modified G-function \citep{davis2007methods} defined as $g(\bfx)=\Big(\sqrt{2}\prod_{i=1}^{5}\dfrac{|4x_i-1|+a_i}{1+a_i} +0.05 \Big)^2$, where $a_i=(i-1)/2$, for $i=1, \dots, 5$. Both functions are defined on the hypercube $x_i\in[0,1]$, for $i=1, \dots, 5$. We generate a set of $|\mathcal{D}| = 900$ candidate design points using a maximin-distance LHD. The noise magnitude ($\sqrt{g(\bfx)/R_f}$) varies between 1\% and 194\% with an average of 29\%, which is moderate but with strong heteroscedasticity across some regions. The initial design with $n_0=50$ points (no replicates) is chosen from $\mathcal{D}$ based on the maximin distance criterion. The total budget is $B=200$ sequential iterations. 

\begin{figure}[!htbp]
	\centering
	\includegraphics[width=1\linewidth]{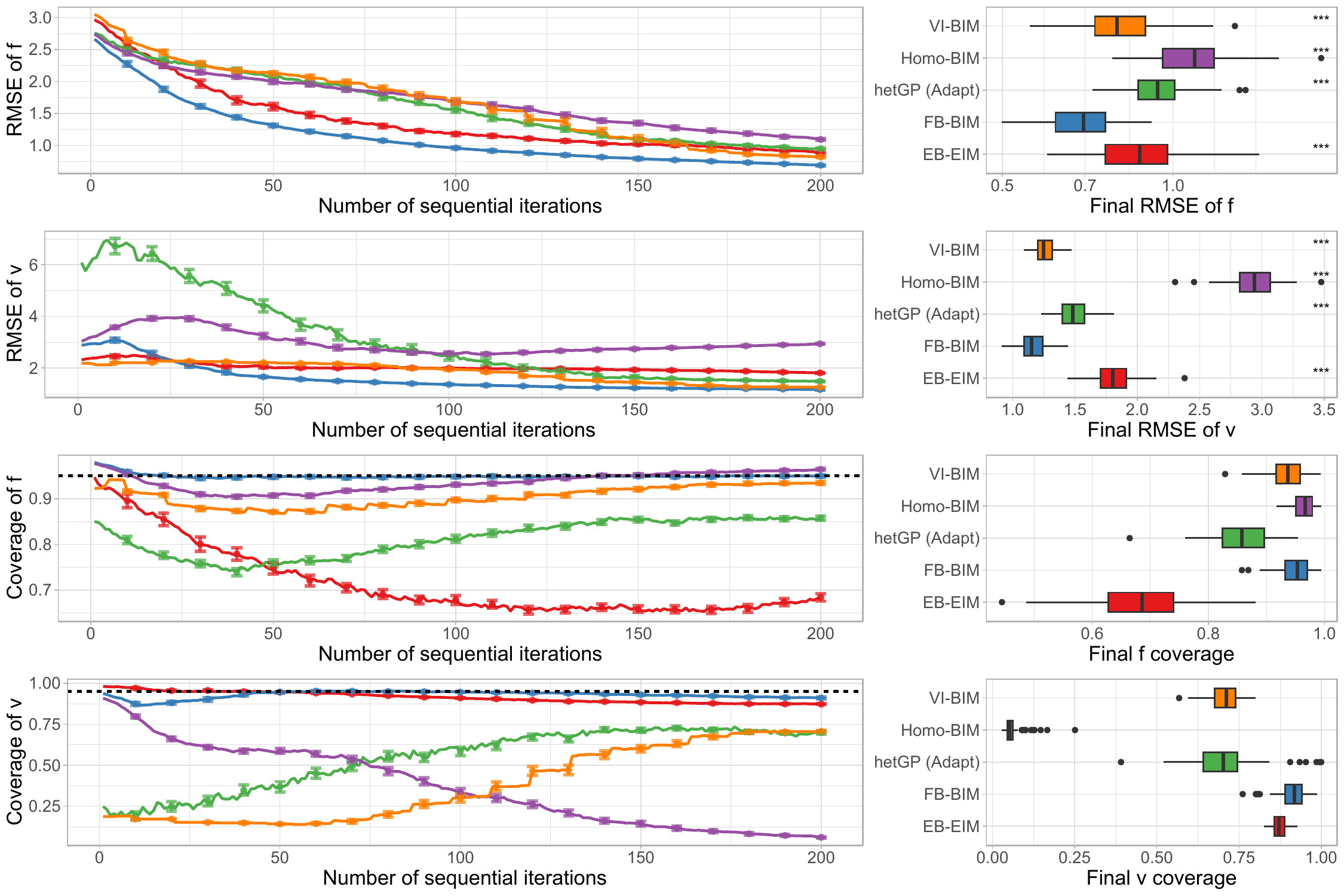}
	\caption{Line plots of prediction performance (left) and final-iteration boxplots (right) for VI-BIM, Homo-BIM, hetGP (Adapt), FB-BIM, and EB-EIM on the 5-D example, denoted by orange, purple, green, blue, and red respectively. The left column shows the trajectories of four metrics over iterations: RMSE for the mean response, RMSE for the log-noise variance, and empirical coverages for the mean response and log-noise variance. The black dashed line in the coverage plots indicates the 95\% nominal level. Error bars (mean $\pm$ standard deviation) are displayed every 10 iterations. The right column presents boxplots of the corresponding metric values at the final iteration. A two-sample t-test is conducted on the results at the final iteration to compare the mean RMSE metrics between FB-BIM  and each alternative strategy. Significance levels are indicated beside the boxplots: ``ns''(p$>$0.05), ``*''(p$<=$0.05), ``**''(p$<=$0.01), ``***''(p$<=$0.001).}
	\label{5-d RMSE boxplot comparisons}
\end{figure}

Figure \ref{5-d RMSE boxplot comparisons} shows performance results summarized over the 100 macro-replications. Overall, FB-BIM demonstrates the most favorable performance across all four metrics. It consistently achieves the lowest RMSE for both the mean response and the log-noise variance, while also maintaining $f$ and $v$-coverages close to the nominal level. The final-iteration boxplots further support these findings. To ensure the robustness of our results, we also evaluated the final-iteration RMSE and coverage on an independent test set, where test points were sampled uniformly from the continuous $[0,1]^5$ design space; the resulting performance metrics are similar to those obtained on $\mathcal{D}$ (see Figure S7 in Supplement B.4). In contrast to the lower-dimensional cases, the performance gains of FB-BIM are more pronounced in this 5-D example. A plausible explanation is that the number of hyperparameters to be inferred increases with the dimension $d$ under the anisotropic assumption, while the sequential budget $B=200$ is only slightly larger. Furthermore, the number of design points under consideration, $|\mathcal{D}| = 900$, is also notably larger. In this sense, the 5-D example provides the smallest effective budget relative to the complexity of the problem. Consequently, the advantages of a fully Bayesian treatment and its uncertainty quantification become more strongly evident. This finding is echoed by Supplement Figure~S10, where the contribution of the C2 term to the expected BIMSPE is still above 15\% by the end of the experiment.

Figure S13 in Supplement E presents histograms of samples drawn from the priors and the final FB-BIM posteriors (after all $B$ sequential iterations) for each GP hyperparameter, aggregated across all macro-replications. For each of the 1-D, 2-D, and 5-D examples, these plots indicate that the posteriors for both lengthscale and variance parameters often shift notably from their priors. Moreover, the baseline noise parameter $\mu_0$ becomes highly concentrated relative to its prior. These patterns confirm that the prior specification is only weakly informative and allows the model to effectively learn from a limited amount of data. A further investigation of prior sensitivity is provided in Supplement B.1.

Table~S2 in Supplement G reports the runtime of all strategies considered in our experiments, which follow the expected pattern in each example: EB-EIM and hetGP are comparably fast with their use of point estimates; Homo-BIM and VI-BIM have intermediate runtimes, and FB-BIM incurs the highest overhead. Crucially, FB-BIM's runtime remains reasonable, requiring 2, 10, and 81 minutes on average for the 1-D, 2-D, and 5-D examples, respectively. Supplement Table~S3 breaks down these runtimes into lookahead acquisition and posterior update wall-clock times, confirming that single-core MCMC sampling dominates FB-BIM's overall runtime compared to the parallelized lookahead step.

As an intermediate between point estimation and full MCMC, VI-BIM has greater computational efficiency but underperforms FB-BIM across all examples, likely due to inherent variational approximation error. In the 5-D example, where the budget is most limited relative to the dimensionality, VI-BIM outperforms the point-estimate-based strategies. Conversely, in the 1- and 2-D examples, where the budget is more generous relative to the design space and problem complexity, the approximation error of VI appears to offset its potential advantages.

Finally, across all three examples, we observe that inference on the noise process affects estimation of the mean response surface, although the impact may be limited. This is illustrated by the Homo-BIM results: while the strategy yields substantially worse $RMSE_v$ and $v$-coverage due to its misspecified homoscedasticity assumption, its $RMSE_f$ and $f$-coverage are relatively robust compared to strategies that explicitly model heteroskedastic noise. For example, in the 1-D setting, Homo-BIM outperforms VI-BIM, and in the 2-D example, Homo-BIM achieves lower $RMSE_f$ than both VI-BIM and EB-EIM despite having much higher $RMSE_v$.  In Figure~S14 of Supplement H, which illustrates final 2-D designs for different strategies, Homo-BIM differs from FB-BIM in that it strongly favors exploration over replication. This behavior may benefit the estimation of the mean response surface, even though Homo-BIM does not have the flexibility to learn the heteroscedastic noise structure.  Conversely, we investigate if FB-BIM is prone to overfitting in examples with true homoscedastic noise in Supplement B.2, and find that our results remain robust.

\section{Application to Seismic Design of Podium Buildings}\label{sec-real-life-example}
Podium buildings consist of a wood frame atop a one- or two-story concrete podium base \citep{triggs2015review}. Podium buildings are increasingly popular in the US and Canada due to their cost-efficiency, design flexibility, and ability to maximize usable floor space within height limits \citep{ni2015midrise}. Due to differences in stiffness and mass between the wood and podium sections, podium buildings require complex time-history dynamic analyses for their seismic designs to ensure reliability of the structure during earthquakes, according to National Building Code of Canada (NBCC). The 2015 edition of the NBCC and the 2016 edition of the American Society of Civil Engineering Standard (ASCE 7) allow engineers to design podium buildings with a simple two-step analysis procedure, subject to specific criteria related to buildings' mass and stiffness. When these criteria are satisfied, the two-step analysis procedure simplifies seismic assessment by treating the upper wood-frame structure as an independent building on a fixed base and the lower concrete structure as a separate building subjected to downward forces from the wood-frame structure above. This allows both structures to be analyzed individually using the equivalent static force procedure \citep{nrc2015nbcc, asce2016minimum, chopra2020earthquake}. 

Time-history dynamic analysis, i.e., via an earthquake simulation model, is used to assess seismic reliability of podium building designs by choosing a set of representative earthquakes for a given region and simulating their impact on the building. To examine the conditions under which the two-step analysis procedure can be applied, \cite{chen2020criterion} designed podium buildings by varying parameters recognized by NBCC and ASCE 7 and evaluated their seismic performance using an earthquake simulation model. Specifically, their study investigated the relationship between the mass and stiffness ratios (input variables) and the maximum inter-story drift (output variable) during earthquakes to distinguish if the podium building designs meet performance-based seismic standards. However, their input choices were relatively \textit{ad hoc}, and the sample size was quite limited. In subsequent work, \cite{huang2023kriging} proposed a sequential algorithm using ordinary kriging to model inter-story drift; however, that method was limited by the assumption of a deterministic simulation model, whereas earthquake simulations are stochastic in practice. Thus, as an illustrative application, we use FB-BIM to predict the mean seismic response of podium buildings that consist of a six-story wood-frame structure atop a one-story concrete podium (or $6 + 1$ podium building for short), while accounting for random noise in the earthquake simulation model. The mean response surface of maximum inter-story drift is a quantity of primary interest in performance-based seismic design. Beyond investigating the relationship between inter-story drift and the mass and stiffness ratios, accurate estimation of inter-story drift supports a range of downstream analyses, including peak transient floor acceleration, floor velocity and story drift ratio \citep{applied2018seismic}. 

The seismic response of the podium buildings under maximum considered earthquake (MCE) level events is analyzed with a modified macroelement model to simulate the wood-frame shear walls \citep{xu2009development, chen2014simulation}, via the general-purpose finite-element software, ABAQUS version 6.21-1. A building design is considered to meet the performance objective if the maximum inter-story drift is within 4\%, with a nonexceedance probability of 80\% under MCE \citep{van2010experimental, pang2010simplified}. We focus on modeling this maximum inter-story drift for the $6+1$ podium building design as the response variable, and closely follow the experimental setup of \citet{huang2023kriging}. The main difference is that, whereas their study considered 15 earthquakes, here we focus on the mean inter-story drift response surface for a single earthquake. Specifically, we use the 1971 San Fernando earthquake (magnitude 6.61), recorded at the CDMG 24271 Lake Hughes station, as our simulation model. The earthquake is scaled to match the target response spectrum for Vancouver, British Columbia, Canada, based on NBCC 2015 and Geological Survey of Canada data \citep{earthquakes2015}. The design grid, $\mathcal{D}$, is spanned by the normalized mass $[M]$ and stiffness ratios $[K]$. Specifically, $[M]$ ranges from 0.5 to 6 with a stride of 0.5, and $[K]$ ranges from 1 to 60 with a stride of 1, yielding a total of 720 design points across the region of $[0.5,6]\times[1,60]$. We follow the same prior specifications as in Section \ref{prior_specification}. For this industrial application, the Mat\'ern kernel may be preferred over the squared exponential kernel because it allows for more realistic modeling of physical phenomena by accommodating less smooth functions \citep{stein1999interpolation}. We choose the Mat\'ern kernel function with $\nu=3/2$ for both the $f$- and $v$- processes. The initial design consists of 20 evaluations with no replicates, where design points were selected using the LHD, and a total budget $B=200$ is allocated for sequential iterations. 

\begin{figure}[!htbp]
	\captionsetup[subfigure]{labelformat=empty}
	\begin{subfigure}[b]{0.43\linewidth}
		\centering
		\includegraphics[width=1\linewidth]{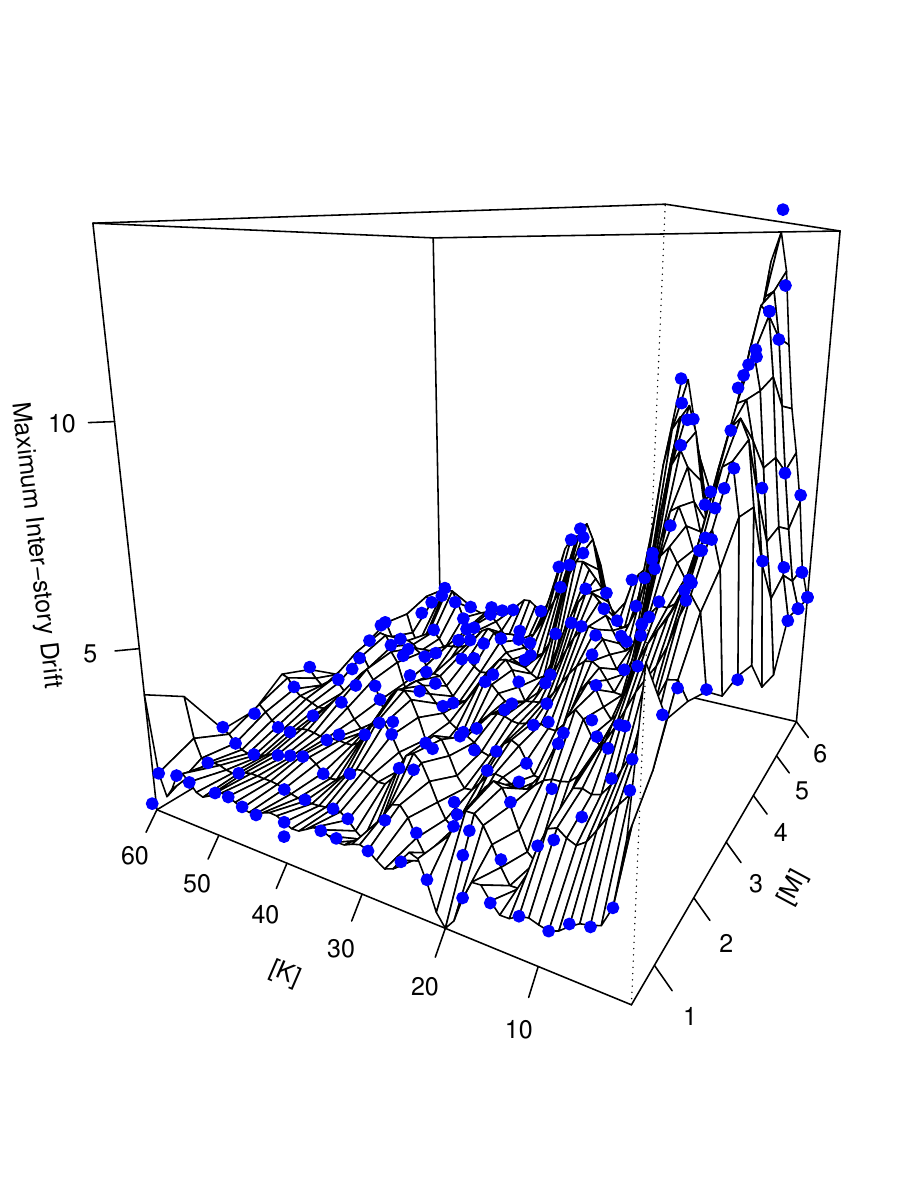} 
		\caption{(a) Predicted mean response surface with 3-D scatterplot of observations}\label{mean_response_surface}
	\end{subfigure} 
	\hfill
	\begin{subfigure}[b]{0.45\linewidth}
		\centering
		\includegraphics[width=1\linewidth]{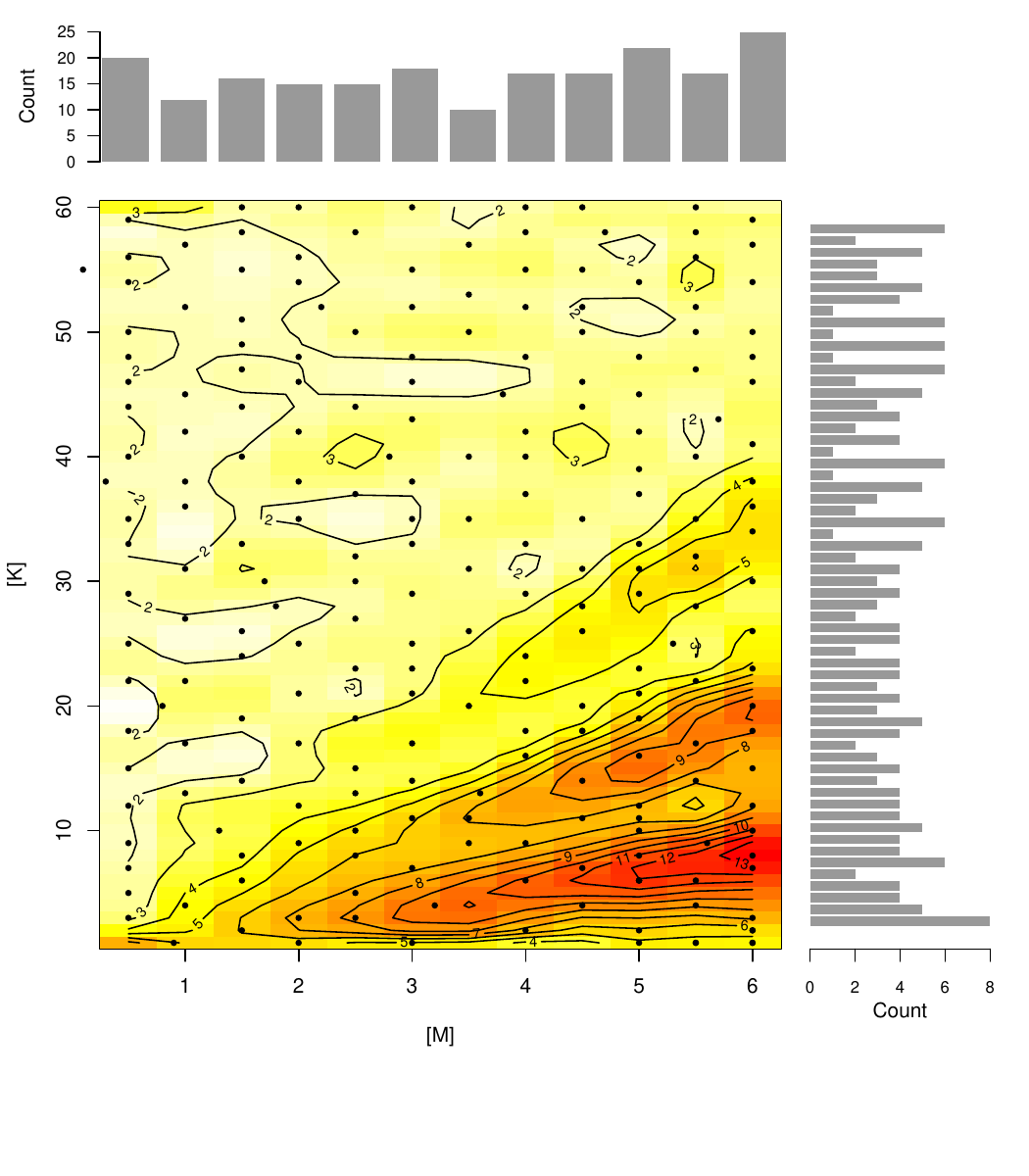}
		\caption{(b) Predicted mean surface with contour lines and labeled design points}\label{contour}
	\end{subfigure}
	\hfill
	\begin{subfigure}[t]{0.32\linewidth}
		\centering
		\includegraphics[width=1\linewidth]{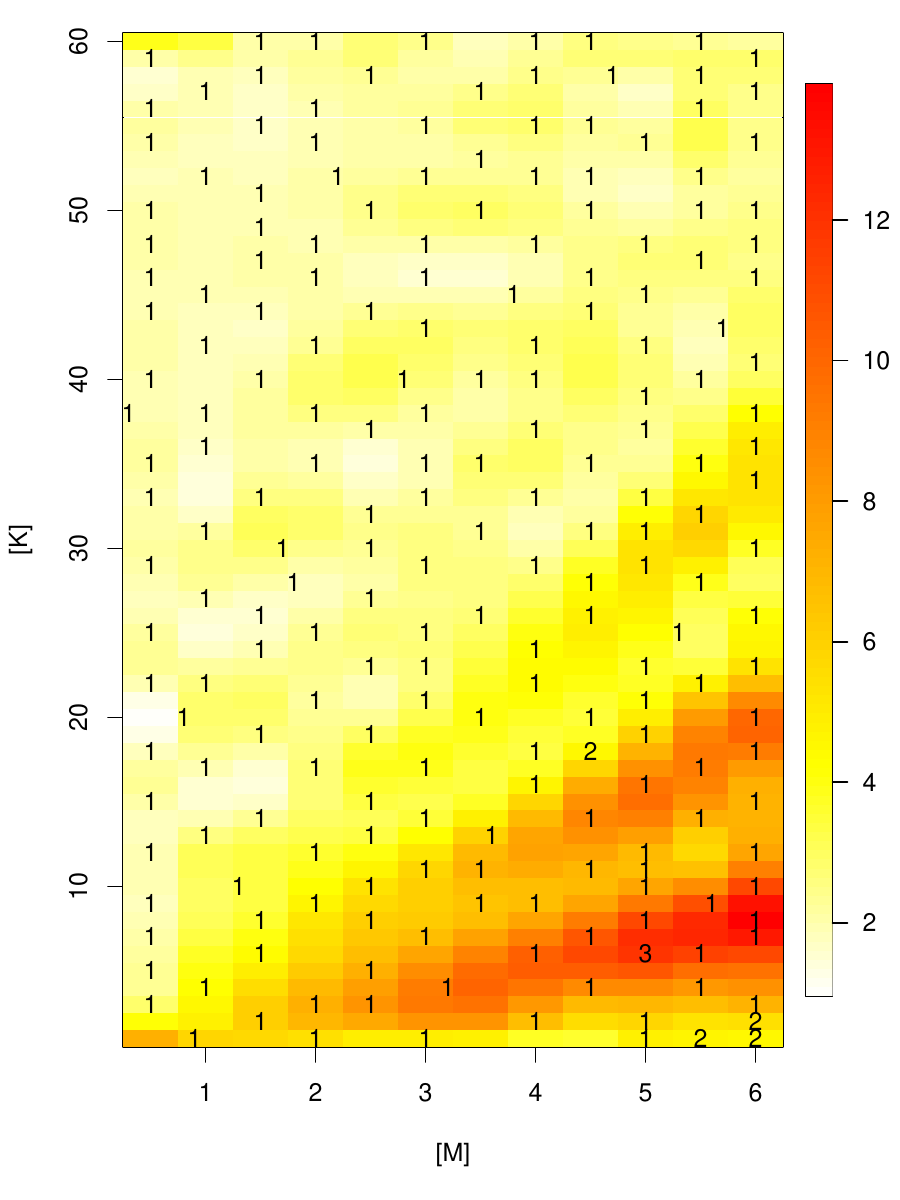}
		\caption{(c) Predicted mean surface with observation frequency}\label{repetition}
	\end{subfigure}
	\hfill
	\begin{subfigure}[t]{0.32\linewidth}
		\centering
		\includegraphics[width=1\linewidth]{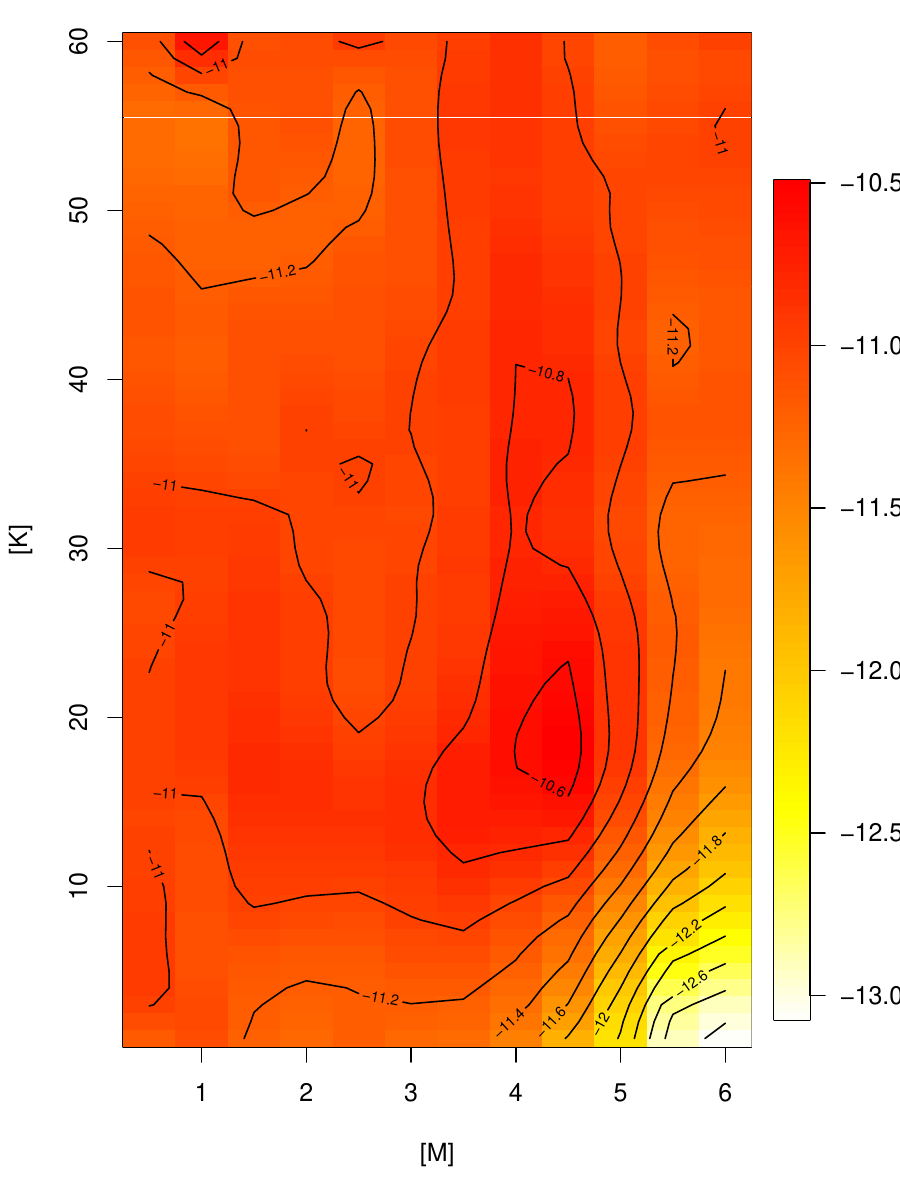}
		\caption{(d) Fitted log noise $\hat{v}(\bfx)$}\label{fitted_v(x)}
	\end{subfigure}
	\hfill
	\begin{subfigure}[t]{0.32\linewidth}
		\centering
		\includegraphics[width=1\textwidth]{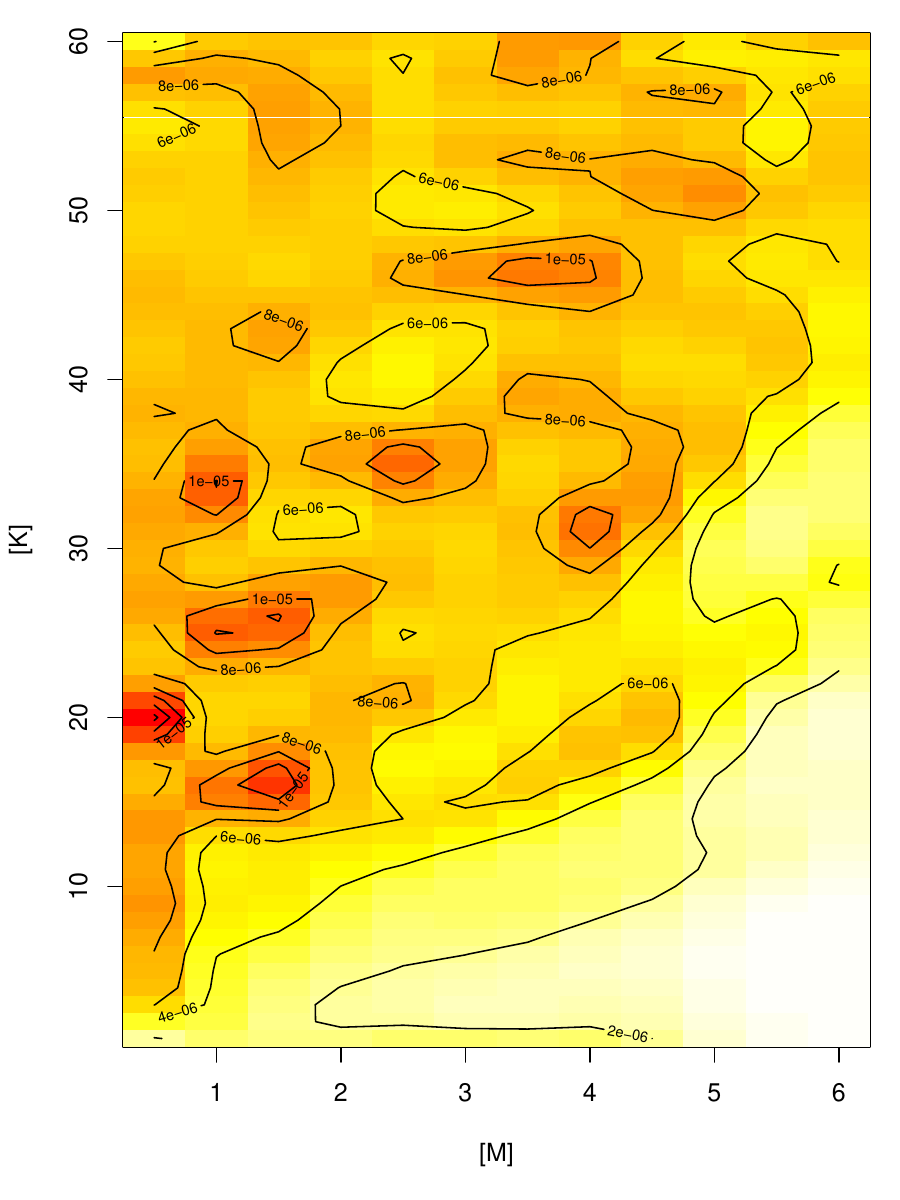}
		\caption{(e) Ratio of $\hat{v}(\bfx)/\hat{Y}(\bfx)$ }\label{fitted_ratio}
	\end{subfigure}
	\caption{Metamodeling results for the seismic response of $6+1$ podium buildings subjected to the simulated earthquake, based on 200 sequential iterations (220 total observations) guided by the FB-BIM strategy.
	} \label{simulation plot}
\end{figure}

Figure \ref{mean_response_surface} illustrates the predicted mean response surface of the final surrogate model constructed using FB-BIM  after 200 iterations, with the observed design points denoted by blue dots. For a more detailed view, Figure \ref{contour} presents contour lines of the predicted mean response surface, with observations shown as scatter points. The sampling frequencies are summarized along the $x$- and $y$-axes to illustrate how observations are distributed across the input space. Figure \ref{repetition} shows the number of evaluations at each observed design point, Figure \ref{fitted_v(x)} presents the fitted logarithmic noise level $\hat{\textbf{v}}(\x)$ accompanied by its contour lines, and Figure \ref{fitted_ratio} displays the ratio of the fitted logarithmic noise level to the predicted mean response surface. 

These results reveal some notable findings. First, as shown in Figure~\ref{mean_response_surface}, the observed evaluations lie close to the predicted mean response surface, with only small deviations across the input space. This suggests that the simulation noise is heteroscedastic but low in magnitude overall. Figure~\ref{fitted_v(x)}, which shows the fitted variance contours, further supports this observation. A possible explanation is that the selected earthquake is moderate in magnitude and that the considered podium building design is relatively robust under this event. Because the fitted log-noise remains small, FB-BIM tends to favor exploration over replication, as uncertainty in the inter-story drift is driven primarily by lack of coverage over the input space rather than by the need to separate signal from noise. This behavior is reflected in Figure~\ref{repetition}, where most design points are evaluated only once, but some replicates are chosen by the expected BIMSPE criterion. It also highlights the advantage of FB-BIM for learning the mean response surface under limited simulation budgets, where repeated evaluations at the same input locations may be impractical.

Second, we observe that the selected earthquake induces larger maximum inter-story drift in the lower-right region of the input space, corresponding to high mass ratios and low stiffness ratios. Over most of the domain, the maximum inter-story drift remains below $4\%$, which is consistent with the safety
regions reported by \citet{huang2023kriging} and \citet{chen2020criterion} based on 15 earthquake records. All figure panels indicate that the predicted mean response surface becomes steeper in regions with relatively small stiffness values (i.e., $0.5 < [K] < 6$), while the fitted noise levels in those regions remain small. This suggests that designs in these parts of the input space very likely will have unstable inter-story drift responses during earthquakes. Such regions may therefore be less desirable from a design perspective and could be excluded from future consideration.

\section{Conclusion and Discussion}\label{sec-conclusion}

In this paper, we propose the FB-BIM sequential strategy for predicting the mean response and log-noise variance surfaces of heteroscedastic stochastic simulations under a fully Bayesian framework. We develop the expected BIMSPE criterion to guide the selection of the next design point, and incorporate sequential importance sampling to improve computational efficiency. The proposed FB-BIM strategy demonstrates robust and promising performance across all synthetic examples. We then applied the method to estimate the mean inter-story drift response surface for podium buildings. We conducted multiple robustness experiments which indicate that FB-BIM continues to work well under varying diffuse prior specifications, kernel misspecification, and can avoid overfitting homoscedastic noise. We also performed an ablation study which suggests that the proposed fully Bayesian BIMSPE criterion provides practical advantages by propagating uncertainty more effectively through the sequential design selection process, compared to using the plug-in empirical IMSPE criterion. Finally, we assessed the capability of FB-BIM to correctly recover the true values of the GP hyperparameters given a known generating distribution.  

Comparisons with alternative strategies highlight the conditions under which the FB-BIM strategy is the most effective. Our results suggest that weakly informative priors can help regularize parameter inference, enabling Bayesian strategies to perform robustly with limited data. Thus, Bayesian strategies (FB-BIM, EB-EIM, and Homo-BIM) can be particularly effective during the early stages of sequential design, when MLE-based methods may be less stable. Among these, FB-BIM has the most reliable predictive accuracy and maintains empirical coverages close to nominal levels for both mean-surface and log-noise estimation. This suggests that the improved uncertainty quantification of FB-BIM plays an impactful role in sequential design, which helps drive better designs and in turn better predictions. This feature is particularly valuable for applications involving expensive simulation models or otherwise constrained budgets for simulation model evaluations. Conversely, the benefits of accounting for hyperparameter uncertainty may diminish as the sample size grows. Hence, a potential adaptation of FB-BIM would be to monitor the magnitude of the C2 term from the Bayesian treatment in \eqref{eq:seqBIMSPE} and omit its evaluation once it becomes negligible, thereby reducing computational overhead in subsequent iterations.

While this paper contributes to establishing a fully Bayesian sequential strategy for stochastic simulations, avenues for future research remain. First, as we utilize Monte Carlo approximations for calculating the expected Bayesian IMSPE, other surrogates besides GP models could be considered; e.g., \cite{zhu2021emulation} employ generalized lambda surrogates that outperform GPs in some examples where a GP assumption might be too strong. It could thus be interesting to pair the FB-BIM strategy with other surrogate models. Second, extending FB-BIM to  batch sequential design, where multiple evaluations are added to the observed dataset in each iteration, offers another possible direction \citep{chevalier2015fast, Beek2020ScalableAB, zhang2022batch, Surer02012026}. Third, more efficient alternatives to MCMC sampling could be explored, so that fully Bayesian sequential design remains viable for larger budgets. Recent developments in Vecchia approximations \citep{patil2025vecchia} could be a promising direction. Fourth, an extension of FB-BIM to continuous design spaces via gradient-based optimization -- which likely requires specialized computational techniques -- would improve the scalability of the strategy to higher-dimensional inputs. We encourage further research into these areas and aim to contribute to addressing these challenges in the future.

\section*{Acknowledgements}
We are grateful to Zhiyong Chen from FPInnovations for introducing us to the problem of seismic design for podium buildings. This work was partially supported by the Natural Sciences and Engineering Research Council of Canada under Discovery Grant RGPIN-2019-04771.

\bibliographystyle{apalike}
\bibliography{references}

@article{chen2020criterion,
  title={Criterion for Applying Two-Step Analysis Procedure to Seismic Design of Wood-Frame Buildings on Concrete Podium},
  author={Chen, Zhiyong and Ni, Chun},
  journal={Journal of Structural Engineering},
  volume={146},
  number={1},
  pages={04019178},
  year={2020},
  publisher={American Society of Civil Engineers}
}

@article{van2010experimental,
  title={Experimental seismic response of a full-scale six-story light-frame wood building},
  author={Van de Lindt, John W and Pei, Shiling and Pryor, Steven E and Shimizu, H and Isoda, H},
  journal={Journal of Structural Engineering},
  volume={136},
  number={10},
  pages={1262--1272},
  year={2010},
  publisher={American Society of Civil Engineers}
}

@article{pang2010simplified,
  title={Simplified direct displacement design of six-story woodframe building and pretest seismic performance assessment},
  author={Pang, Weichiang and Rosowsky, David V and Pei, Shiling and van de Lindt, John W},
  journal={Journal of Structural Engineering},
  volume={136},
  number={7},
  pages={813--825},
  year={2010},
  publisher={American Society of Civil Engineers}
}

@inproceedings{pellegrino2010fea,
  title={{FEA-based multi-objective optimization of IPM motor design including rotor losses}},
  author={Pellegrino, G and Cupertino, F},
  booktitle={2010 IEEE Energy Conversion Congress and Exposition},
  pages={3659--3666},
  year={2010},
  organization={IEEE}
}

@article{kleijnen2009kriging,
  title={Kriging metamodeling in simulation: A review},
  author={Kleijnen, Jack PC},
  journal={European Journal of Operational Research},
  volume={192},
  number={3},
  pages={707--716},
  year={2009},
  publisher={Elsevier}
}

@article{fuhg2021state,
  title={State-of-the-art and comparative review of adaptive sampling methods for kriging},
  author={Fuhg, Jan N and Fau, Am{\'e}lie and Nackenhorst, Udo},
  journal={Archives of Computational Methods in Engineering},
  volume={28},
  number={4},
  pages={2689--2747},
  year={2021},
  publisher={Springer}
}

@article{sacks1989design,
  title={Design and analysis of computer experiments},
  author={Sacks, Jerome and Welch, William J and Mitchell, Toby J and Wynn, Henry P},
  journal={Statistical Science},
  volume={4},
  number={4},
  pages={409--423},
  year={1989},
  publisher={Institute of Mathematical Statistics}
}

@inproceedings{barton1994metamodeling,
  title={Metamodeling: a state of the art review},
  author={Barton, Russell R},
  booktitle={Proceedings of Winter Simulation Conference},
  pages={237--244},
  year={1994},
  organization={IEEE}
}

@article{kleijnen2017regression,
  title={Regression and Kriging metamodels with their experimental designs in simulation: A review},
  author={Kleijnen, Jack PC},
  journal={European Journal of Operational Research},
  volume={256},
  number={1},
  pages={1--16},
  year={2017},
  publisher={Elsevier}
}

@article{jones1998efficient,
  title={Efficient global optimization of expensive black-box functions},
  author={Jones, Donald R and Schonlau, Matthias and Welch, William J},
  journal={Journal of Global Optimization},
  volume={13},
  number={4},
  pages={455--492},
  year={1998},
  publisher={Springer}
}

@inproceedings{goldberg1997regression,
  title={Regression with input-dependent noise: A {Gaussian} process treatment},
  author={Goldberg, Paul and Williams, Christopher and Bishop, Christopher},
  booktitle={Advances in Neural Information Processing Systems},
  volume={10},
  year={1997}
}

@article{gramacy2012cases,
  title={Cases for the nugget in modeling computer experiments},
  author={Gramacy, Robert B and Lee, Herbert KH},
  journal={Statistics and Computing},
  volume={22},
  pages={713--722},
  year={2012},
  publisher={Springer},
 url={https://link.springer.com/article/10.1007/s11222-010-9224-x#citeas}
}

@misc{molga2005test,
    author ={Molga, Marcin and Smutnicki, Czes{\l}aw},
    title ={Test functions for optimization needs},
    year = {2005},
   note={Available at {https://robertmarks.org/Classes/ENGR5358/Papers/functions.pdf}}
}

@article{loeppky2009choosing,
  title={Choosing the sample size of a computer experiment: A practical guide},
  author={Loeppky, Jason L and Sacks, Jerome and Welch, William J},
  journal={Technometrics},
  volume={51},
  number={4},
  pages={366--376},
  year={2009},
  publisher={Taylor \& Francis},
 url={https://www.tandfonline.com/doi/abs/10.1198/TECH.2009.08040}
}

@article{hoffman2014no,
  title={{The No-U-Turn sampler: adaptively setting path lengths in Hamiltonian {Monte Carlo}}},
  author={Hoffman, Matthew D and Gelman, Andrew and others},
  journal={Journal of Machine Learning Research},
  volume={15},
  number={1},
  pages={1593--1623},
  year={2014},
 url={https://dl.acm.org/doi/10.5555/2627435.2638586}
}

@inproceedings{huang2023kriging,
  title={A Kriging Metamodel with Adaptive Sampling for Seismic Evaluation of Podium Buildings},
  author={Huang, Yuying and Chen, Zhiyong and Wong, Samuel WK},
  booktitle={14th International Conference on Application of Statistics and Probability in Civil Engineering (ICASP14)},
  year={2023},
}

@article{ankenman2010stochastic,
  title={Stochastic kriging for simulation metamodeling},
  author={Ankenman, Bruce and Nelson, Barry L and Staum, Jeremy},
  journal={Operations Research},
  volume={58},
  number={2},
  pages={371--382},
  year={2010},
  publisher={INFORMS}
}

@article{chen2017sequential,
  title={Sequential design strategies for mean response surface metamodeling via stochastic kriging with adaptive exploration and exploitation},
  author={Chen, Xi and Zhou, Qiang},
  journal={European Journal of Operational Research},
  volume={262},
  number={2},
  pages={575--585},
  year={2017},
  publisher={Elsevier}
}

@article{kleijnen2005robustness,
  title={Robustness of Kriging when interpolating in random simulation with heterogeneous variances: some experiments},
  author={Kleijnen, Jack PC and van Beers, Wim CM},
  journal={European Journal of Operational Research},
  volume={165},
  number={3},
  pages={826--834},
  year={2005},
  publisher={Elsevier}
}

@article{cohn1996active,
  title={Active learning with statistical models},
  author={Cohn, David A and Ghahramani, Zoubin and Jordan, Michael I},
  journal={Journal of Artificial Intelligence Research},
  volume={4},
  number={1},
  pages={129--145},
  year={1996},
url={https://dl.acm.org/doi/10.5555/1622737.1622744}
}

@article{gramacy2009adaptive,
  title={Adaptive design and analysis of supercomputer experiments},
  author={Gramacy, Robert B and Lee, Herbert KH},
  journal={Technometrics},
  pages={130--145},
  year={2009},
  publisher={JSTOR},
volume={51},
number={2},
url={https://www.tandfonline.com/doi/abs/10.1198/TECH.2009.0015}
}

@article{yuan2013sequential,
  title={A sequential approach for stochastic computer model calibration and prediction},
  author={Yuan, Jun and Ng, Szu Hui},
  journal={Reliability Engineering \& System Safety},
  volume={111},
  pages={273--286},
  year={2013},
  publisher={Elsevier},
url={https://www.sciencedirect.com/science/article/abs/pii/S0951832012002335}
}

@article{yuan2015calibration,
  title={Calibration, validation, and prediction in random simulation models: {Gaussian} process metamodels and a {Bayesian} integrated solution},
  author={Yuan, Jun and Ng, Szu Hui},
  journal={ACM Transactions on Modeling and Computer Simulation (TOMACS)},
  volume={25},
  number={3},
  pages={1--25},
  year={2015},
  publisher={ACM New York, NY, USA}
}

@book{stein1999interpolation,
  title={Interpolation of spatial data: some theory for kriging},
  author={Stein, Michael L},
  year={1999},
  publisher={Springer}
}

@incollection{williams1998prediction,
  title={Prediction with {Gaussian} processes: From linear regression to linear prediction and beyond},
  author={Williams, Christopher KI},
  booktitle={Learning in Graphical Models},
  pages={599--621},
  year={1998},
  publisher={Springer}
}

@book{santner2003design,
  title={The design and analysis of computer experiments},
  author={Santner, Thomas J and Williams, Brian J and Notz, William I and Williams, Brain J},
edition={1st},
  year={2003},
  publisher={Springer}
}

@incollection{ginsbourger2010kriging,
  title={Kriging is well-suited to parallelize optimization},
  author={Ginsbourger, David and Le Riche, Rodolphe and Carraro, Laurent},
  booktitle={Computational Intelligence in Expensive Optimization Problems},
  pages={131--162},
  year={2010},
  publisher={Springer}
}

@article{binois2021hetgp,
  title={{hetGP: Heteroskedastic {Gaussian} Process Modeling and Sequential Design in R}},
  author={Binois, Micka{\"e}l and Gramacy, Robert B},
  year={2021},
  journal={Journal of Statistical Software},
  pages={1--44},
  volume={98},
  number={13},
}

@article{binois2019replication,
  title={Replication or exploration? {Sequential} design for stochastic simulation experiments},
  author={Binois, Micka{\"e}l and Huang, Jiangeng and Gramacy, Robert B and Ludkovski, Mike},
  journal={Technometrics},
  volume={61},
  number={1},
  pages={7--23},
  year={2019}, 
  publisher={Taylor \& Francis}
}

@article{binois2018practical,
  title={{Practical Heteroscedastic {Gaussian} Process Modeling for Large Simulation Experiments}},
  author={Binois, Mickael and Gramacy, Robert B and Ludkovski, Mike},
  journal={Journal of Computational and Graphical Statistics},
  volume={27},
  number={4},
  pages={808--821},
  year={2018},
  publisher={Taylor \& Francis}
}

@article{chevalier2015fast,
  title={Fast update of conditional simulation ensembles},
  author={Chevalier, Cl{\'e}ment and Emery, Xavier and Ginsbourger, David},
  journal={Mathematical Geosciences},
  volume={47},
  pages={771--789},
  year={2015},
  publisher={Springer},
 url={https://link.springer.com/article/10.1007/s11004-014-9573-7#citeas}
}

@article{hao2021novel,
  title={A novel Nested Stochastic Kriging model for response noise quantification and reliability analysis},
  author={Hao, Peng and Feng, Shaojun and Liu, Hao and Wang, Yutian and Wang, Bo and Wang, Bin},
  journal={Computer Methods in Applied Mechanics and Engineering},
  volume={384},
  pages={113941},
  year={2021},
  publisher={Elsevier},
 url={https://www.sciencedirect.com/science/article/abs/pii/S0045782521002784}
}

@article{chen2018bayesian,
  title={A {Bayesian} stochastic kriging optimization model dealing with heteroscedastic simulation noise for freeway traffic management},
  author={Chen, Xiqun and He, Xiang and Xiong, Chenfeng and Zhu, Zheng and Zhang, Lei},
  journal={Transportation Science},
  volume={53},
  number={2},
  pages={545--565},
  year={2018},
  publisher={INFORMS}
}

@inproceedings{chen2014sequential,
  title={Sequential experimental designs for stochastic kriging},
  author={Chen, Xi and Zhou, Qiang},
  booktitle={Proceedings of the Winter Simulation Conference 2014},
  pages={3821--3832},
  year={2014},
  organization={IEEE}
}

@article{kleijnen2004application,
  title={Application-driven sequential designs for simulation experiments: Kriging metamodelling},
  author={Kleijnen, Jack PC and van Beers, Wim CM},
  journal={Journal of the Operational Research Society},
  volume={55},
  pages={876--883},
  year={2004},
  number={8},
  publisher={Springer},
url={https://www.tandfonline.com/doi/citedby/10.1057/palgrave.jors.2601747?scroll=top&needAccess=true}
}

@inproceedings{wang2016effects,
  title={The effects of estimation of heteroscedasticity on stochastic kriging},
  author={Wang, Wenjing and Chen, Xi},
  booktitle={2016 Winter Simulation Conference (WSC)},
  pages={326--337},
  year={2016},
  organization={IEEE}
}

@article{sacks1989designs,
  title={Designs for computer experiments},
  author={Sacks, Jerome and Schiller, Susannah B and Welch, William J},
  journal={Technometrics},
  volume={31},
  number={1},
  pages={41--47},
  year={1989},
  publisher={Taylor \& Francis}
}

@article{leatherman2018computer,
  title={Computer experiment designs for accurate prediction},
  author={Leatherman, Erin R and Santner, Thomas J and Dean, Angela M},
  journal={Statistics and Computing},
  volume={28},
  pages={739--751},
  year={2018},
  publisher={Springer}
}

@inproceedings{kersting2007most,
  title={Most likely heteroscedastic {Gaussian} process regression},
  author={Kersting, Kristian and Plagemann, Christian and Pfaff, Patrick and Burgard, Wolfram},
  booktitle={Proceedings of the 24th International Conference on Machine Learning},
  pages={393--400},
  year={2007}
}

@article{robinson2010semi,
  title={A semi-parametric approach to dual modeling when no replication exists},
  author={Robinson, Timothy J and Birch, Jeffrey B and Starnes, B Alden},
  journal={Journal of Statistical Planning and Inference},
  volume={140},
  number={10},
  pages={2860--2869},
  year={2010},
  publisher={Elsevier}
}

@article{yin2011kriging,
  title={Kriging metamodel with modified nugget-effect: The heteroscedastic variance case},
  author={Yin, Jun and Ng, Szu Hui and Ng, Kien Ming},
  journal={Computers \& Industrial Engineering},
  volume={61},
  number={3},
  pages={760--777},
  year={2011},
  publisher={Elsevier}
}

@book{gramacy2020surrogates,
  title = {Surrogates: {G}aussian Process Modeling, Design and \
    Optimization for the Applied Sciences},
  author = {Robert B. Gramacy},
  publisher = {Chapman Hall/CRC},
  year = {2020}
}

@INPROCEEDINGS{seo2000gaussian,
  author={Sambu Seo and Wallat, M. and Graepel, T. and Obermayer, K.},
  booktitle={Proceedings of the IEEE-INNS-ENNS International Joint Conference on Neural Networks. IJCNN 2000. Neural Computing: New Challenges and Perspectives for the New Millennium }, 
  title={{Gaussian} process regression: active data selection and test point rejection}, 
  year={2000},
  volume={3},
  number={},
  pages={241-246},
  doi={10.1109/IJCNN.2000.861310},
url={https://ieeexplore.ieee.org/document/861310}}

@inproceedings{cohn1993neural,
    author = {Cohn, David},
    title = {Neural network exploration using optimal experiment design},
    booktitle = {Advances in Neural Information Processing Systems},
    volume={6},
    year = {1993},
url={https://proceedings.neurips.cc/paper/1993/hash/d840cc5d906c3e9c84374c8919d2074e-Abstract.html}
}

@book{williams2006gaussian,
  title={{Gaussian} processes for machine learning},
  author={Williams, Christopher KI and Rasmussen, Carl Edward},
  year={2006},
  publisher={The MIT press}
}

@incollection{doucet2001introduction,
    author = {Doucet, Arnaud and De Freitas, Nando and Gordon, Neil},
    title = {An introduction to sequential {Monte Carlo} methods},
    booktitle = {Sequential {Monte Carlo} Methods in Practice},
    pages={3--14},
    publisher = {Springer},
    year = {2001},
    url={https://link.springer.com/book/10.1007/978-1-4757-3437-9}
}

@article{johnson1990minimax,
  title={Minimax and maximin distance designs},
  author={Johnson, Mark E and Moore, Leslie M and Ylvisaker, Donald},
  journal={Journal of Statistical Planning and Inference},
  volume={26},
  number={2},
  pages={131--148},
  year={1990},
  publisher={Elsevier},
 url={https://www.sciencedirect.com/science/article/abs/pii/037837589090122B}
}

@article{kong1994sequential,
  title={Sequential imputations and {Bayesian} missing data problems},
  author={Kong, Augustine and Liu, Jun S and Wong, Wing Hung},
  journal={Journal of the American Statistical Association},
  volume={89},
  number={425},
  pages={278--288},
  year={1994},
  publisher={Taylor \& Francis},
  url={https://www.jstor.org/stable/2291224}
}

@inproceedings{svensson2015marginalizing,
  title={Marginalizing {Gaussian} process hyperparameters using sequential {Monte Carlo}},
  author={Svensson, Andreas and Dahlin, Johan and Sch{\"o}n, Thomas B},
  booktitle={6th IEEE International Workshop on Computational Advances in Multi-Sensor Adaptive Processing (CAMSAP)},
  pages={477--480},
  year={2015}
}

@article{liu1998sequential,
  title={Sequential {Monte Carlo} methods for dynamic systems},
  author={Liu, Jun S and Chen, Rong},
  journal={Journal of the American Statistical Association},
  volume={93},
  number={443},
  pages={1032--1044},
  year={1998},
  publisher={Taylor \& Francis},
url={https://www.jstor.org/stable/2669847}
}

@article{yerramilli2023fully,
  title={Fully {Bayesian} inference for latent variable {Gaussian} process models},
  author={Yerramilli, Suraj and Iyer, Akshay and Chen, Wei and Apley, Daniel W},
  journal={SIAM/ASA Journal on Uncertainty Quantification},
  volume={11},
  number={4},
  pages={1357--1381},
  year={2023},
  publisher={SIAM}
}

@article{helbert2009assessment,
  title={Assessment of uncertainty in computer experiments from Universal to {Bayesian} Kriging},
  author={Helbert, C{\'e}line and Dupuy, Delphine and Carraro, Laurent},
  journal={Applied Stochastic Models in Business and Industry},
  volume={25},
  number={2},
  pages={99--113},
  year={2009},
  publisher={Wiley Online Library},
  url={https://onlinelibrary.wiley.com/doi/abs/10.1002/asmb.743}
}

@inproceedings{lalchand2020approximate,
  title={Approximate inference for fully {Bayesian} {Gaussian} process regression},
  author={Lalchand, Vidhi and Rasmussen, Carl Edward},
  booktitle={Proceedings of The 2nd Symposium on Advances in Approximate {Bayesian} Inference},
  pages={1--12},
  volume={118},
  year={2020},
  url={https://proceedings.mlr.press/v118/lalchand20a.html}
}

@article{karagiannis2019bayesian,
  title={On the {Bayesian} calibration of expensive computer models with input dependent parameters},
  author={Karagiannis, Georgios and Konomi, Bledar A and Lin, Guang},
  journal={Spatial Statistics},
  volume={34},
  pages={100258},
  year={2019},
  publisher={Elsevier},
url={https://www.sciencedirect.com/science/article/abs/pii/S2211675317301215}
}

@article{park2013bayesian,
  title={{Bayesian} active learning for drug combinations},
  author={Park, Mijung and Nassar, Marcel and Vikalo, Haris},
  journal={IEEE Transactions on Biomedical Engineering},
  volume={60},
  number={11},
  pages={3248--3255},
  year={2013},
  publisher={IEEE}
}

@article{xu2014fully,
  title={A fully {Bayesian} method for jointly fitting instrumental calibration and X-ray spectral models},
  author={Xu, Jin and van Dyk, David A and Kashyap, Vinay L and Siemiginowska, Aneta and Connors, Alanna and Drake, Jeremy and Meng, XiaoLi and Ratzlaff, Pete and Yu, Yaming},
  journal={The Astrophysical Journal},
  volume={794},
  number={2},
  pages={97},
  year={2014},
  publisher={IOP Publishing}
}

@article{kennedy2001bayesian,
  title={{Bayesian} calibration of computer models},
  author={Kennedy, Marc C and O'Hagan, Anthony},
  journal={Journal of the Royal Statistical Society Series B: Statistical Methodology},
  volume={63},
  number={3},
  pages={425--464},
  year={2001},
  publisher={Wiley Online Library}
}

@techreport{settles2009active,
  title={Active Learning Literature Survey},
  author={Settles, Burr},
  institution={University of Wisconsin--Madison, Department of Computer Sciences},
  year={2009},
}

@article{liu2018survey,
  title={A survey of adaptive sampling for global metamodeling in support of simulation-based complex engineering design},
  author={Liu, Haitao and Ong, Yew Soon and Cai, Jianfei},
  journal={Structural and Multidisciplinary Optimization},
  volume={57},
  pages={393--416},
  year={2018},
  publisher={Springer},
  url={https://link.springer.com/article/10.1007/s11222-017-9760-8#citeas}
}

@article{ait2025bayesian,
  title={{Bayesian} sequential design of computer experiments for quantile set inversion},
  author={Ait Abdelmalek-Lomenech, Romain and Bect, Julien and Chabridon, Vincent and Vazquez, Emmanuel},
  journal={Technometrics},
  volume={67},
  number={1},
  pages={112--121},
  year={2025},
  publisher={Taylor \& Francis}
}

@article{xu2009development,
  title={{Development of a wood-frame shear wall model in ABAQUS}},
  author={Xu, Jian and Dolan, J Daniel},
  journal={Journal of Structural Engineering},
  volume={135},
  number={8},
  pages={977--984},
  year={2009},
  publisher={American Society of Civil Engineers}
}

@inproceedings{chen2014simulation,
  title={Simulation of the lateral drift of multi-storey light wood frame buildings based on a modified macro-element model},
  author={Chen, Zhiyong and Chui, Ying H and Doudak, Ghasan and Ni, Chun and Mohammad, Mohammad},
  booktitle={Proceedings of the 13th World Conference on Timber Engineering (WCTE 2014)},
  year={2014}
}

@misc{earthquakes2015,
  author       = {{Earthquakes Canada}},
  title        = {Earthquakes in {Canada}},
  year         = {2025},
  note         = {Accessed January 25, 2025},
  howpublished ={\url{https://www.earthquakescanada.ca}},
}

@book{forrester2008engineering,
  title={Engineering design via surrogate modelling: a practical guide},
  author={Forrester, Alexander and Sobester, Andras and Keane, Andy},
  year={2008},
  publisher={John Wiley \& Sons}
}

@article{kennedy2023multilevel,
  title={Multilevel emulation for stochastic computer models with application to large offshore wind farms},
  author={Kennedy, Jack C and Henderson, Daniel A and Wilson, Kevin J},
  journal={Journal of the Royal Statistical Society Series C: Applied Statistics},
  volume={72},
  number={3},
  pages={608--627},
  year={2023},
  publisher={Oxford University Press US}
}

@article{mckinley2018approximate,
  title={Approximate {Bayesian} computation and simulation-based inference for complex stochastic epidemic models},
  author={McKinley, Trevelyan J and Vernon, Ian and Andrianakis, Ioannis and McCreesh, Nicky and Oakley, Jeremy E and Nsubuga, Rebecca N and Goldstein, Michael and White, Richard G},
journal={Statistical Science},
  volume={33},
  number={1},
  pages={4--18},
  year={2018},
url={https://projecteuclid.org/journals/statistical-science/volume-33/issue-1/Approximate-{Bayesian}-Computation-and-Simulation-Based-Inference-for-Complex-Stochastic/10.1214/17-STS618.full}
}

@article{peleg2017advanced,
  title={{An advanced stochastic weather generator for simulating 2-D high-resolution climate variables}},
  author={Peleg, Nadav and Fatichi, Simone and Paschalis, Athanasios and Molnar, Peter and Burlando, Paolo},
  journal={Journal of Advances in Modeling Earth Systems},
  volume={9},
  number={3},
  pages={1595--1627},
  year={2017},
  publisher={Wiley Online Library}
}

@article{richardson1981stochastic,
  title={Stochastic simulation of daily precipitation, temperature, and solar radiation},
  author={Richardson, Clarence W},
  journal={Water Resources Research},
  volume={17},
  number={1},
  pages={182--190},
  year={1981},
  publisher={Wiley Online Library}
}

@article{baker2022analyzing,
  title={Analyzing stochastic computer models: A review with opportunities},
  author={Baker, Evan and Barbillon, Pierre and Fadikar, Arindam and Gramacy, Robert B and Herbei, Radu and Higdon, David and Huang, Jiangeng and Johnson, Leah R and Ma, Pulong and Mondal, Anirban and others},
  journal={Statistical Science},
  volume={37},
  number={1},
  pages={64--89},
  year={2022},
  publisher={Institute of Mathematical Statistics}
}

@book{chernoff1992sequential,
  title={Sequential design of experiments},
  author={Chernoff, Herman},
  year={1992},
  publisher={Springer}
}

@book{harville1998matrix,
    author = {Harville, David A},
    title = {Matrix algebra from a statistician's perspective},
    publisher = {Springer},
    year = {1998},
url={https://link.springer.com/book/10.1007/b98818}
}

@book{ni2015midrise,
  title     = {Mid-Rise Wood-Frame Construction Handbook},
  author    = {Ni, Chun and Popovski, Marjan},
  year      = {2015},
  publisher = {FPInnovations}
}

@techreport{triggs2015review,
  author      = {Triggs, Greg},
  title       = {{Review of Building Code Approaches for Podium Structures—Western US Examples}},
  year        = {2015},
  institution = {BC Advisory Group on Advanced Wood Design Solutions}
}

@book{chopra2020earthquake,
  title={Earthquake engineering for concrete dams: analysis, design, and evaluation},
  author={Chopra, Anil K},
  year={2020},
  publisher={John Wiley \& Sons}
}

@book{nrc2015nbcc,
  author    = {{National Research Council}},
  title     = {National Building Code of Canada},
  year      = {2015},
  address   = {Ottawa},
  publisher = {National Research Council}
}

@book{asce2016minimum,
  title     = {Minimum Design Loads and Associated Criteria for Buildings and Other Structures},
  author    = {{ASCE}},
  year      = {2016},
  publisher = {ASCCE/SEI 7},
  address   = {Reston, VA}
}

@inproceedings{chevalier2014corrected,
  title={Corrected kriging update formulae for batch-sequential data assimilation},
  author={Chevalier, Cl{\'e}ment and Ginsbourger, David and Emery, Xavier},
  booktitle={Mathematics of Planet Earth: Proceedings of the 15th Annual Conference of the International Association for Mathematical Geosciences},
  pages={119--122},
  year={2014},
  organization={Springer},
url={https://link.springer.com/chapter/10.1007/978-3-642-32408-6_29#citeas}
}

@book{applied2018seismic,
  title={Seismic Performance Assessment of Buildings},
  author={{Applied Technology Council}},
  year={2018},
edition={2nd},
volume={1},
  publisher={Federal Emergency Management Agency}
}

@article{lyu2021evaluating,
  title={Evaluating Gaussian process metamodels and sequential designs for noisy level set estimation},
  author={Lyu, Xiong and Binois, Micka{\"e}l and Ludkovski, Michael},
  journal={Statistics and computing},
  volume={31},
  number={43},
  year={2021},
  publisher={Springer},
url={https://link.springer.com/article/10.1007/s11222-021-10014-w#citeas}
}

@article{friedman1991multivariate,
  title={Multivariate adaptive regression splines},
  author={Friedman, Jerome H},
  journal={The Annals of Statistics},
  volume={19},
  number={1},
  pages={1--67},
  year={1991},
  publisher={Institute of Mathematical Statistics},
  url={https://projecteuclid.org/journals/annals-of-statistics/volume-19/issue-1/Multivariate-Adaptive-Regression-Splines/10.1214/aos/1176347963.full}
}

@book{davis2007methods,
  title={Methods of numerical integration},
  author={Davis, Philip J and Rabinowitz, Philip},
  year={2007},
  publisher={Courier Corporation},
}

@inproceedings{lazaro2011variational,
  title={Variational Heteroscedastic {Gaussian} Process Regression},
  author={L{\'a}zaro-Gredilla, Miguel and Titsias, Michalis K},
  booktitle={Proceedings of the 28th International Conference on Machine Learning},
  pages={841--848},
  year={2011},
  url={https://dl.acm.org/doi/10.5555/3104482.3104588}
}

@article{zhu2021emulation,
author = {Zhu, Xujia and Sudret, Bruno},
title = {Emulation of Stochastic Simulators Using Generalized Lambda Models},
journal = {SIAM/ASA Journal on Uncertainty Quantification},
volume = {9},
number = {4},
pages = {1345-1380},
year = {2021},
doi = {10.1137/20M1337302},
URL = {https://doi.org/10.1137/20M1337302}
}

@article{Surer02012026,
author = {Özge Sürer},
title = {Batch Sequential Experimental Design for Calibration of Stochastic Simulation Models},
journal = {Technometrics},
volume = {68},
number = {1},
pages = {1--13},
year = {2026},
publisher = {Taylor \& Francis},
doi = {10.1080/00401706.2025.2520860},
URL = {    https://doi.org/10.1080/00401706.2025.2520860}
}

@article{Beek2020ScalableAB,
  title={Scalable Adaptive Batch Sampling in Simulation-Based Design With Heteroscedastic Noise},
  author={Anton van Beek and Umar Farooq Ghumman and Joydeep Munshi and Siyu Tao and TeYu Chien and Ganesh Balasubramanian and Matthew Plumlee and Daniel W. Apley and Wei Chen},
  journal={Journal of Mechanical Design},
  year={2021},
  volume = {143},
number = {3},
  pages = {031709},
  url={https://api.semanticscholar.org/CorpusID:228865887}
}

@article{zhang2022batch,
  title={Batch-sequential design and heteroskedastic surrogate modeling for delta smelt conservation},
  author={Zhang, Boya and Gramacy, Robert B and Johnson, Leah R and Rose, Kenneth A and Smith, Eric},
  journal={The Annals of Applied Statistics},
  volume={16},
  number={2},
  pages={816--842},
  year={2022},
  publisher={JSTOR}
}

@article{Jordan1999,
  title   = {An Introduction to Variational Methods for Graphical Models},
  author  = {Jordan, Michael I. and Ghahramani, Zoubin and Jaakkola, Tommi S. and Saul, Lawrence K.},
  journal = {Machine Learning},
  year    = {1999},
  volume  = {37},
  number  = {2},
  pages   = {183--233}
}

@article{WainwrightJordan2008,
  title   = {Graphical Models, Exponential Families, and Variational Inference},
  author  = {Wainwright, Martin J. and Jordan, Michael I.},
  journal = {Foundations and Trends in Machine Learning},
  year    = {2008},
  volume  = {1},
  number  = {1--2},
  pages   = {1--305}
}

@article{Blei2017VI,
  title   = {Variational Inference: A Review for Statisticians},
  author  = {Blei, David M. and Kucukelbir, Alp and McAuliffe, Jon D.},
  journal = {Journal of the American Statistical Association},
  year    = {2017},
  volume  = {112},
  number  = {518},
  pages   = {859--877},
  url={https://www.tandfonline.com/doi/full/10.1080/01621459.2017.1285773#}
}

@inproceedings{Hensman2013,
  title     = {Gaussian Processes for Big Data},
  author    = {Hensman, James and Fusi, Nicolo and Lawrence, Neil D.},
  booktitle = {Proceedings of the Conference on Uncertainty in Artificial Intelligence (UAI)},
  year      = {2013}
}

@inproceedings{Hensman2015,
  title     = {Scalable Variational Gaussian Process Classification},
  author    = {Hensman, James and Matthews, Alexander G. de G. and Ghahramani, Zoubin},
  booktitle = {Proceedings of the International Conference on Artificial Intelligence and Statistics (AISTATS)},
  year      = {2015}
}

@inproceedings{Kersting2007,
  title     = {Most Likely Heteroscedastic Gaussian Process Regression},
  author    = {Kersting, Kristian and Plagemann, Christian and Pfaff, Patrick and Burgard, Wolfram},
  booktitle = {Proceedings of the International Conference on Machine Learning (ICML)},
  year      = {2007}
}

@INPROCEEDINGS{QuadriantoMAP2009,
  author={Quadrianto, Novi and Kersting, Kristian and Reid, Mark D. and Caetano, Tibério S. and Buntine, Wray L.},
  booktitle={2009 Ninth IEEE International Conference on Data Mining}, 
  title={Kernel Conditional Quantile Estimation via Reduction Revisited}, 
  year={2009},
  volume={},
  number={},
  pages={938-943},
  url={https://ieeexplore.ieee.org/document/5360337},
  doi={10.1109/ICDM.2009.82}}

@ARTICLE{GPflow2017,
    author = {Matthews, Alexander G. de G. and {van der Wilk}, Mark and Nickson, Tom and Fujii, Keisuke. and {Boukouvalas}, Alexis and {Le{\'o}n-Villagr{\'a}}, Pablo and Ghahramani, Zoubin and Hensman, James},
    title = "{{GP}flow: A {G}aussian process library using {T}ensor{F}low}",
    journal = {Journal of Machine Learning Research},
    year    = {2017},
    month = {apr},
    volume  = {18},
    number  = {40},
    pages   = {1-6},
    url     = {http://jmlr.org/papers/v18/16-537.html}
}

@InProceedings{Titsias2009,
  title = 	 {Variational Learning of Inducing Variables in Sparse Gaussian Processes},
  author = 	 {Titsias, Michalis},
  booktitle = 	 {Proceedings of the Twelfth International Conference on Artificial Intelligence and Statistics},
  pages = 	 {567--574},
  year = 	 {2009},
  editor = 	 {van Dyk, David and Welling, Max},
  volume = 	 {5},
  series = 	 {Proceedings of Machine Learning Research},
  address = 	 {Hilton Clearwater Beach Resort, Clearwater Beach, Florida USA},
  month = 	 {16--18 Apr},
  publisher =    {PMLR},
  url = 	 {https://proceedings.mlr.press/v5/titsias09a.html},
}

@article{van2009adaptive,
  title={Adaptive Bayesian estimation using a {Gaussian} random field with inverse Gamma bandwidth},
  author={van der Vaart, AW and van Zanten, JH},
  journal={The Annals of Statistics},
  volume={37},
  number={1},
  pages={2655--2675},
  year={2009}
}

@article{gelman2006prior,
  title={Prior distributions for variance parameters in hierarchical models},
  author={Gelman, Andrew},
  journal={Bayesian Analysis},
  volume={1},
  number={3},
  pages={515--533},
  year={2006}
}

@article{patil2025vecchia,
  title={Vecchia approximated {Bayesian} heteroskedastic {Gaussian} processes},
  author={Patil, Parul V and Gramacy, Robert B and Carey, Cayelan C and Thomas, R Quinn},
  journal={arXiv preprint arXiv:2507.07815},
  year={2025}
}

\FloatBarrier

\newpage

\begin{center}
{\large\bf SUPPLEMENTARY MATERIAL}
\end{center}
\supplementarysection

\renewcommand{\thefigure}{S\arabic{figure}}
\renewcommand{\thetable}{S\arabic{table}}
\renewcommand{\thealgorithm}{S\arabic{algorithm}}
\renewcommand{\theequation}{S\arabic{equation}}

\section{Practical Effect of ESS Threshold on the Performance of FB-BIM}\label{appendix: tau_comparison}

\begin{figure}[!htbp]
	\captionsetup[subfigure]{labelformat=empty}
	\begin{subfigure}[b]{1\linewidth}
		\centering
		\includegraphics[width=0.9\linewidth]{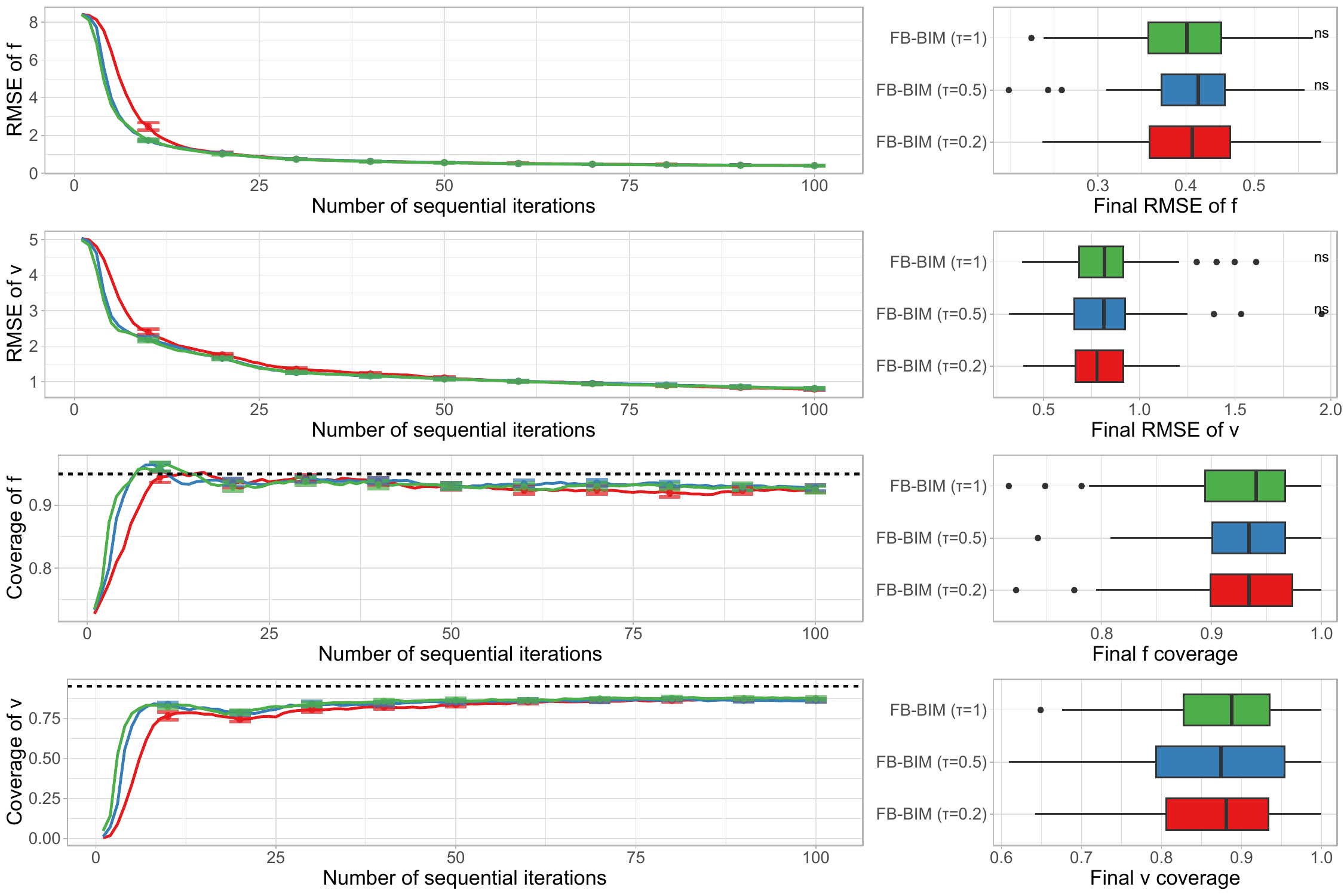}
		\caption{(a) 1-D example.}
		\label{Appendix: 1-d RMSE boxplot comparisons.a}
	\end{subfigure} 
\end{figure}
\begin{figure}[ht]\ContinuedFloat
	\begin{subfigure}[b]{1\linewidth}
		\centering
		\includegraphics[width=0.9\linewidth]{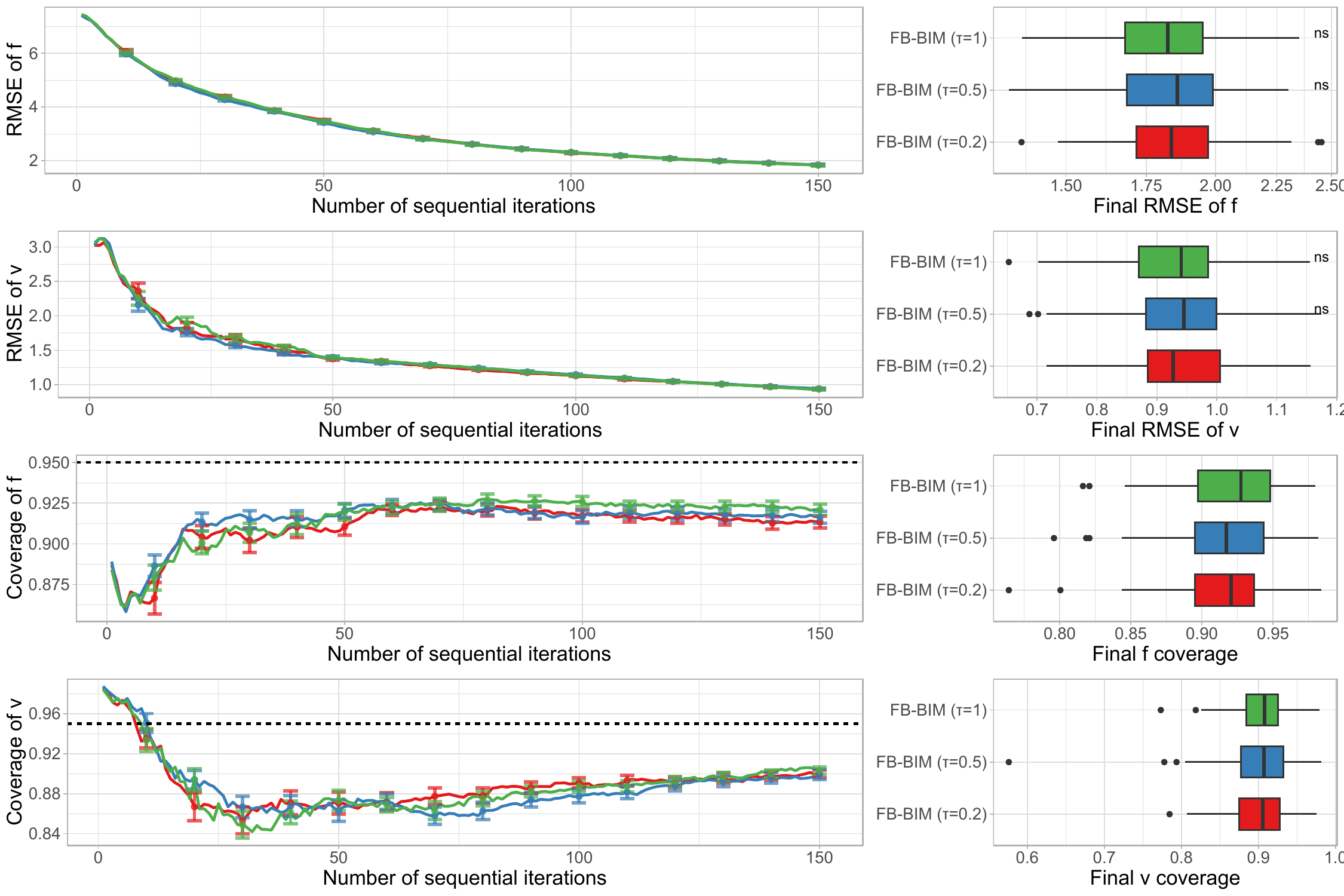} 
		\caption{2-D example. }\label{Appendix: 2-d RMSE boxplot comparisons.b}
	\end{subfigure} 
	\caption{Line plots of prediction performance (left) and final-iteration boxplots (right) for FB-BIM with different values of $\tau$, denoted by green ($\tau=1$), blue ($\tau=0.5$), and red ($\tau=0.2$), respectively. The black dashed line in the coverage plots indicates the 95\% nominal level. Error bars (mean $\pm$ standard deviation) are displayed every 10 iterations. A two-sample t-test is conducted on the final results to compare the mean RMSE metrics between FB-BIM ($\tau=0.2$) and each alternative strategy. Significance levels are indicated beside the boxplots: ``ns''(p$>$0.05), ``*''(p$<=$0.05), ``**''(p$<=$0.01), ``***''(p$<=$0.001). \vspace{1in}} 
\end{figure}

Figures \ref{Appendix: 1-d RMSE boxplot comparisons.a} and \ref{Appendix: 2-d RMSE boxplot comparisons.b} compare the performance of the FB-BIM strategy with different ESS thresholds $\tau M$ (with $\tau=0.2, 0.5, 1$) on the 1-D and 2-D examples, when taking $M=500$ posterior samples.
In the 1-D case, the performance results of all metrics suggest no significant differences in prediction performance across $\tau$ values in the FB-BIM strategy. Across the 100 macro-replications, for FB-BIM with $\tau=0.2$, MCMC needed to be run a median of 9 times (SD = 1.4) out of the 100 sequential iterations, while for the strategy with $\tau=0.5$, MCMC needed to be run a median of 17 times (SD = 1.9). 
Similarly, we observe that the FB-BIM strategy shows no significant differences in prediction performances across different values of $\tau$ in the 2-D example. Across the 100 macro-replications, for the FB-BIM strategy with $\tau=0.2$, MCMC needed to be run a median of 14 times (SD = 2.1) out of 150 sequential iterations, while for the strategy with $\tau=0.5$, MCMC needed to be run a median of 27 times (SD = 3.1). 

Figure~\ref{MCMC-freq} depicts the frequency of MCMC being run across the 100 macro-replications as sequential iterations progress. Since MCMC is triggered whenever the importance weights are reset, this also reflects the frequency of weight resetting. We observe that the frequency trajectories are very similar across different $\tau$ values. In both examples, the frequency of MCMC peaks during early iterations and then gradually decreases as the sequential design proceeds, as the posteriors are impacted less by each new observation.

\begin{figure}[!htbp]
	\captionsetup[subfigure]{labelformat=empty}
	\begin{subfigure}[b]{1\linewidth}
		\centering
		\includegraphics[width=0.9\linewidth]{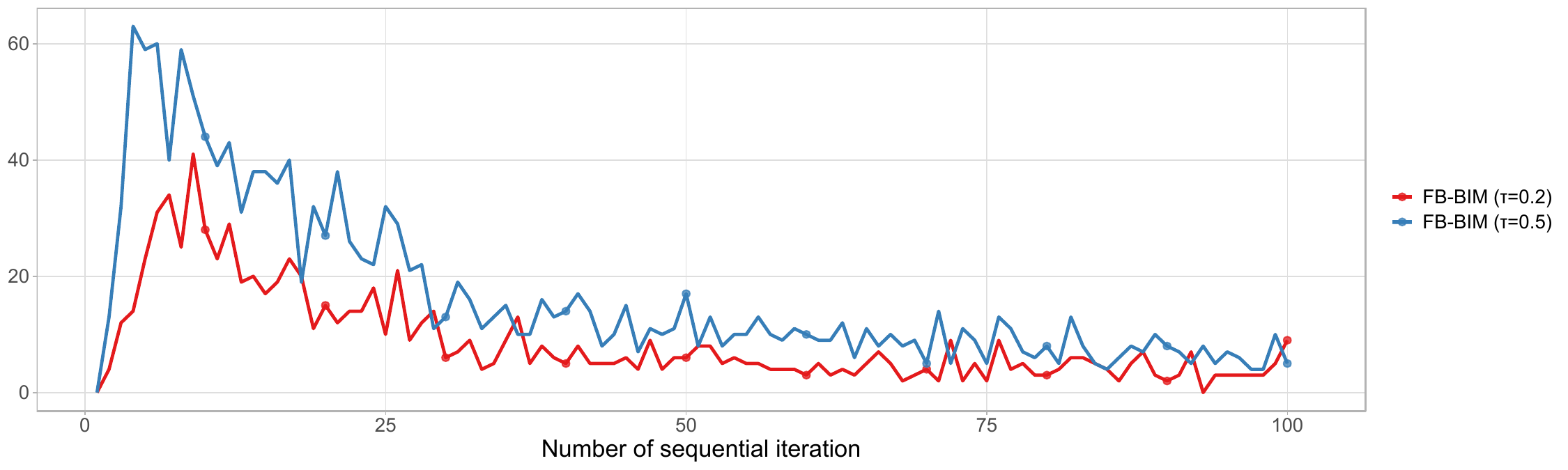}
		\caption{(a) 1-D example.}
	\end{subfigure} 
\end{figure}
\begin{figure}[ht]\ContinuedFloat
	\begin{subfigure}[b]{1\linewidth}
		\centering
		\includegraphics[width=0.9\linewidth]{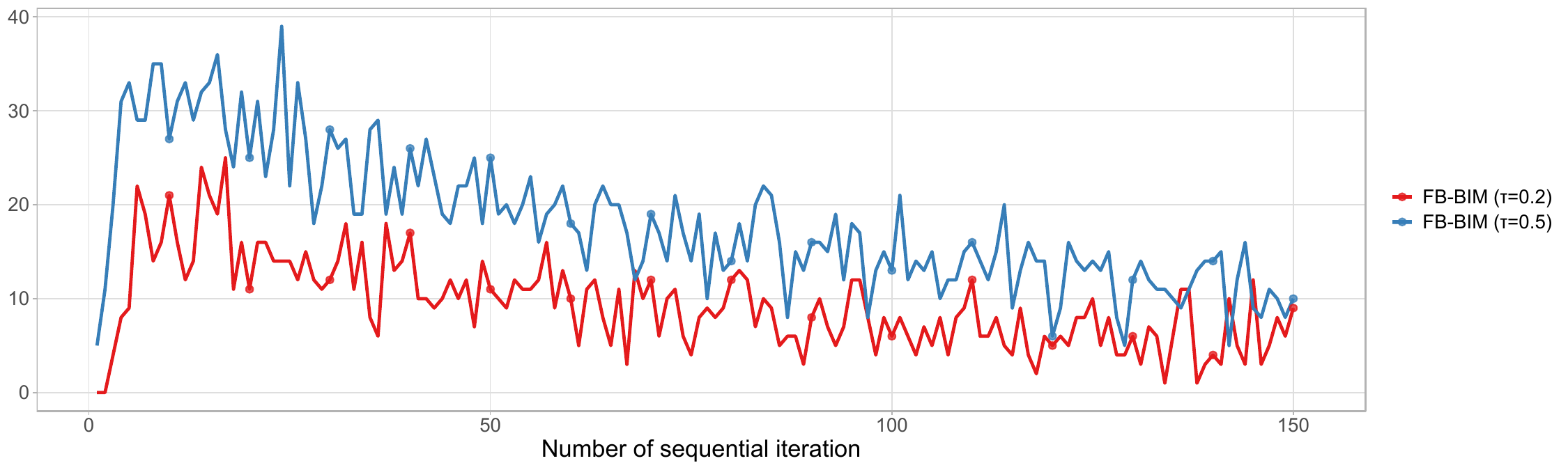} 
		\caption{2-D example. }
	\end{subfigure} 
	\caption{Line plots of MCMC frequency of FB-BIM with different values of $\tau$, denoted by blue ($\tau=0.5$), and red ($\tau=0.2$), respectively.
	} \label{MCMC-freq}
\end{figure}

The empirical results provided above suggest that sequential importance sampling maintains the robust prediction performance of FB-BIM across a broad range of $\tau$ values. The computational savings realized by reducing MCMC re-runs are particularly significant at later sequential iterations, since each gradient evaluation in NUTS scales as $\mathcal{O}(N^3)$. Moreover, a single chain of MCMC is inherently serial and cannot be easily parallelized, unlike the lookahead step of expected BIMSPE calculations that are independent across candidate points. Consequently, reducing the number of MCMC re-runs by $90\%$ can translate to a $60$-$90\%$ reduction in the wall-clock time of FB-BIM on a multi-core workstation. Given these results, which demonstrate substantial computational savings with negligible impact on accuracy, we choose $\tau=0.2$ as the SIS threshold in FB-BIM for all of our examples and applications.

\section{Additional Robustness Experiments}
\subsection{Prior Sensitivity \label{Section2:sensitivity}}

To assess the robustness of our results, we examine how varying the prior hyperparameters may impact the prediction performance of the fitted model. While the default priors suggested in the main text Section 2.5.2 are designed to be weakly informative, we investigate the sensitivity of FB-BIM to these choices by considering hyperparameter values that lead to more diffuse priors, and compare the results with those obtained under the original priors. We do this analysis for each of the 1-, 2-, and 5-D examples in Sections 3.1, 3.2, and 3.3 of the main text, with the following modifications: (a) we assign a more diffuse prior for $\sigma_v$, namely $\sigma_v \sim \text{Half-Normal}(0,2)$;
(b) we assign a more diffuse prior to $\sigma_f$, namely $\sigma_f \sim \text{Cauchy}(0,10)$; 
(c) we shift the prior mass towards larger lengthscales of both $f$- and $v$-processes: specifically, we keep $\alpha_{fh} = \alpha_{vh} = 3$ for each $h = 1, \ldots, d$ to ensure the priors have finite variance; we set the scale parameters $\beta_{vh} = \beta_{fh}$ to be the range of the design space in the $h$-th dimension of $\mathcal{X}$; overall, this yields flatter lengthscale priors that shift the prior modes and prior means to 0.25 and 0.5 times the range of the design grid, respectively.

\begin{figure}[!htbp]
	\captionsetup[subfigure]{labelformat=empty}
	\begin{subfigure}[b]{1\linewidth}
		\centering
		\includegraphics[width=0.9\textwidth]{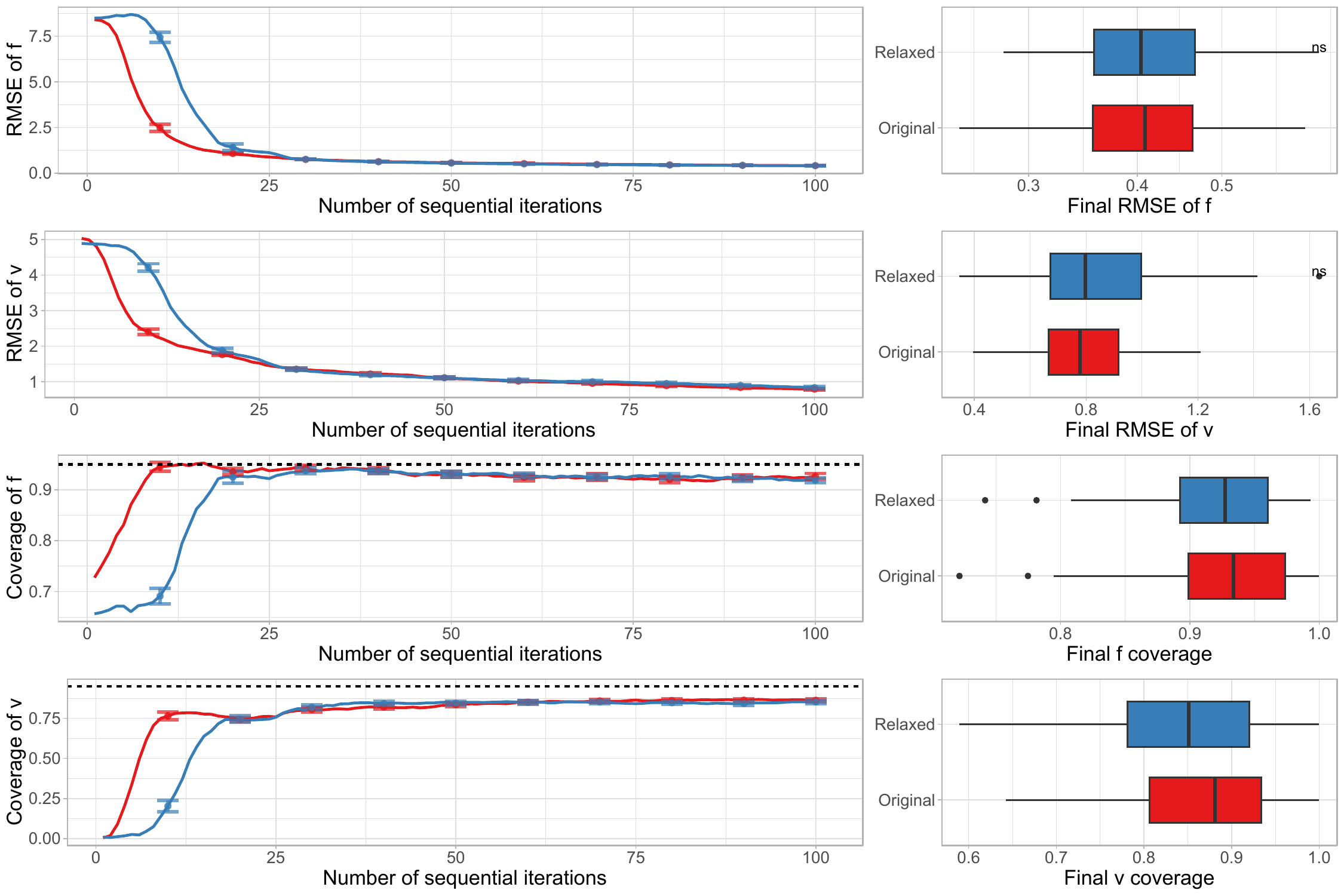}
		\caption{(a) 1-D example}
	\end{subfigure} 
\end{figure}
\begin{figure}[ht]\ContinuedFloat
	\begin{subfigure}[b]{1\linewidth}
		\centering
		\includegraphics[width=0.9\textwidth]{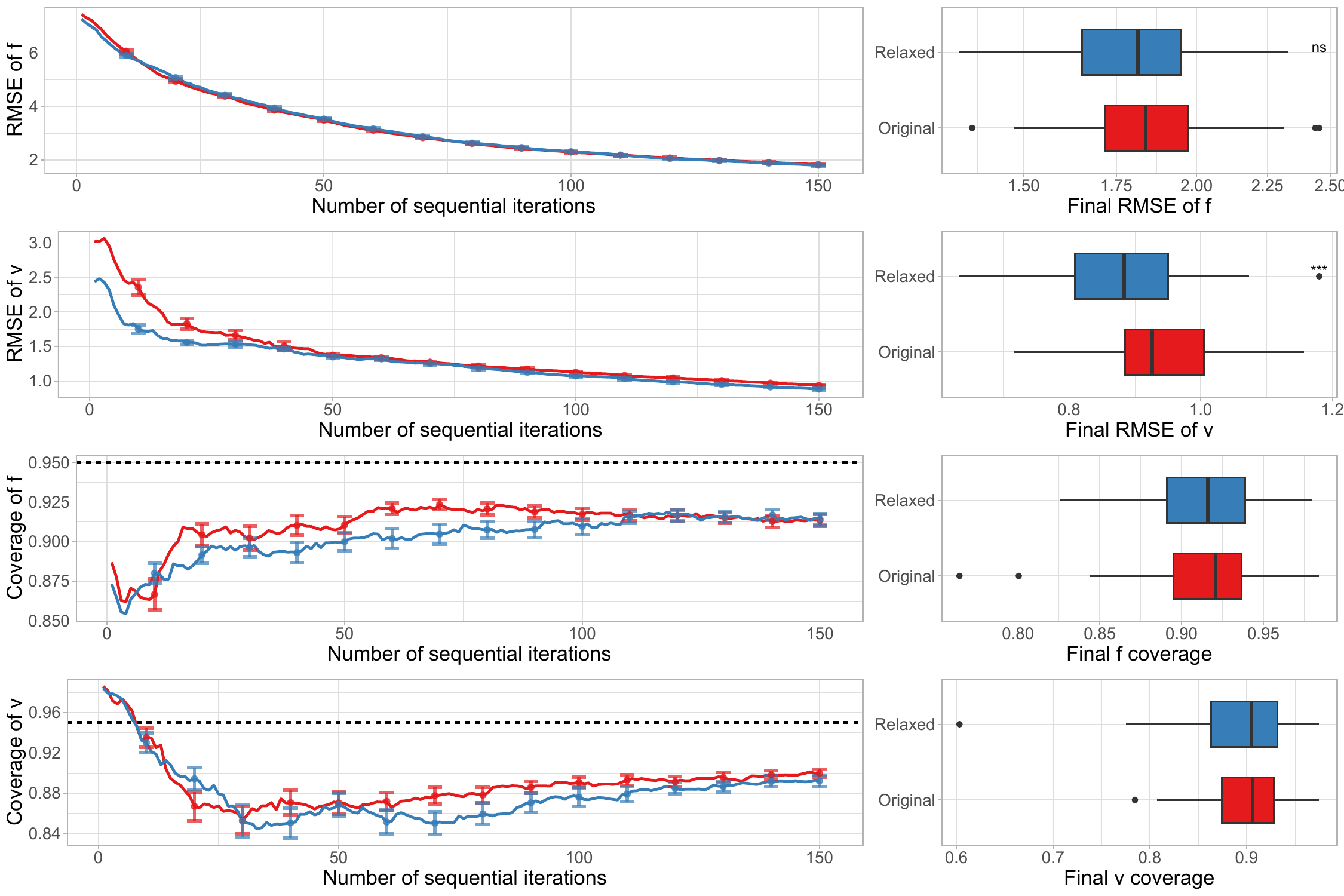}
		\caption{2-D example}
	\end{subfigure}
\end{figure}
\begin{figure}[ht]\ContinuedFloat
	\begin{subfigure}[b]{1\linewidth}
		\centering
		\includegraphics[width=0.9\textwidth]{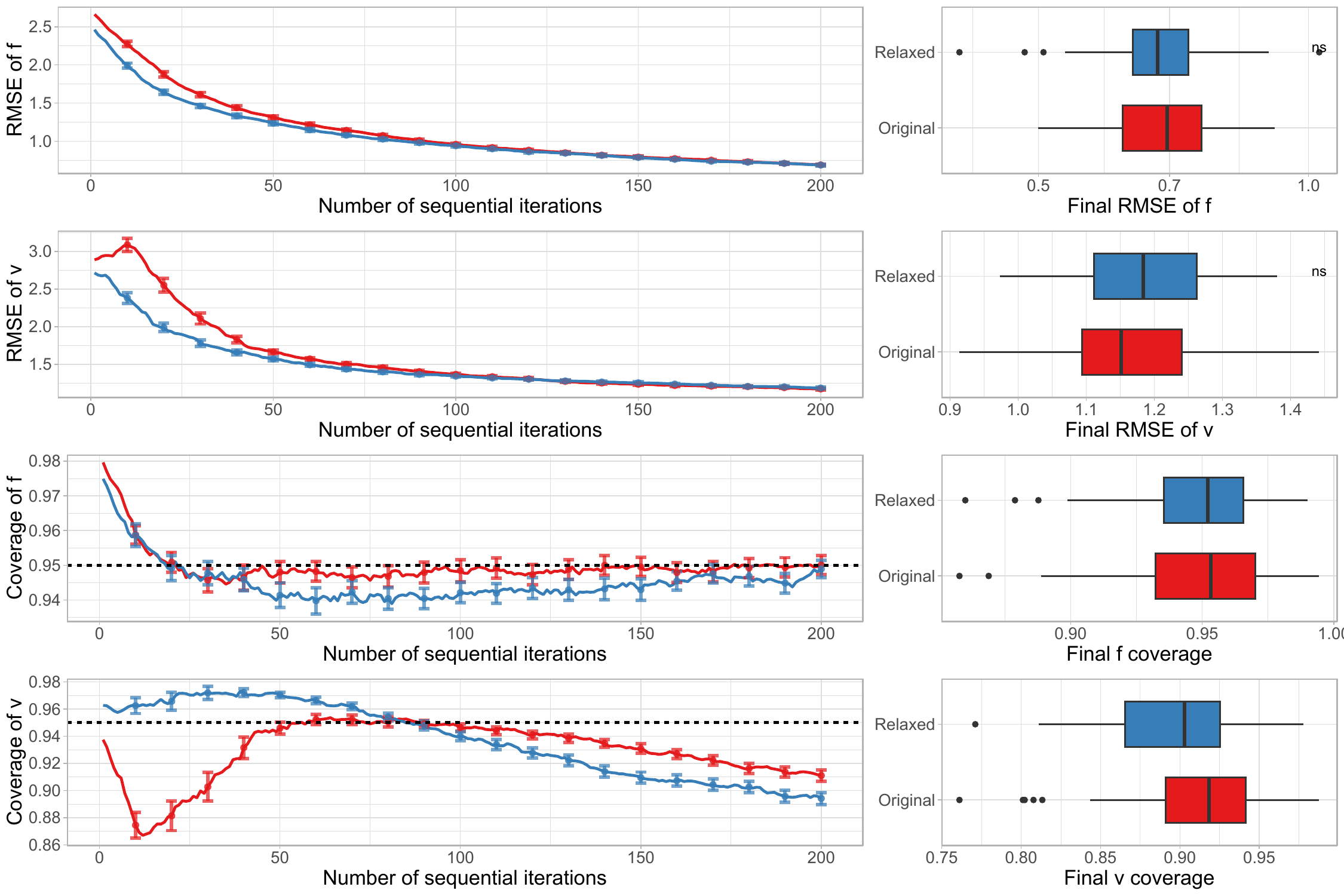}
		\caption{5-D example}
	\end{subfigure}
	\caption{Line plots of prediction performance (left) and final-iteration boxplots (right) for FB-BIM under the original and the more diffuse prior settings. The black dashed line in the coverage plots indicates the 95\% nominal level. Error bars (mean $\pm$ standard deviation) are displayed every 10 iterations. A two sample t-test is conducted on the final results to compare the mean RMSE metrics between the two prior settings. Significance levels are indicated beside the boxplots: ``ns''(p$>$0.05), ``*''(p$<=$0.05), ``**''(p$<=$0.01), ``***''(p$<=$0.001).
	}  \label{func sensitivity}
\end{figure}

Figure \ref{func sensitivity} compares the prediction performances of FB-BIM under the original and the more diffuse prior settings. The trajectories of the metrics exhibit notable differences mainly during the early sequential iterations, when there is limited observed data and the priors have a stronger influence; e.g., the relaxed priors may be better calibrated than the original ones for the $v$-process in the 2-D and 5-D examples, as evidenced by lower log-noise RMSEs in the very early iterations (see also the final posterior samples in Figures \ref{2-d: prior_posterior} and \ref{5-d: prior_posterior}, which support this observation). As data are accumulated through sequential iterations, the performance of FB-BIM under both prior configurations largely align. These findings demonstrate that the strategy is relatively robust to prior specification; the likelihood quickly dominates both sets of priors, leading to consistent predictive performance.

\FloatBarrier

\subsection{Overfit Robustness to Homoscedastic Noise \label{sec-overfit-robustness}}

To assess the robustness of our results against potential overfitting, we evaluate the predictive performance of FB-BIM, which models the log-noise variance using a second (latent) GP, in scenarios where the ground-truth noise is actually homoscedastic. We consider three examples that parallel the 1-, 2-, and 5-dimensional heteroscedastic examples presented in Sections~3.1, 3.2, and 3.3 of the main text. In each example, the homoscedastic noise variance is set to the mean of the corresponding heteroscedastic noise variance over $\mathcal{D}$, to maintain the same average noise level. Specifically, the constant noise variances (corresponding to the average of $g(\x)$ in the heteroscedastic version of the examples) are set to 1.1, 2.9, and 2.3 in the 1-, 2-, and 5-dimensional examples, respectively. In addition, we include results from Homo-BIM, which correctly specifies the homoscedastic noise structure, as a baseline for comparison in each example; see Supplement \ref{e:homo} for implementation details of the Homo-BIM strategy.

As shown in the panels of Figure \ref{overfit-robustness} for each example, the trajectories of the mean-response and log-noise RMSEs largely overlap between FB-BIM and Homo-BIM. The final-iteration boxplots show no significant differences between Homo-BIM and FB-BIM in terms of mean-response RMSE. Homo-BIM consistently achieves a slightly lower log-noise RMSE, which is expected given its correct specification of a homoscedastic noise structure, but this advantage does not extend to mean-surface prediction. Note that the empirical coverage of the log-noise surface under Homo-BIM for each macro-replication is effectively binary in this setup: it is $1$ when the true constant log-noise lies within its corresponding credible interval, and $0$ otherwise. Consequently, the final-iteration boxplots for Homo-BIM have a large point mass at $1$ with a small number of macro-replications appearing as outliers at $0$. The line plots, which instead display the mean and standard deviation across all macro-replications, indicate that Homo-BIM has close to nominal coverage for the 1-D and 5-D examples, while FB-BIM tends to exhibit some overcoverage of the log-noise in this homoscedastic setting. 
Overall, these results suggest that while FB-BIM is designed for heteroscedastic simulation models, it remains robust in homoscedastic settings, without any notable overfitting or performance decrease relative to a correctly specified homoscedastic model.

\begin{figure}[!htbp]
	\captionsetup[subfigure]{labelformat=empty}
	\begin{subfigure}[b]{1\linewidth}
		\centering
		\includegraphics[width=0.9\textwidth]{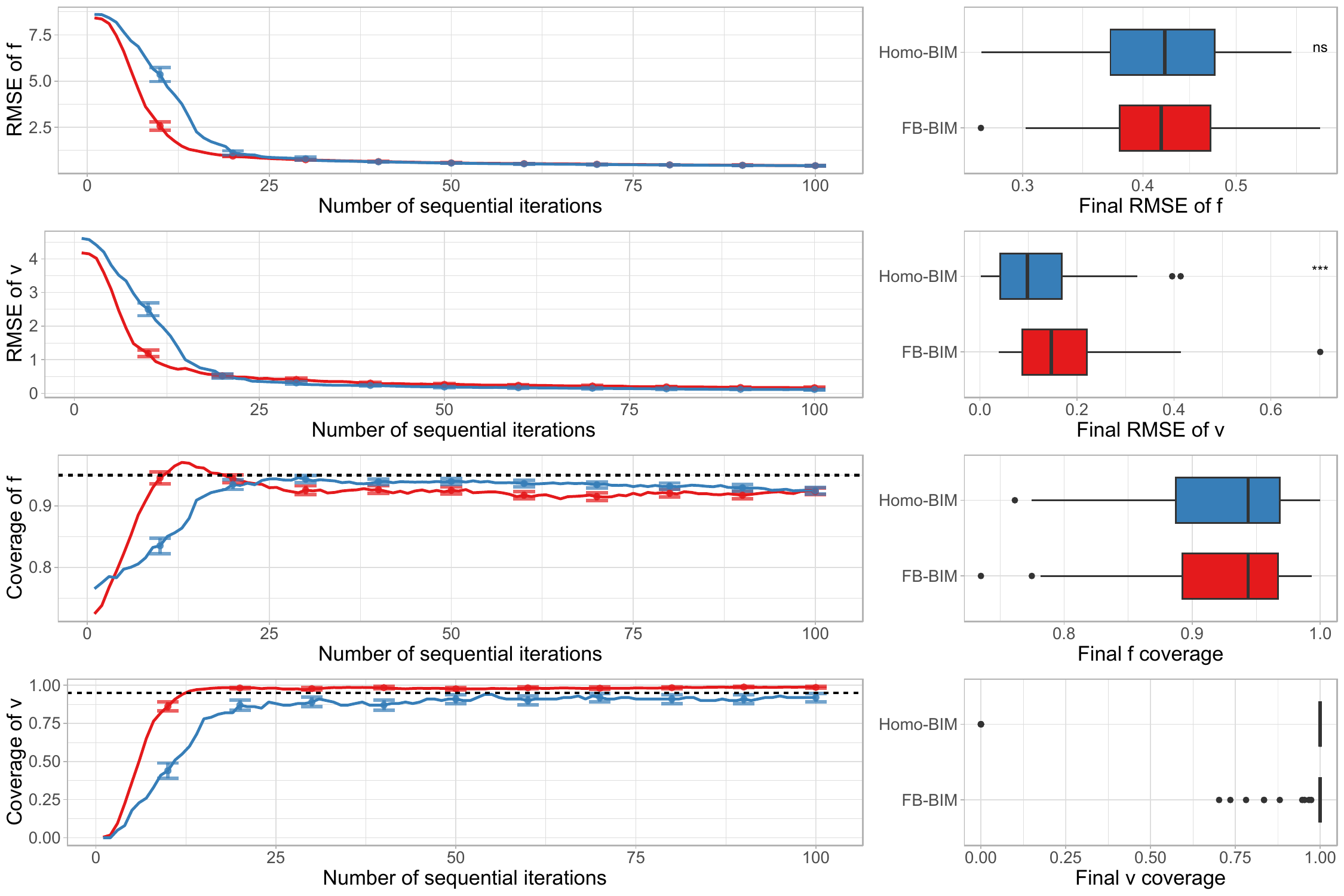}
		\caption{(a) homoscedastic 1-D example}
	\end{subfigure} 
\end{figure}
\begin{figure}[ht]\ContinuedFloat
	\begin{subfigure}[b]{1\linewidth}
		\centering
		\includegraphics[width=0.9\textwidth]{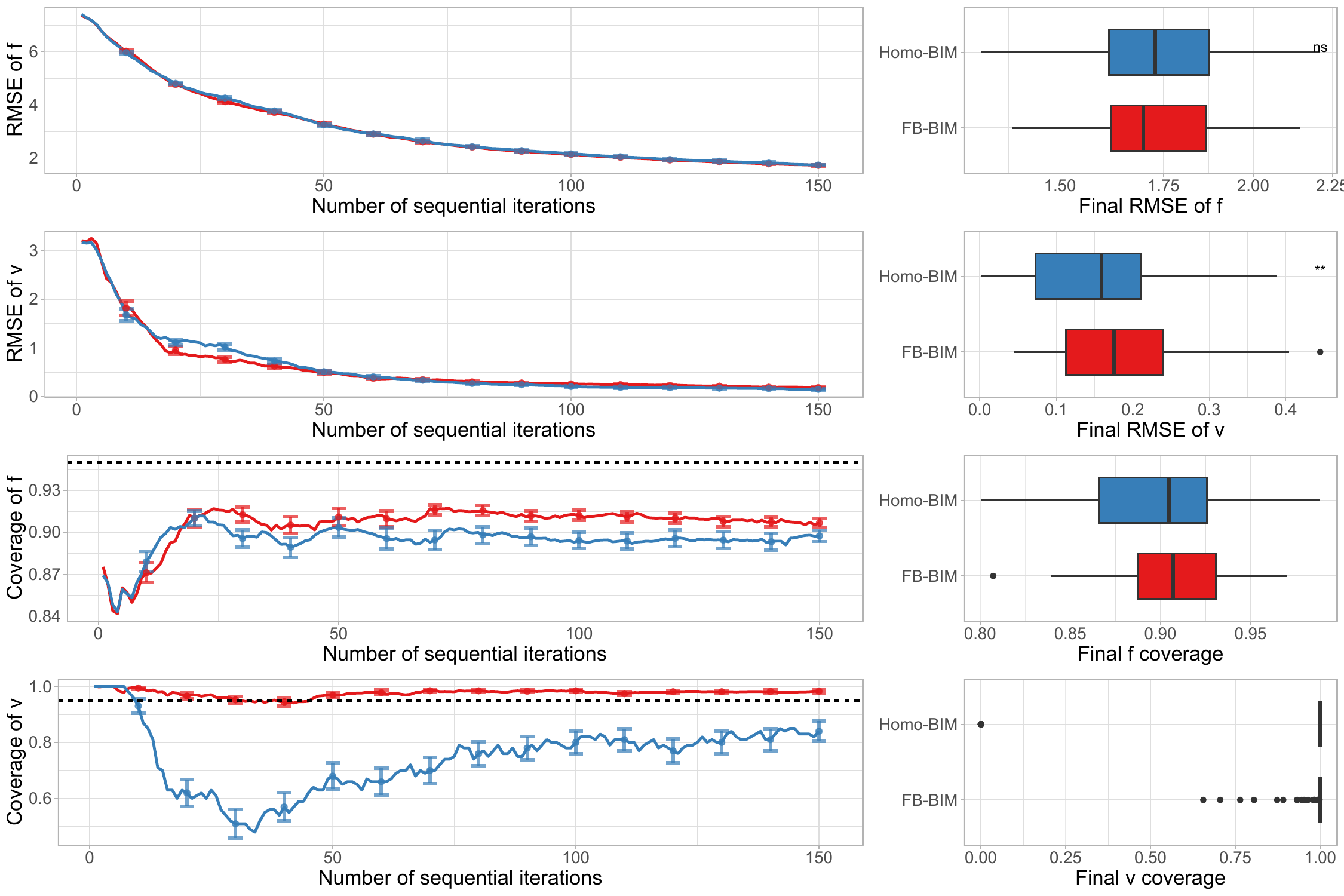}
		\caption{homoscedastic 2-D example}
	\end{subfigure}
\end{figure}
\begin{figure}[ht]\ContinuedFloat
	\begin{subfigure}[b]{1\linewidth}
		\centering
		\includegraphics[width=0.9\textwidth]{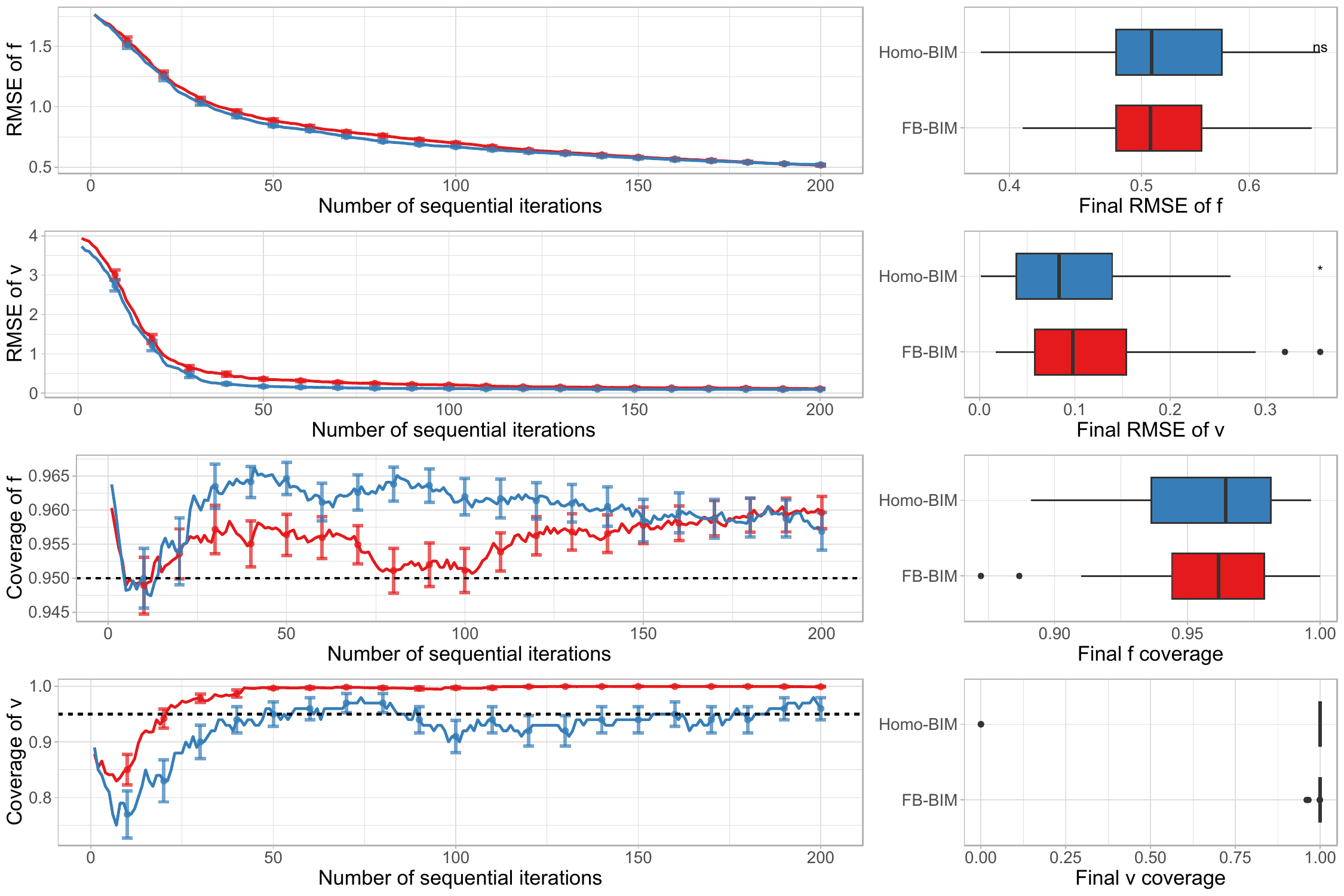}
		\caption{homoscedastic 5-D example}
	\end{subfigure}
	\caption{Line plots of prediction performance (left) and final-iteration boxplots (right) for FB-BIM and Homo-BIM under the homoscedastic simulations. The black dashed line in the coverage plots indicates the 95\% nominal level. Error bars (mean $\pm$ standard deviation) are displayed every 10 iterations. A two sample t-test is conducted on the final results to compare the mean RMSE metrics between the two strategies. Significance levels are indicated beside the boxplots: ``ns''(p$>$0.05), ``*''(p$<=$0.05), ``**''(p$<=$0.01), ``***''(p$<=$0.001).} \label{overfit-robustness}
\end{figure}

\FloatBarrier
\subsection{Sensitivity to Number of $\tilde{y}_{N+1}$ Samples $K$\label{sensitivity_tilde_y}} 

In Section~2.3 of the main text, the evaluation of expected BIMSPE in (7) involves an outer expectation with respect to the posterior predictive distribution of $\tilde{y}_{N+1}$. For computational efficiency, our proposed implementation approximates this expectation using a single Monte Carlo draw ($K=1$). Here, we assess the impact of this approximation by comparing the predictive performance of FB-BIM with $K=1$ against a more computationally intensive version that uses $K=20$ samples of $\tilde{y}_{N+1}$, denoted as FB-BIM20. For this robustness experiment, we considered the same 1- and 2-dimensional examples presented in Sections~3.1 and~3.2 of the main text, keeping all other settings constant.

\begin{figure}
	\centering
	\includegraphics[width=0.8\linewidth]{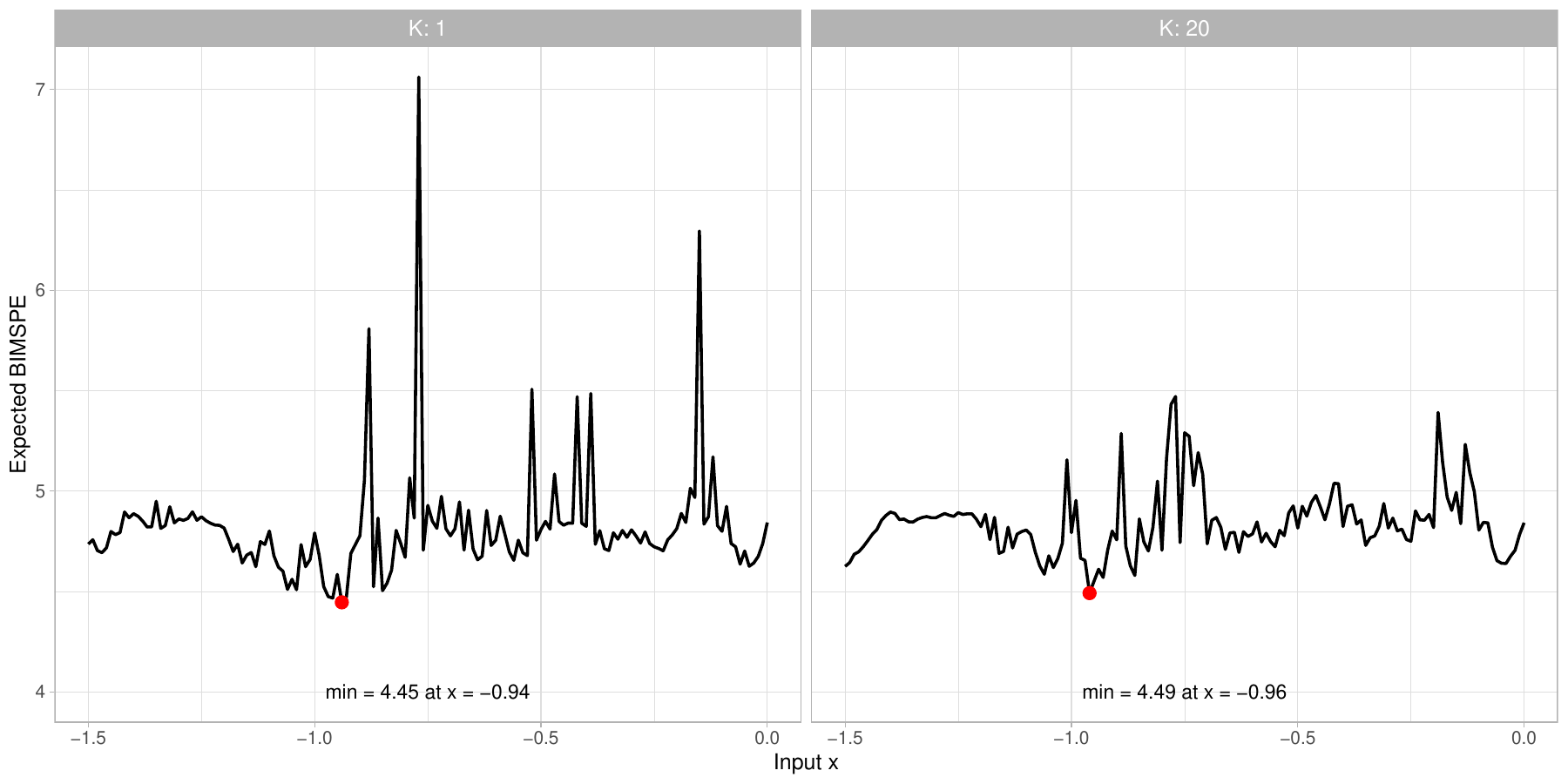}
	\caption{Line plots of the acquisition surfaces for FB-BIM with $K=1$ (left) and with $K=20$ (right) given the same observed dataset in the 1-dimensional example of one macro-replication, using the same set of MCMC draws. The red dot indicates the candidate point that minimizes the selection criterion.}
	\label{bimspe_by_K}
\end{figure}

Across both examples, the trajectories of all performance metrics are very similar under FB-BIM and FB-BIM20, as shown in Figure \ref{ytilde_robustness}. The final-iteration boxplots indicate no significant difference in the mean-response RMSE, while FB-BIM shows marginally better performance in terms of log-noise RMSE. Meanwhile, as illustrated in Figure~\ref{bimspe_by_K}, the acquisition surface obtained with $K=20$ is smoother, with little change in the location of the minimum. Overall, these results suggest that the performance of FB-BIM is largely insensitive to the number of $\tilde{y}_{N+1}$ draws used for approximating the expected BIMSPE selection criterion. This supports our choice of $K=1$ for computational efficiency, as the complexity of the expected BIMSPE calculation scales linearly with $K$. These findings correspond with the intuition that $K$ only influences the lookahead acquisition step (rather than posterior inference), so that $K=1$ is sufficient for effectively selecting good subsequent design points.

\begin{figure}[!htbp]
	\captionsetup[subfigure]{labelformat=empty}
	\begin{subfigure}[b]{1\linewidth}
		\centering
		\includegraphics[width=0.9\linewidth]{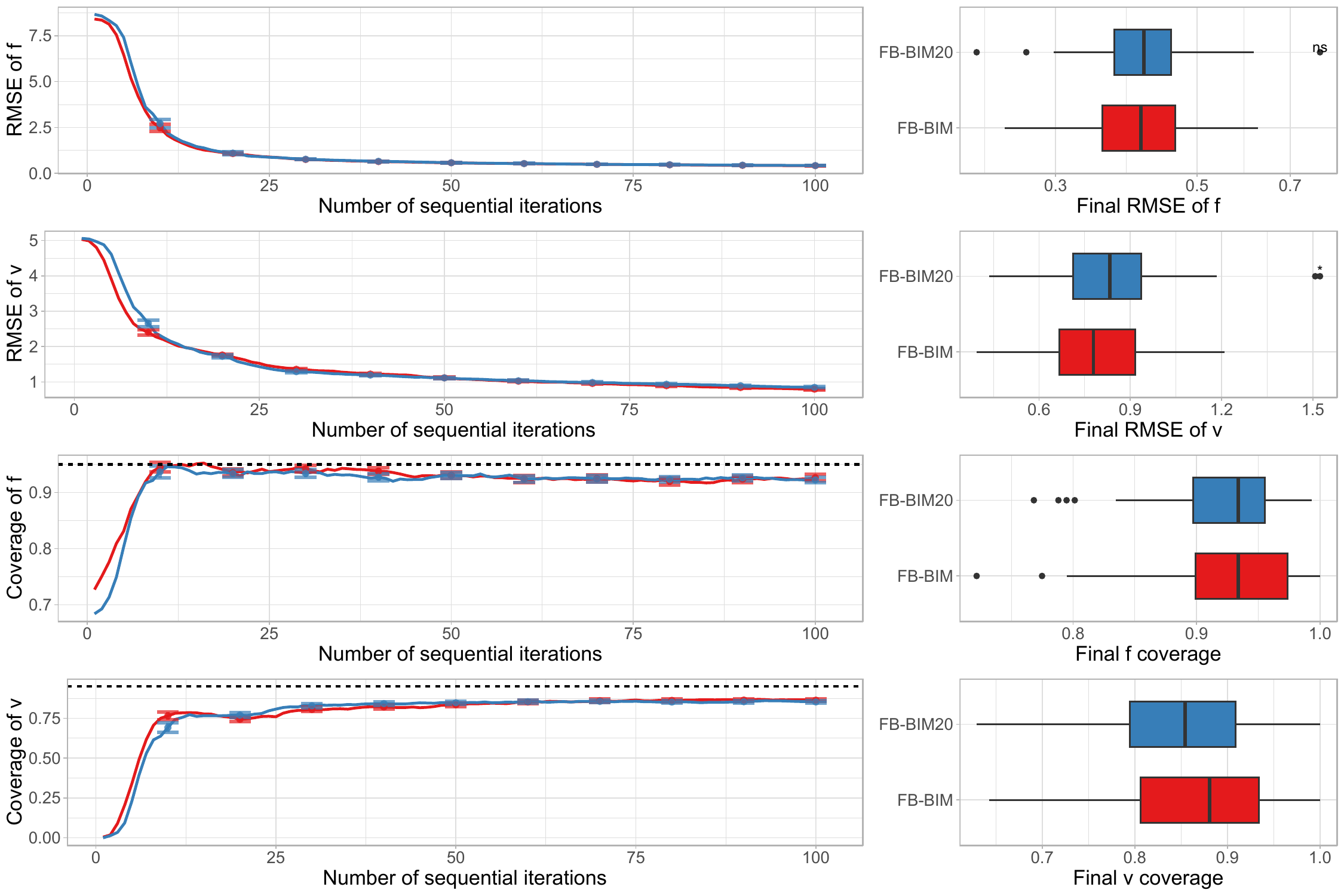}
		\caption{(a) 1-D example.}
		\label{Appendix: 1-d sensitivity yhat}
	\end{subfigure} 
\end{figure}
\begin{figure}[ht]\ContinuedFloat
	\begin{subfigure}[b]{1\linewidth}
		\centering
		\includegraphics[width=0.9\linewidth]{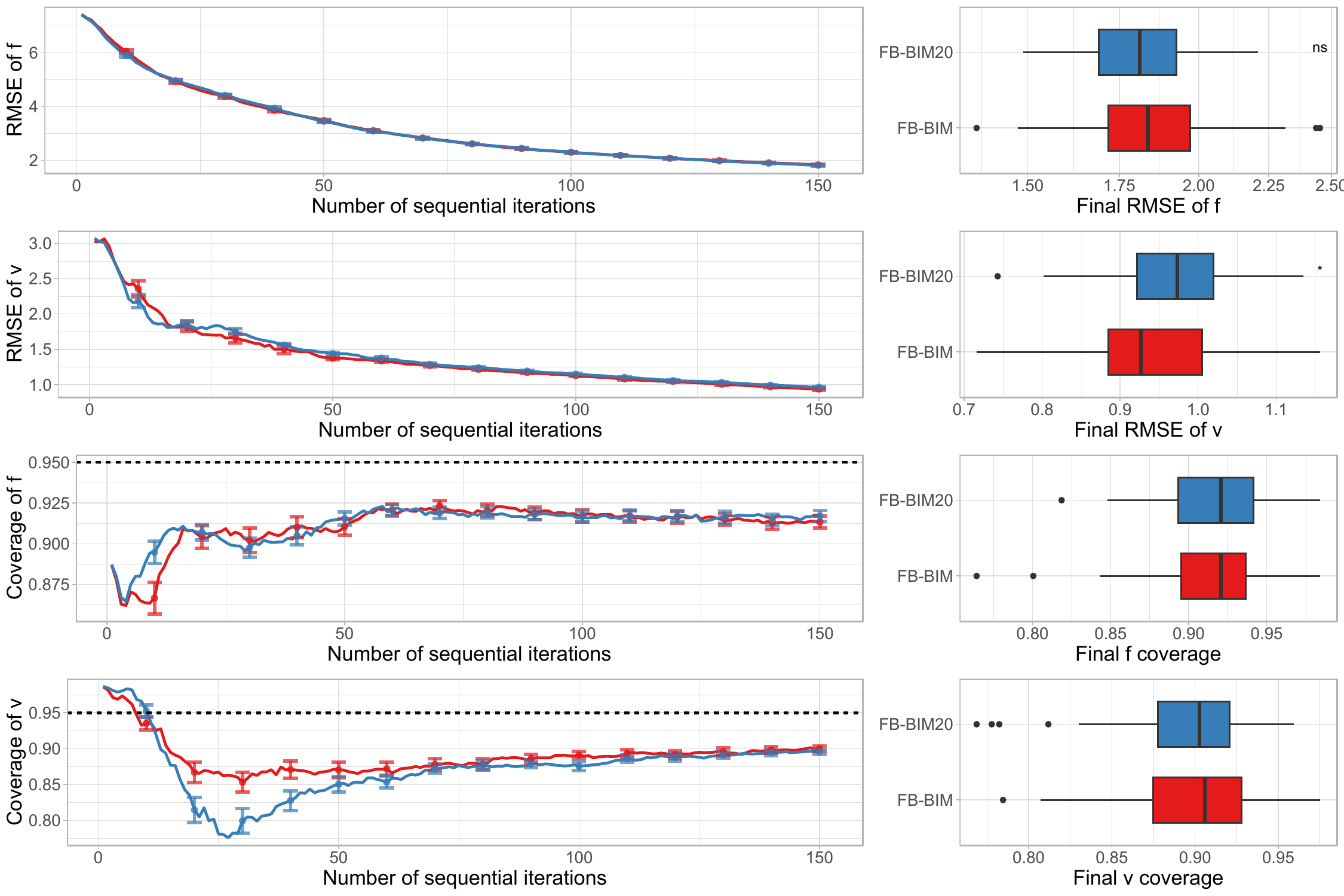} 
		\caption{2-D example. }\label{Appendix: 2-d sensitivity yhat}
	\end{subfigure} 
	\caption{Line plots of prediction performance (left) and final-iteration boxplots (right) for FB-BIM ($K=1$) and FB-BIM20 ($K=20$). The black dashed line in the coverage plots indicates the 95\% nominal level. Error bars (mean $\pm$ standard deviation) are displayed every 10 iterations.     A two sample t-test is conducted on the final results to compare the mean RMSE metrics between the two strategies. Significance levels are indicated beside the boxplots: ``ns''(p$>$0.05), ``*''(p$<=$0.05), ``**''(p$<=$0.01), ``***''(p$<=$0.001).} \label{ytilde_robustness}
\end{figure}

\FloatBarrier

\subsection{Robustness to Test Set Selection} \label{random_grid}

\begin{figure}[!htbp]
	\centering
	\includegraphics[width=0.8\linewidth]{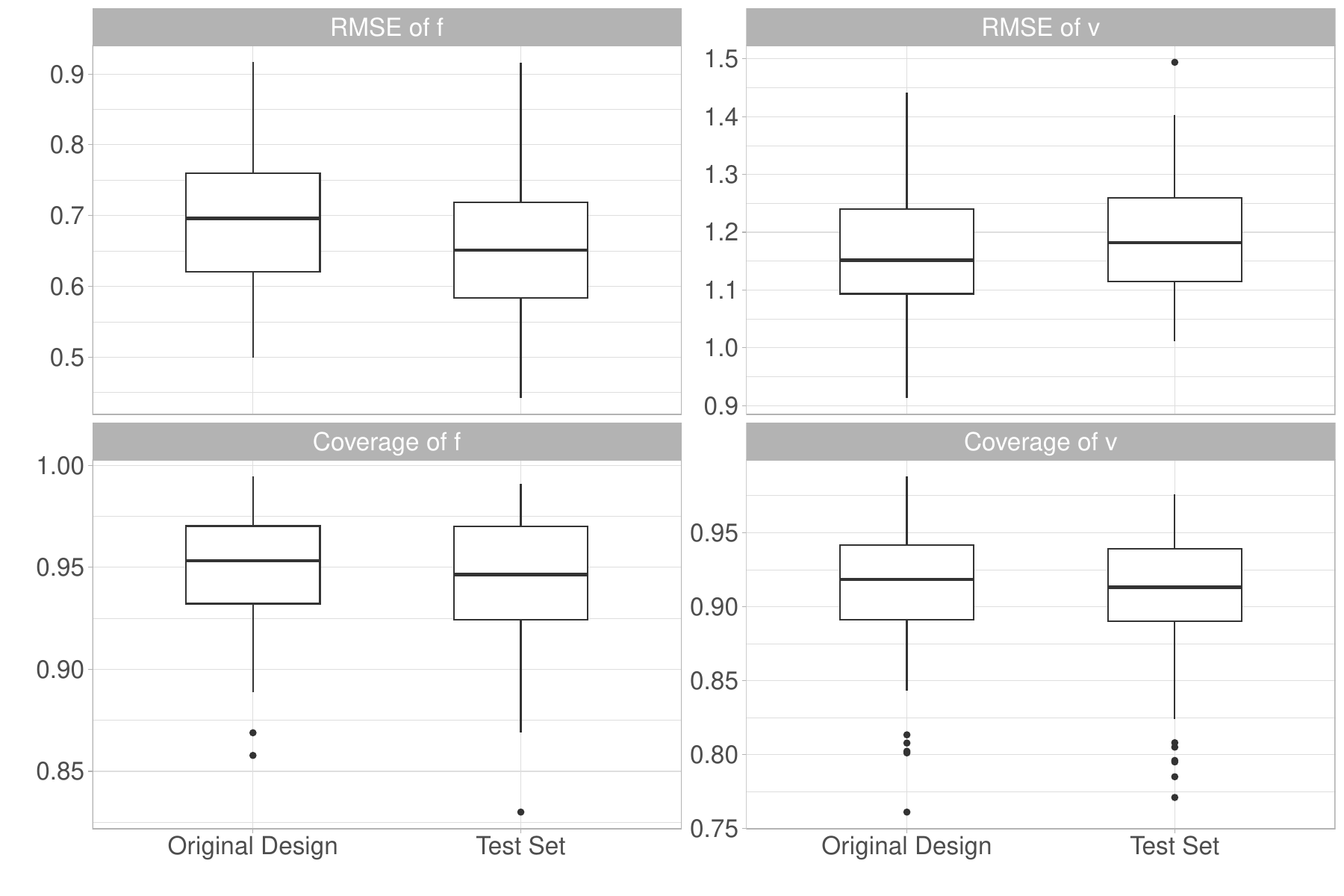}
	\caption{Final-iteration boxplots of RMSE and coverage metrics for FB-BIM on the original design $\mathcal{D}$ and an independent test set in the 5-D example.}
	\label{random_grid_fig}
\end{figure}

Recall that in our experiments, performance metrics were evaluated on the discrete candidate set $\mathcal{D}$. To ensure the robustness of our results to the choice of test points, here we additionally evaluate the final-iteration RMSE and coverage for the 5-dimensional example on an independent test set. Specifically, we generate a new set of $|\mathcal{D}| = 1000$ test points sampled uniformly from the continuous design space $[0,1]^5$. For each of the 100 macro-replications, we then use the fitted model based on FB-BIM's final design to make predictions on the new test set and compute the corresponding performance metrics. As shown in Figure \ref{random_grid_fig}, the resulting RMSE and coverage metrics exhibit no notable differences between the two sets of evaluation points, which support the robustness of our results beyond the specific candidate set $\mathcal{D}$.

\FloatBarrier

\section{BIMSPE Ablation Study} \label{sec-ablation}

As discussed in Section~2.2 of the main text, BIMSPE provides a Bayesian analogue of classic IMSPE by integrating over the posterior distribution of the hyperparameters and latent noise variables. By the law of total variance, BIMSPE includes not only the expected conditional variance of the mean response surface, but also a second term that captures uncertainty in the posterior mean induced by $\boldsymbol{\theta}$ and $\mathbf{v}_{0:N}$. Consequently, in the sequential design setting, the expected BIMSPE provides more comprehensive uncertainty quantification (UQ) as a selection criterion.

To assess the importance of this comprehensive UQ, we compare FB-BIM with a simplified variant, FB-EIM. The latter remains fully Bayesian in the sense that it obtains MCMC draws from the posterior distribution for model fitting, but its selection criterion is based on the standard plug-in empirical IMSPE. Specifically, FB-EIM plugs the posterior median of the hyperparameters and latent noise variables into the expected BIMSPE criterion. Under these point estimates, the C2 term in the law of total variance of (7) in the main text vanishes, i.e., the expected BIMSPE criterion simplifies to the standard empirical (discretized) IMSPE, with only the conditional variance term remaining as in traditional ALC.
Otherwise, the setup of FB-EIM is identical to FB-BIM; see Supplement \ref{appendix: alternative_details} for further implementation details.

We evaluate FB-EIM on the 1-, 2-, and 5-dimensional examples presented in Sections~3.1, 3.2, and 3.3 of the main text, with results summarized in Figure \ref{ablation}. Across all three examples, we observe that FB-EIM yields a higher $RMSE_f$, and the performance gap becomes more pronounced as the input dimension increases. This suggests that incorporating UQ in the selection criterion becomes increasingly important for effective sequential design in higher dimensions. For the log-noise process, FB-EIM is also inferior in the 1- and 2-dimensional examples. These results suggest that the fully Bayesian BIMSPE criterion provides a practical advantage by propagating uncertainty more effectively through the sequential design selection process. Note that coverage is omitted from this ablation study; since both methods use the same MCMC-based posterior framework for inference, their coverage results are expected to vary only slightly due to differences in the selected design points.

\begin{figure}[!htbp]
	\centering
	\includegraphics[width=1\linewidth]{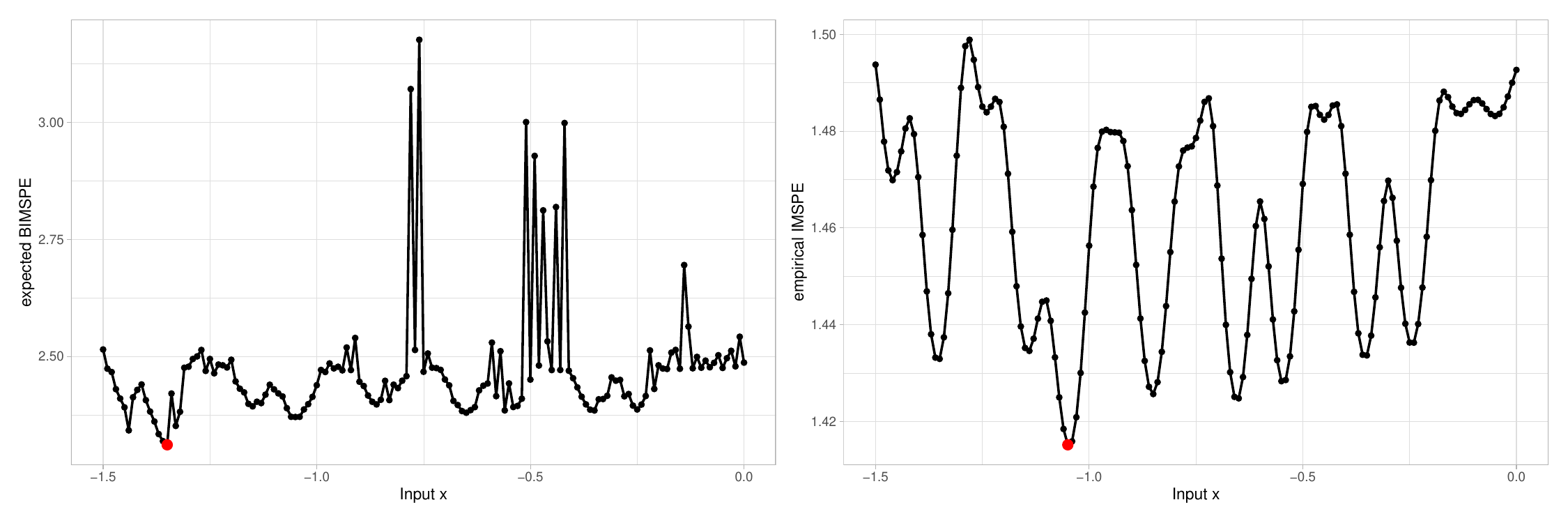}
	\caption{Line plots of the acquisition surfaces for FB-BIM (left) and FB-EIM (right) given the same observed dataset in the 1-D example of one macro-replication. The red dot indicates the candidate point that minimizes the selection criterion.}
	\label{acquisition}
\end{figure}

Figure \ref{acquisition} provides a visualization of the expected BIMSPE criterion and the empirical IMSPE criterion over the design grid given the same observed dataset. The BIMSPE value is greater than that of the empirical IMSPE given the incorporation of the uncertainty in the posterior mean, and the two criteria lead to different selection of $\x_{N+1}$.

\begin{figure}[!htbp]
	\centering
	\includegraphics[width=1\linewidth]{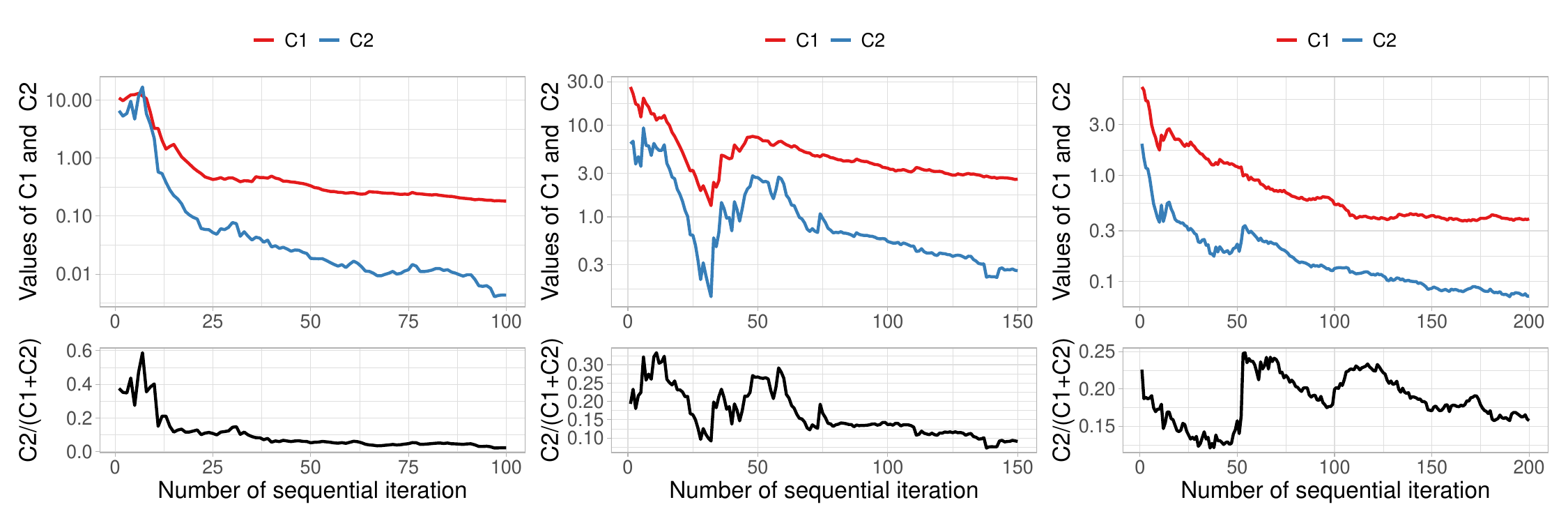}
	\caption{Line plots (top) of C1 (red) and C2 (blue) in the expected BIMSPE criterion for one macro-replication of the 1-D example (left), 2-D example (middle), and 5-D example (right), along with the corresponding plots of the ratio C2/(C1+C2) (bottom panels). At each iteration, C1 and C2 are computed and then averaged over the entire candidate set $\mathcal{D}$. Values in the top panels are shown on a $\log_{10}(\cdot)$ scale.}
	\label{C1C2}
\end{figure}

Figure~\ref{C1C2} provides a visualization of averaged C1 and C2 components over the grid in the expected BIMSPE criterion in 1-D, 2-D, and 5-D examples respectively as the sequential iterations progress. We observe that the C1 and C2 terms start at similar magnitudes, but C2 decreases more rapidly than C1, so that the ratio of C2 to the expected BIMSPE criterion, C2/(C1+C2), declines as the iterations proceed.

\begin{figure}[!h]
	\captionsetup[subfigure]{labelformat=empty}
	\begin{subfigure}[b]{1\linewidth}
		\centering
		\includegraphics[width=0.9\textwidth]{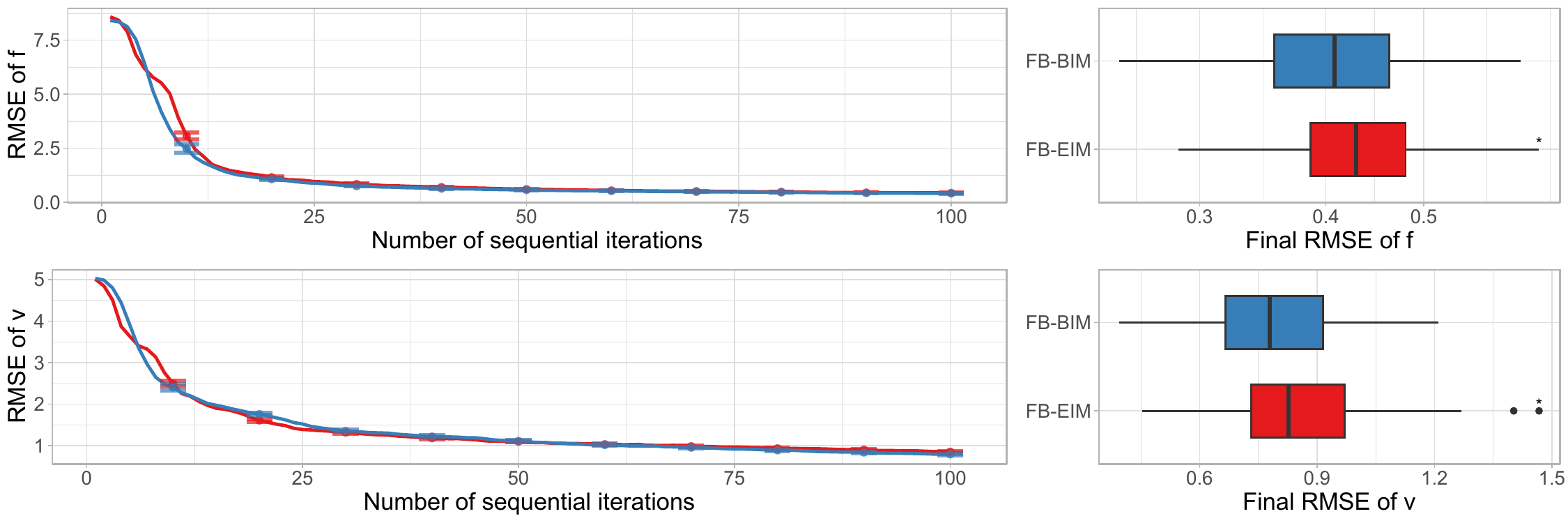}
		\caption{(a) 1-D example}
	\end{subfigure} 
\end{figure}
\begin{figure}[ht]\ContinuedFloat
	\begin{subfigure}[b]{1\linewidth}
		\centering
		\includegraphics[width=0.9\textwidth]{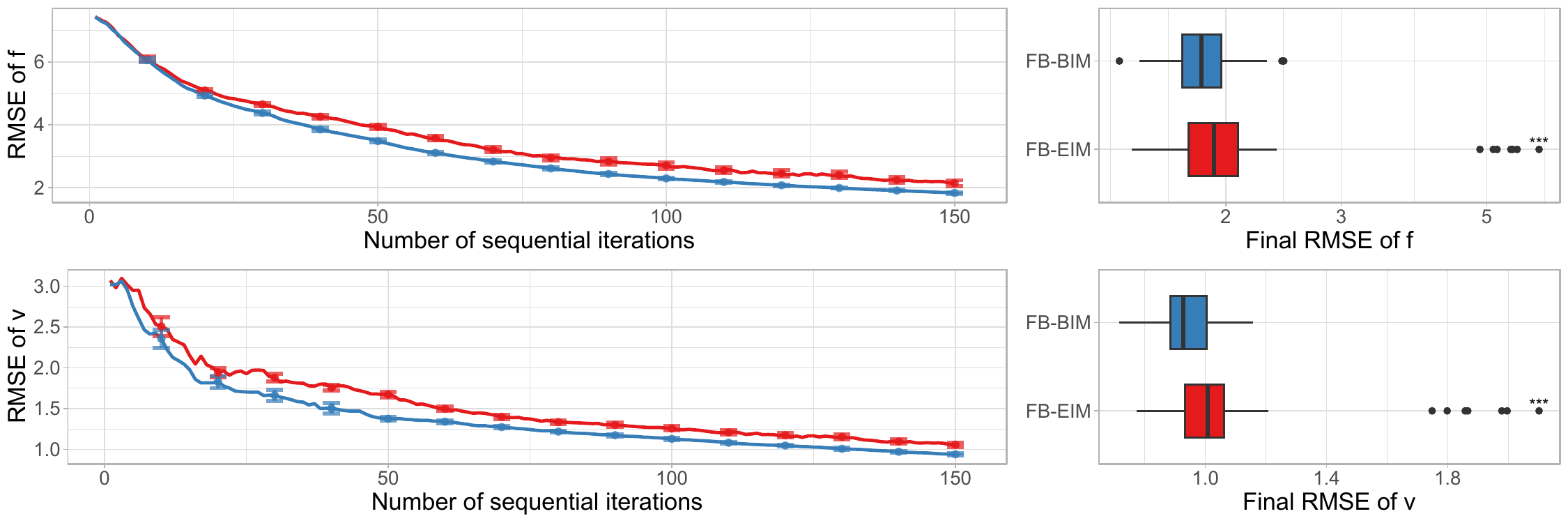}
		\caption{2-D example}
	\end{subfigure}
\end{figure}
\begin{figure}[ht]\ContinuedFloat
	\begin{subfigure}[b]{1\linewidth}
		\centering
		\includegraphics[width=0.9\textwidth]{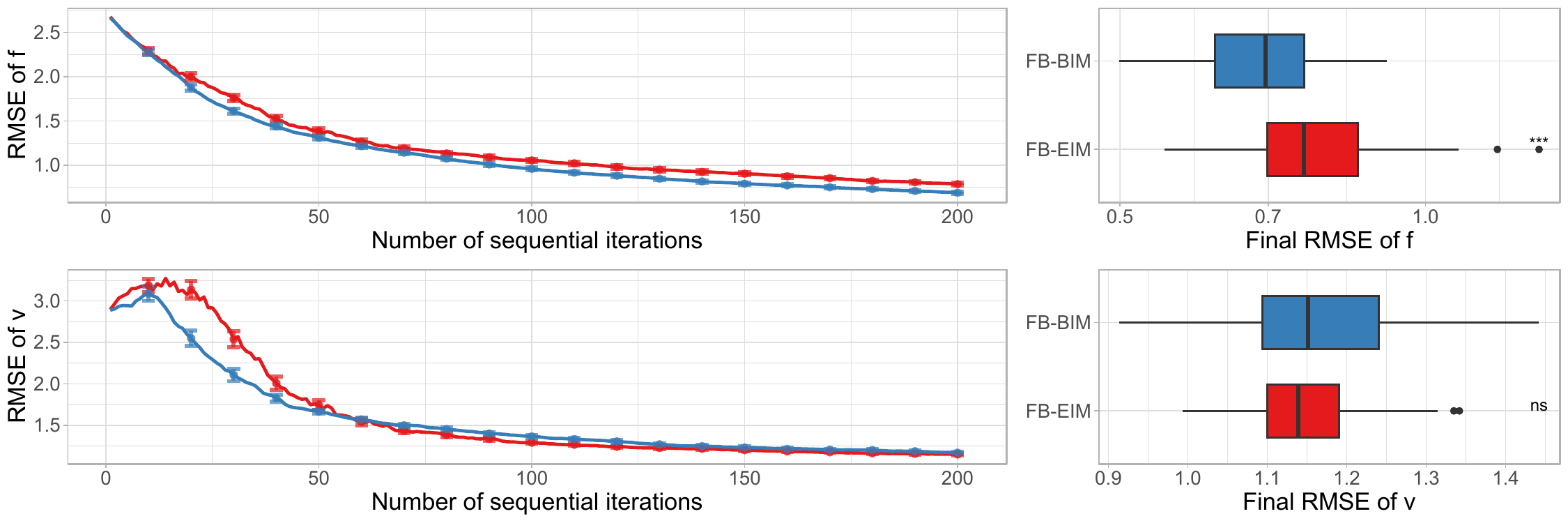}
		\caption{5-D example}
	\end{subfigure}
	\caption{Line plots of prediction performance (left) and final-iteration boxplots (right) for the FB-BIM and FB-EIM strategies. The black dashed line in the coverage plots indicates the 95\% nominal level. Error bars (mean $\pm$ standard deviation) are displayed every 10 iterations. A two sample t-test is conducted on the final results to compare the mean RMSE metrics between the two strategies. Significance levels are indicated beside the boxplots: ``ns''(p$>$0.05), ``*''(p$<=$0.05), ``**''(p$<=$0.01), ``***''(p$<=$0.001).} \label{ablation}
\end{figure}

\FloatBarrier

\section{Synthetic Simulation with Known Generating Models \label{known-generating-dist}}
Extending the illustrative 2-D example, here we explore hyperparameter estimation and robustness to model misspecification of FB-BIM under data generating mechanisms with known underlying truth on the same input space $\x \in [-2, 2] \times [-1, 1]$ with 441 design points of a uniform grid. We consider a process with noise sampled as $v(\x) \sim \mathcal{GP}(\mu_0, k_v(\x, \x^{'}))$, where $k_v(\x, \x^{'})$ is the squared exponential kernel. Then observations are drawn via $Y(\bfx)|\textbf{g}_{0:N} \sim \mathcal{GP}(\textbf{0}, \boldsymbol{K}_N)$, where $\boldsymbol{K}_N = k_f(\x_i, \x_j)+\delta_{ij}g_i$, for $1\leq i, j \leq N$ with $\delta_{ij}$ being the Kronecker delta function and $k_f(\x, \x^{'})$ is also squared exponential. We set $\sigma_f=1, l_{f1}=0.8, l_{f2}=0.6$ for the $f$-process, and $\sigma_g=0.8, l_{v1}=0.6, l_{v2}=0.4$ for the $v$-process. More specifically, we control the magnitude of log-noise $v$ by varying the mean value of the $v$-process. We set $\mu_0=-2,~-0.2,~1$ to represent low, medium, and high noise levels, respectively.

We consider 100 independent macro-replications with initial designs based on the maximin LHD with a size of 21 points and a budget for the sequential design of $B=300$. The priors are the same as those in the illustrative 2-D example. We study FB-BIM  with squared exponential kernels for both $f-$ and $v-$ processes and investigate if the hyperparameters can be correctly estimated by the final designs. We also study FB-BIM  with the ``misspecified'' less smooth Mat\'ern kernels with $\nu=3/2$ for both processes to investigate if the performances are robust and comparable.

We first report the average noise magnitude over the entire grid, defined as the mean of $\sqrt{g(\mathbf{x})/R_f}$, computed across the 100 macro-replications under the low-, medium-, and high-noise settings. Under the low-noise setting, the average noise magnitude ranges from $0.14$ to $0.35$, with a mean of $0.21$. For the medium-noise setting, the range is $0.34$ to $0.85$, with an average of $0.51$. Under the high-noise setting, the noise magnitude ranges from $0.61$ to $1.6$, with an average of $0.93$. These values are clearly separated across the three scenarios and confirm that the intended noise levels are well represented.

\begin{figure}[!htbp]
	\centering
	\includegraphics[width=1\linewidth]{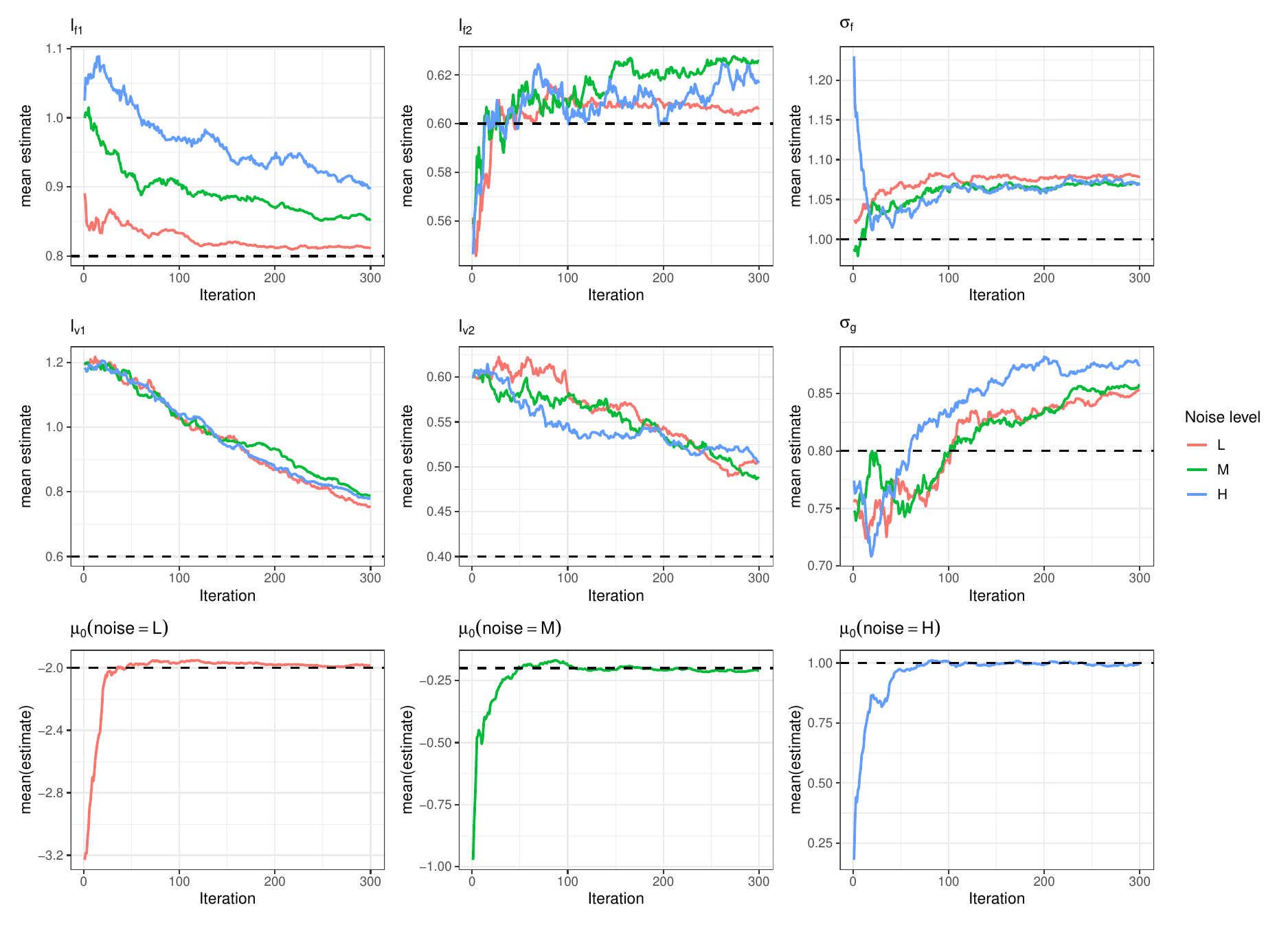}
	\caption{Trajectories of posterior point estimates for key model hyperparameters over the sequential design iterations in the 2-D example with a known generating model under low, medium, and high noise level settings, respectively denoted by L (red), M (green), and H (blue). Each subplot reports the mean estimate at each iteration. The black dashed horizontal line in each subplot indicates the corresponding ground-truth value used to generate the data.}
	\label{bench2D-para-fit}
\end{figure}

Figure~\ref{bench2D-para-fit} presents the trajectories of posterior estimates for each hyperparameter across 300 sequential iterations under the three noise levels. For each macro-replication, the posterior estimate is computed as the weighted average of posterior samples, and the trajectories shown in the figure are averaged over 100 macro-replications. Table~\ref{bench2D-para-table} summarizes the hyperparameter estimation results at the final designs (i.e., after the final iteration), including the average estimate (Avg Est), RMSE, and empirical coverage of the nominal $95\%$ credible intervals across the 100 macro-replications. For each noise level, the table also reports the corresponding ground-truth parameter values (True Value).

Overall, the impact of noise varies across hyperparameters. Among the parameters of the $f$-process ($\theta_f = \{l_{f1}, l_{f2}, \sigma_f\}$) and the $v$-process ($\theta_v = \{\mu_0, l_{v1}, l_{v2}, \sigma_g\}$), the parameters in the $f$-process are generally learned more efficiently. In particular, under lower noise levels, the $f$-process parameters tend to converge to their true values more quickly. The lengthscales $(l_{f1}, l_{f2})$ approach the ground truth fastest in the low-noise setting, followed by the medium- and high-noise settings, while $\sigma_f$ is recovered reasonably well across all noise levels. For the log-noise $v$-process, the average estimate of $\mu_0$ converges to its true value after approximately 25, 50, and 75 iterations under low, medium, and high noise levels, respectively. The variance parameter $\sigma_g$ also approaches its true value across all noise settings. As a parameter controlling the overall noise magnitude, $\mu_0$ is relatively easy to learn, whereas lengthscales of the log-noise process are more difficult to recover as we observe that the trajectories of the lengthscales $(l_{v1}, l_{v2})$ continue the tendency to decrease throughout the iterations regardless of the noise levels. Moreover, the noise level appears to have a greater impact on the estimation of the $f$-process lengthscales than those of the $v$-process.

We also compare the performances of the squared exponential (SE) kernel with the mis-specified Mat\'ern kernel across different noise levels. Although the two kernels produce visually similar results in terms of $RMSE_f$, the SE kernel achieves lower $RMSE_f$ under low and medium noise levels. This difference becomes less pronounced as the noise level increases. In absolute terms, $RMSE_f$ improves as the noise level decreases. In addition, under low and medium noise, the empirical coverage for the mean response tends to decline as more observations are collected when using the mis-specified kernel. A plausible explanation is that as $N$ increases, the posterior predictive variance of $f$ shrinks and the resulting credible intervals become narrower, while the bias induced by kernel mis-specification persists. Under higher noise, the predictive variance of $f$ shrinks more slowly, and coverage is therefore less affected. For the noise process, the $RMSE_v$ of the two kernels are visually similar and remain relatively stable across all noise levels (typically around $0.4$), which aligns with our findings from Figure~\ref{bench2D-para-fit}. Finally, coverage results indicate that the SE kernel generally provides better calibration for the $f$-process than the mis-specified Mat\'ern kernel, with differences diminishing as noise increases. For the $v$-process, both kernels achieve coverages that are relatively close to the nominal level, suggesting that inference on the log-noise process is less sensitive to kernel mis-specification.

\begin{table}
	\caption{Summary of final design GP parameters by noise level in the controlled 2-D experiment.}
	\centering
	\begin{tabular}[t]{llrrrr}
		\toprule
		Noise Level & Parameter & True Value & Avg Est & RMSE & Coverage\\
		\midrule
		High & $l_{f1}$ & 0.8 & 0.900 & 0.298 & 1.00\\
		& $l_{f2}$ & 0.6 & 0.617 & 0.205 & 0.97\\
		& $\sigma_f$ & 1.0 & 1.070 & 0.292 & 0.96\\
		& $l_{v1}$ & 0.6 & 0.779 & 0.282 & 0.99\\
		& $l_{v2}$ & 0.4 & 0.505 & 0.203 & 0.98\\
		& $\sigma_g$ & 0.8 & 0.875 & 0.181 & 0.99\\
		& $\mu_0$ & 1.0 & 0.997 & 0.305 & 0.97\\
		\addlinespace
		Medium & $l_{f1}$ & 0.8 & 0.851 & 0.192 & 0.96\\
		& $l_{f2}$ & 0.6 & 0.626 & 0.136 & 0.97\\
		& $\sigma_f$ & 1.0 & 1.070 & 0.263 & 0.91\\
		& $l_{v1}$ & 0.6 & 0.789 & 0.308 & 0.99\\
		& $l_{v2}$ & 0.4 & 0.487 & 0.250 & 0.98\\
		& $\sigma_g$ & 0.8 & 0.857 & 0.192 & 0.95\\
		& $\mu_0$ & -0.2 & -0.209 & 0.313 & 0.96\\
		\addlinespace
		Low & $l_{f1}$ & 0.8 & 0.811 & 0.095 & 0.95\\
		& $l_{f2}$ & 0.6 & 0.607 & 0.078 & 0.98\\
		& $\sigma_f$ & 1.0 & 1.077 & 0.254 & 0.94\\
		& $l_{v1}$ & 0.6 & 0.752 & 0.284 & 0.99\\
		& $l_{v2}$ & 0.4 & 0.504 & 0.221 & 0.99\\
		& $\sigma_g$ & 0.8 & 0.853 & 0.185 & 0.99\\
		& $\mu_0$ & -2.0 & -1.983 & 0.282 & 0.97\\
		\bottomrule
	\end{tabular}\label{bench2D-para-table}
\end{table}

\FloatBarrier

\begin{figure}[!b]
	\captionsetup[subfigure]{labelformat=empty}
	\begin{subfigure}[b]{1\linewidth}
		\centering
		\includegraphics[width=0.9\textwidth]{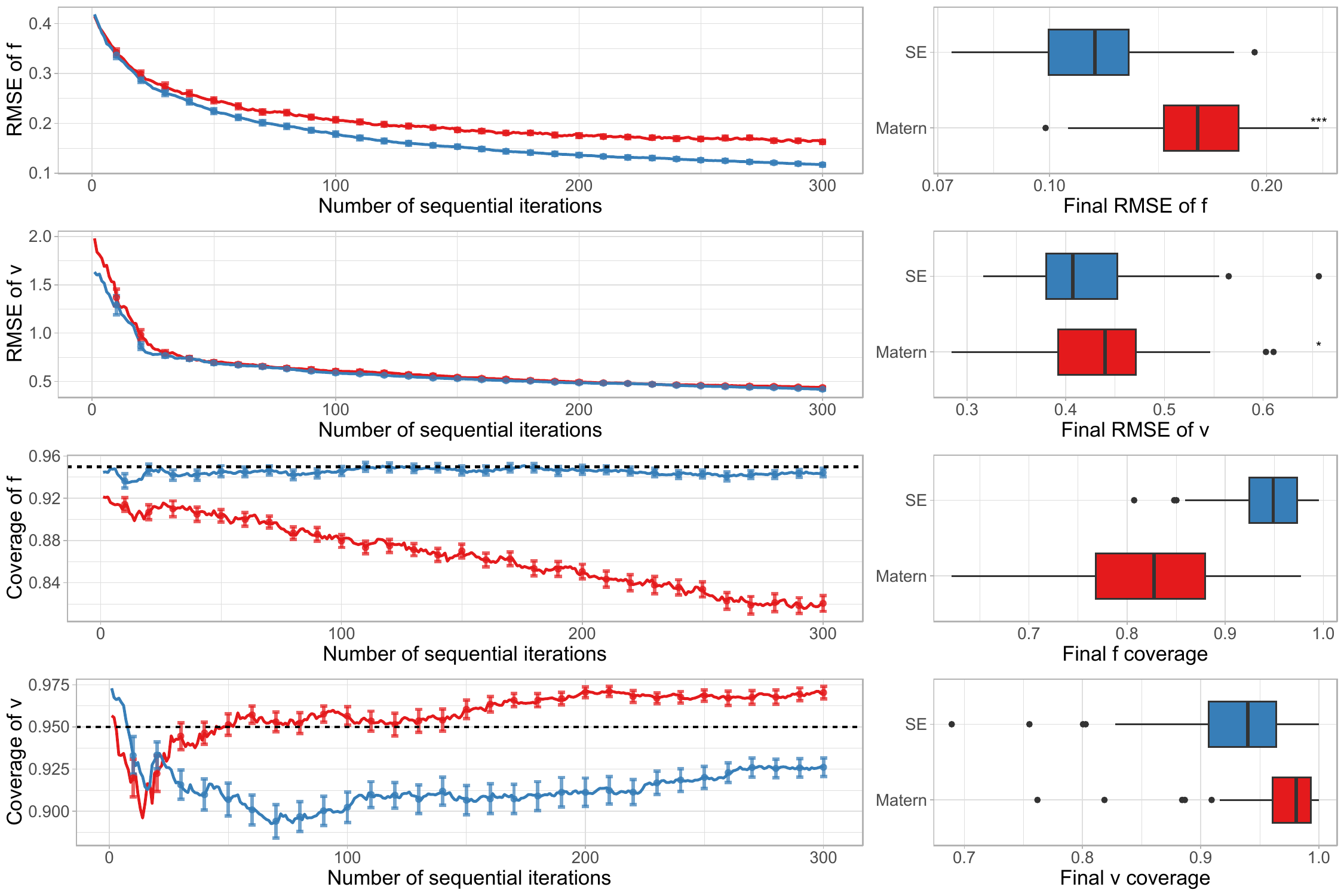}
		\caption{(a) Low noise level}
	\end{subfigure} 
\end{figure}
\begin{figure}[ht]\ContinuedFloat
	\begin{subfigure}[b]{1\linewidth}
		\centering
		\includegraphics[width=0.9\textwidth]{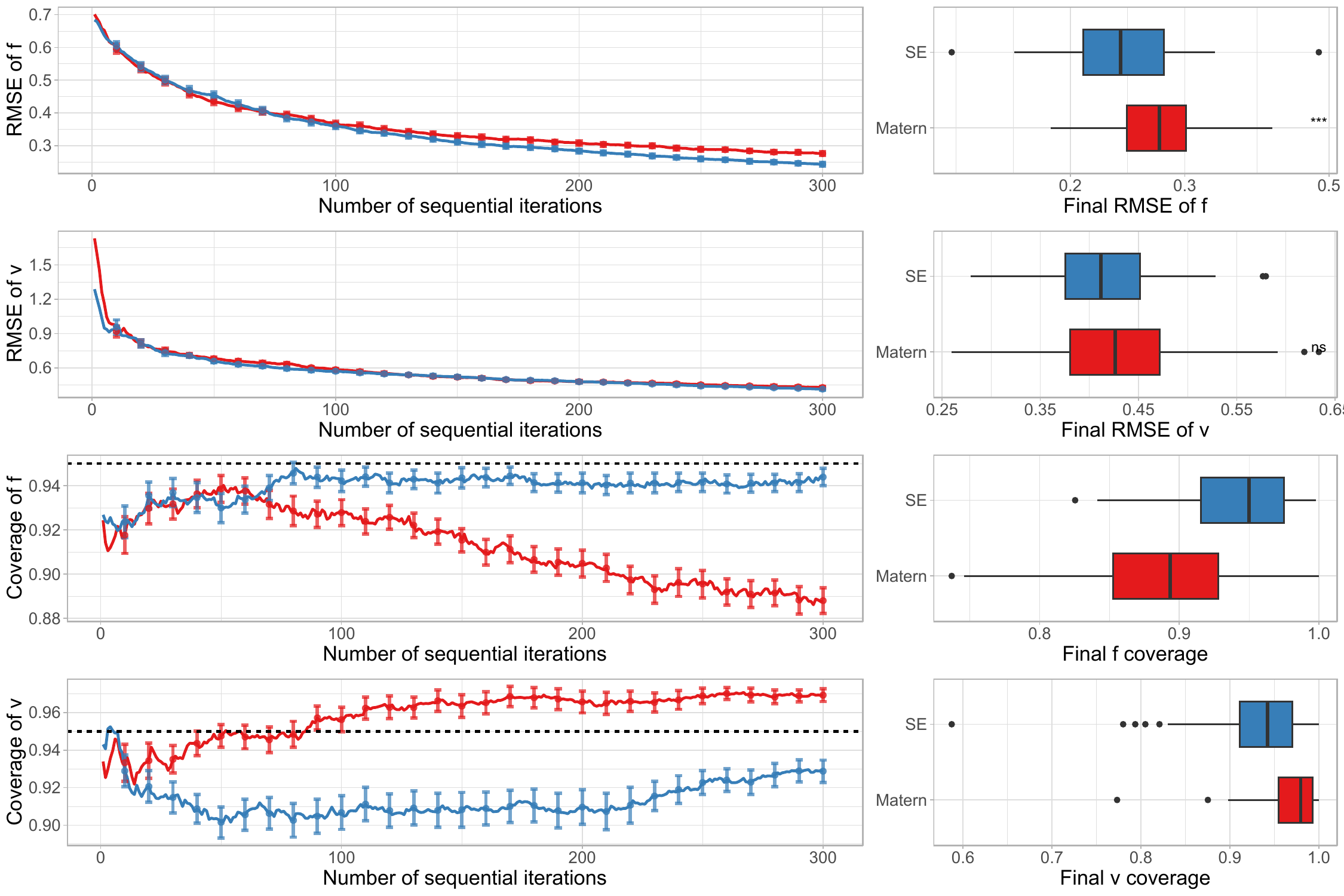}
		\caption{Medium noise level}
	\end{subfigure}
\end{figure}
\begin{figure}[ht]\ContinuedFloat
	\begin{subfigure}[b]{1\linewidth}
		\centering
		\includegraphics[width=0.9\textwidth]{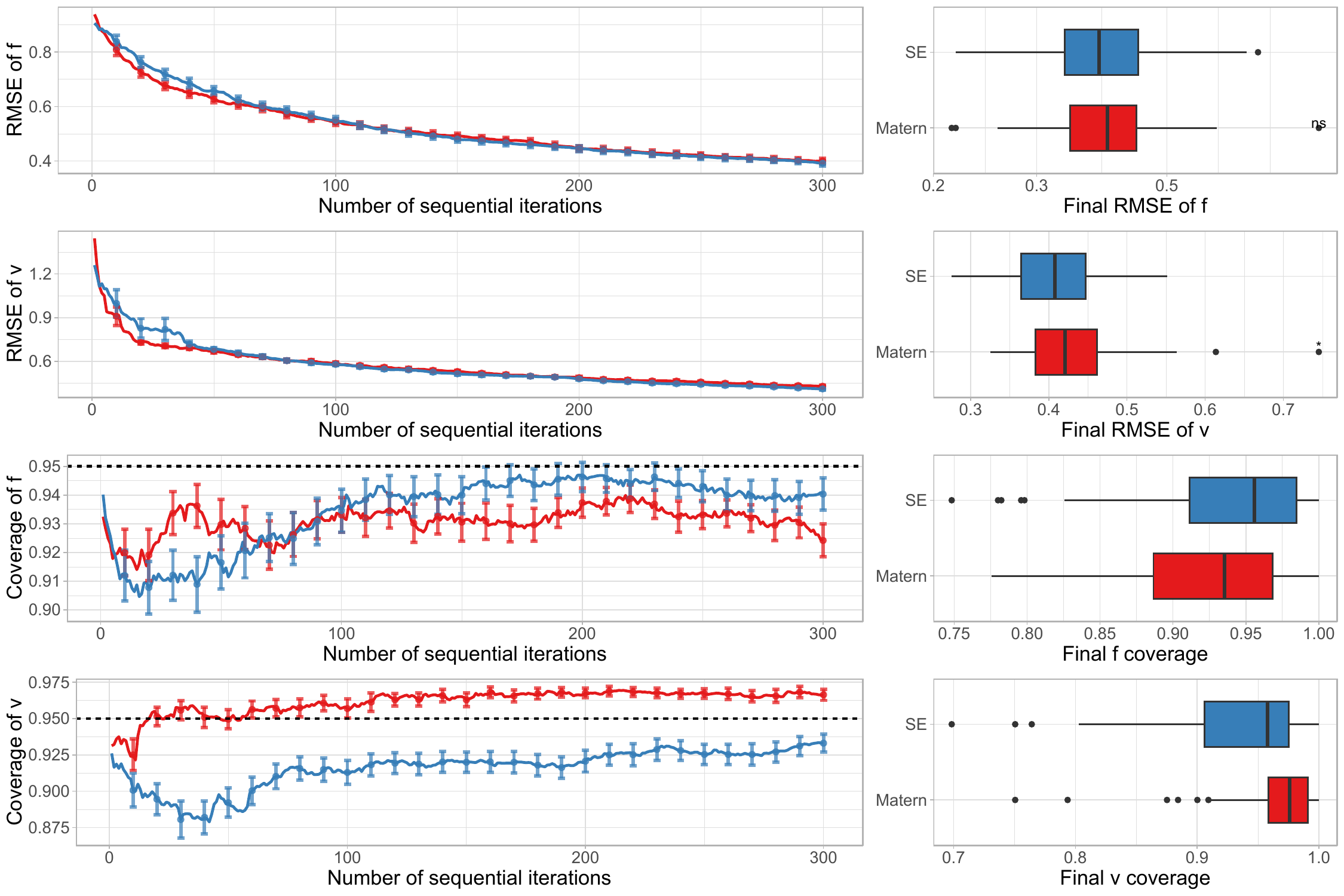}
		\caption{High noise level}
	\end{subfigure}
	\caption{Line plots of prediction performance (left) and final-iteration boxplots (right) for FB-BIM  with correctly specified squared exponential kernel, and FB-BIM  with mis-specified Mat\'ern kernel ($\nu=3/2$) on the extended 2-D example, denoted by blue and red respectively. The left column shows the trajectories of four metrics over iterations: RMSE for the mean response, RMSE for the log-noise variance, and empirical coverages for the mean response and log-noise variance. The black dashed line in the coverage plots indicates the 95\% nominal level. Error bars (mean $\pm$ standard deviation) are displayed every 10 iterations. The right column presents boxplots of the corresponding metric values at the final iteration. A two-sample t-test is conducted on the results at the final iteration to compare the mean RMSE metrics. Significance levels are indicated beside the boxplots: ``ns''(p$>$0.05), ``*''(p$<=$0.05), ``**''(p$<=$0.01), ``***''(p$<=$0.001).}  \label{bench2D-kernel}
\end{figure}

\FloatBarrier

\section{Histograms of Prior and Posterior Samples from FB-BIM \label{appendix:prior&posterior}}
\begin{figure}[!htbp]
	\captionsetup[subfigure]{labelformat=empty}
	\begin{subfigure}[b]{1\linewidth}
		\centering
		\includegraphics[width=1\linewidth]{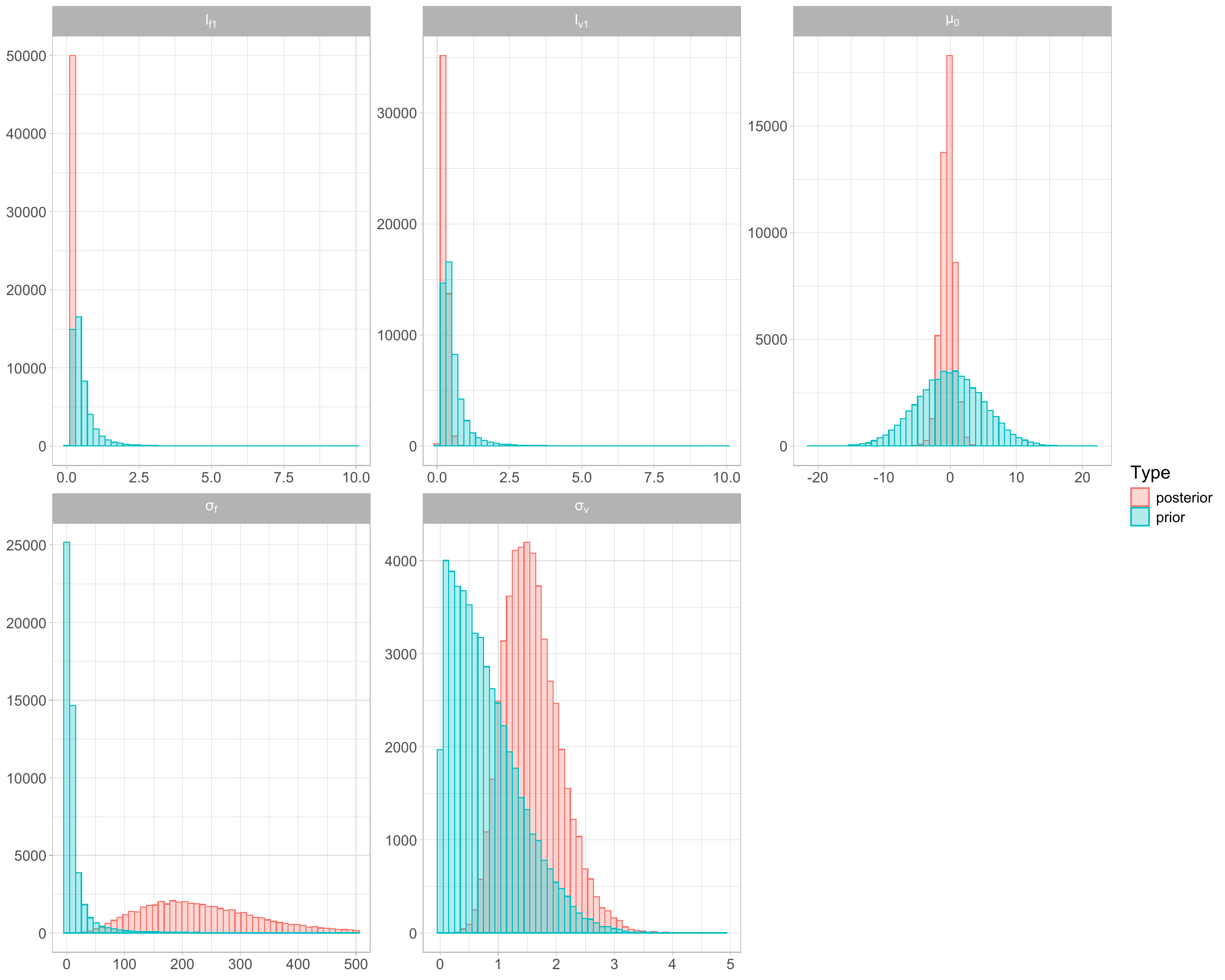}
		\caption{(a) 1-D example}\label{1-d: prior_posterior}
	\end{subfigure} 
\end{figure}
\clearpage  
\begin{figure}[ht]
	\addtocounter{figure}{-1}
	\begin{subfigure}[b]{1\linewidth}
		\addtocounter{subfigure}{1} 
		\centering
		\includegraphics[width=1\linewidth]{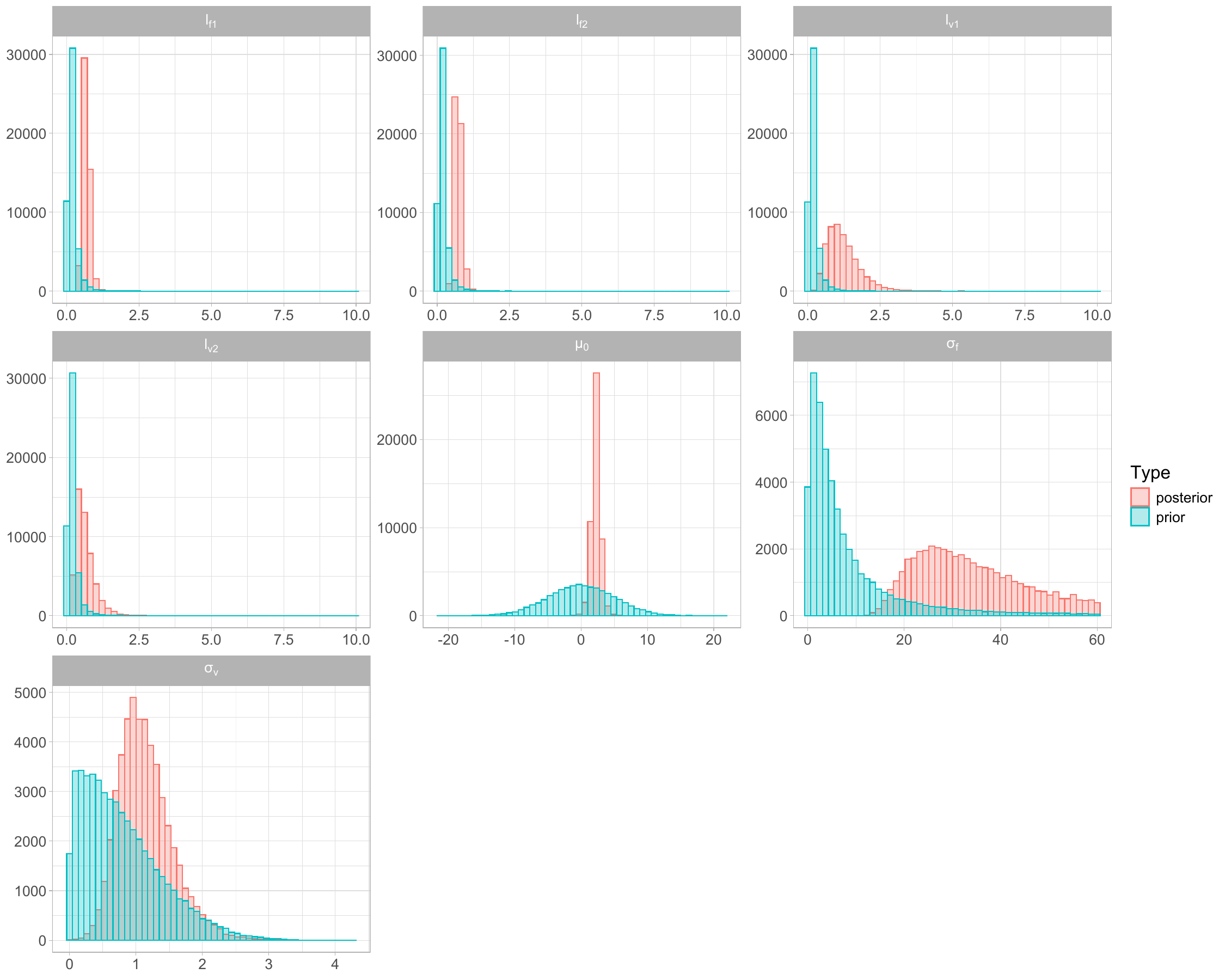}
		\caption{2-D example}\label{2-d: prior_posterior}
	\end{subfigure} 
\end{figure}
\clearpage  
\begin{figure}[ht]
	\addtocounter{figure}{-1}
	\begin{subfigure}[b]{1\linewidth}
		\addtocounter{subfigure}{2} 
		\centering
		\includegraphics[width=1\linewidth]{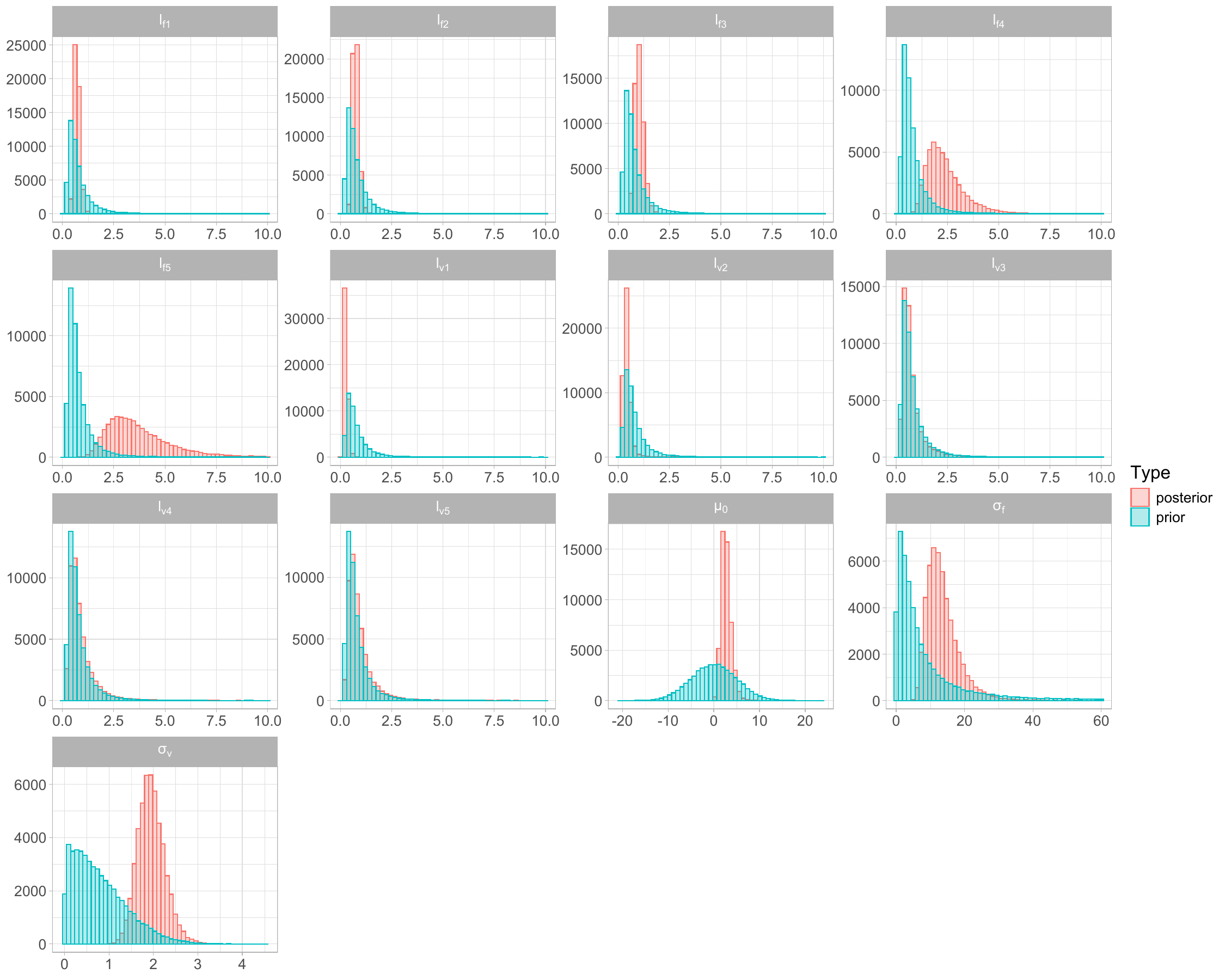}
		\caption{5-D example}\label{5-d: prior_posterior}
	\end{subfigure} 
	\caption{Histograms of 50000 draws from prior and posterior distributions of each hyperparameter. The prior draws are generated from the specified prior distributions. The posterior draws are the MCMC samples generated from the final posterior distributions (i.e., after the allotted sequential iterations), combined across all simulation macro-replications. For visual clarity, posteriors of $\sigma_f$ and $\sigma_v$ are displayed instead of the corresponding variance parameters; all lengthscales (i.e., $l_{f\cdot}, l_{v\cdot}$) are truncated at 10, $\sigma_f$ is truncated at 500 in 1D example and truncated at 60 in 2D and 5D examples.} \label{all_prior_posterior}
\end{figure}

\FloatBarrier

\section{Implementation Details of Alternative Strategies \label{appendix: alternative_details}}

\subsection{Fully Bayesian Inference with Empirical IMSPE (FB-EIM)}

Recall that the computation of  $\mathbb{E}_{\tilde{y}_{N+1} \mid D_N} \left[ \text{BIMSPE} \big(D_N \cup \{(\tilde{\mathbf{x}}_{N+1}, \tilde{y}_{N+1})\}\big) \right]$ involves two terms, C1 and C2, as defined in (7) of the main text. Under the FB-BIM strategy, these expectations are computed with respect to the full posterior distribution $p(\bm{\theta}, \textbf{v}_{0:N}|D_{N})$ and the C2 term also averages over the draws $\{\tilde{y}^{(k)}_{N+1}\}_{k=1}^K$ from $p(\tilde{y}_{N+1}|D_N)$. In contrast, the FB-EIM strategy computes the selection criterion using the posterior medians as point estimates for the hyperparameters $\bm{\theta}$ and latent noise variables $\mathbf{v}_{0:N}$, namely, $\bm{\hat{\theta}}_{median}=\text{median} \big(\{\bm{\theta}^{(m)}\}_{m=1}^{M}|D_{N}\big)$, and $\hat{\textbf{v}}_{0:N+1,median}=\text{median}\big(\{ \mathbf{v}_{0:N+1}^{(m)}\}_{m=1}^{M}|D_{N}\big)$, as obtained from all post burn-in MCMC samples. Consequently, the C2 term in (7) of the main text becomes zero, and C1 reduces to a single posterior variance with plug-in estimates.

Specifically, suppose we obtain the posterior median point estimates $\bm{\hat{\theta}}_{median}$ and $\hat{\textbf{v}}_{0:N, median}$ at the $N$-th iteration. In the lookahead step, if $\tilde{\bfx}_{N+1}$ is a new point, we first compute the posterior median $\hat{v}_{N+1}$ from its predictive distribution $p(v_{N+1}|\hat{\textbf{v}}_{0:N, median}, \bm{\hat{\theta}}_{median})$; if $\tilde{\bfx}_{N+1}$ is a replicate, we assign the known posterior median from $\hat{\textbf{v}}_{0:N, median}$ at $\tilde{\x}_{N+1}$. Using these point estimates, 
the expected BIMSPE simplifies to the standard plug-in empirical (discretized) IMSPE selection criterion:
\begin{equation}
	\mathbb{E}_{\tilde{y}_{N+1} \mid D_N} \left[ \text{BIMSPE} \big(D_N \cup \{(\tilde{\mathbf{x}}_{N+1}, \tilde{y}_{N+1})\}\big) \right] \approx  \frac{1}{|\mathcal{D}|} \sum_{\bfx_* \in \mathcal{D}} \V \left(f(\x_*) | \bm{\hat{\theta}}_{median}, \hat{\textbf{v}}_{0:N, median}, \hat{v}_{N+1}, D_N \right) .\label{C1_median}
\end{equation}

We evaluate the simulation model at the design point where this selection criterion is minimized, and observe $y_{N+1}$. Then MCMC is re-run based on the augmented observed data. The FB-EIM strategy is detailed in Algorithm \ref{algorithm_empirical_median}. All other settings in the FB-EIM strategy are the same as those used in the FB-BIM strategy across all synthetic examples. Other implementation details, i.e., the configuration of the MCMC sampler, similarly follow the specifications outlined in Section 2.5.1 of the main text.

\begin{algorithm}[!htbp]
	\caption{FB-EIM Design Strategy}\label{algorithm_empirical_median}
	\begin{algorithmic}[1] 
		\State \textbf{Input:} Initial design $D_0$, discretized set of design points $\mathcal{D}$, budget $B$, number of samples $M$
		\State Draw $M$ samples $\{(\bm{\theta}^{(m)}, \mathbf{v}_{0}^{(m)})\}_{m=1}^{M}$ from $p(\bm{\theta}, \mathbf{v}_0 \mid D_0)$ via MCMC  
		\State Compute the posterior medians $\bm{\hat{\theta}}_{median}=\text{median} \big(\{\bm{\theta}^{(m)}\}_{m=1}^{M}|D_{0}\big)$, and $\hat{\textbf{v}}_{0,median}=\text{median}\big(\{ \mathbf{v}_{0}^{(m)}\}_{m=1}^{M}|D_{0}\big)$
		\Comment{Initialization}
		\For{$N = 0$ to $B-1$} \Comment{Sequential design iterations}
		\For{$\tilde{\mathbf{x}}_{N+1} \in \mathcal{D}$} \Comment{Lookahead acquisition step}
		\If{$\tilde{\mathbf{x}}_{N+1}$ is a new design point}
		\State Set  $\hat{v}_{N+1}$ as the median of $p(v_{N+1} \mid \hat{\mathbf{v}}_{0:N, median}, \bm{\hat{\theta}}_{median})$ 
		\Else{ if $\tilde{\mathbf{x}}_{N+1}$ is a replicate}
		\State Set $\hat{v}_{N+1}$ to the existing median of $v$ at that design point
		\EndIf
		\State Estimate $\mathbb{E}_{\tilde{y}_{N+1} \mid D_N} \left[ \text{BIMSPE} \big(D_N \cup \{(\tilde{\mathbf{x}}_{N+1}, \tilde{y}_{N+1})\}\big) \right]$ using \eqref{C1_median}
		
		\EndFor
		\State Select $\mathbf{x}_{N+1}$ using (6) in the main text and observe $y_{N+1}$ \Comment{Evaluation step}
		\State $D_{N+1} = D_N \cup \{(\mathbf{x}_{N+1}, y_{N+1})\}$   
		\State Run MCMC to draw fresh $\{(\bm{\theta}^{(m)}, \mathbf{v}_{0:N+1}^{(m)})\}_{m=1}^{M}$  from $p(\bm{\theta}, \mathbf{v}_{0:N+1} \mid D_{N+1})$
		\State Compute the posterior medians $\bm{\hat{\theta}}_{median}=\text{median} \big(\{\bm{\theta}^{(m)}\}_{m=1}^{M}|D_{N+1}\big)$, and $\hat{\textbf{v}}_{0:N+1,median}=\text{median}\big(\{ \mathbf{v}_{0:N+1}^{(m)}\}_{m=1}^{M}|D_{N+1}\big)$
		\EndFor
		\State \Return Final posterior estimates $(\bm{\hat{\theta}}_{median}, \mathbf{\hat{v}}_{0:B, median})$ 
	\end{algorithmic} 
\end{algorithm}

\FloatBarrier

\subsection{Empirical Bayes via MAP with Empirical IMSPE (EB-EIM)}

As a complementary alternative to fully Bayesian inference and variational approximations, we also consider \emph{maximum a posteriori} (MAP) estimation, in which the hyperparameters and latent noise variables are fixed at their posterior modes. Compared to MCMC, MAP estimation is substantially faster since it only requires optimization, and has been adopted in fitting dual-GP surrogate models \citep{Kersting2007, QuadriantoMAP2009}, as well as used as a benchmark against other methods \citep{lazaro2011variational}.

The EB-EIM strategy is closely related to the FB-EIM strategy introduced in the previous subsection. In the lookahead step, both strategies use point estimates to evaluate the expected BIMSPE criterion for selecting the next design point, which reduces to the sum of posterior variances computed from plug-in estimates over the discretized design space. Consequently, we estimate the expected BIMSPE  based on the following, which again simplifies to the standard empirical (discretized) IMSPE:
\begin{equation}
	\mathbb{E}_{\tilde{y}_{N+1} \mid D_N} \left[ \text{BIMSPE} \big(D_N \cup \{(\tilde{\mathbf{x}}_{N+1}, \tilde{y}_{N+1})\}\big) \right] \approx  \frac{1}{|\mathcal{D}|} \sum_{\bfx_* \in \mathcal{D}}  \V \left(f(\x_*) | \bm{\hat{\theta}}_{MAP}, \hat{\textbf{v}}_{0:N, MAP}, \hat{v}_{N+1}, D_N \right) .\label{C1_MAP}
\end{equation}

The distinction between EB-EIM and FB-EIM lies in how the point estimates are obtained. When new data arrives, the FB-EIM strategy re-runs MCMC at each iteration to update the posterior medians. In contrast, the EB-EIM strategy does not require MCMC sampling; instead, it estimates $\boldsymbol{\theta}$ and $\mathbf{v}_{0:N}$ by maximizing the joint posterior density:
\begin{equation}
	\hat{\bm{\theta}}_{MAP}, \hat{\textbf{v}}_{0:N, MAP} = \argmax_{\theta, \textbf{v}_{0:N}}  p(\bm{\theta}, \textbf{v}_{0:N}|\textbf{y}_{0:N}).
	\label{MAP_likelihood}
\end{equation}
To obtain credible intervals for the $f$ and $v$ processes in this case, we use a Laplace approximation at the mode of the posterior distribution. The EB-EIM strategy is detailed in Algorithm \ref{algorithm_empirical_MAP}. The prior settings for all hyperparameters in the EB-EIM strategy are the same as the default priors used in the FB-BIM strategy across all synthetic examples. 

Notably, we found that additional constraints on the noise process parameters are necessary to ensure stable convergence and achieve reasonable MAP results. Specifically, in our implementation for EB-EIM, the lengthscale of the noise GP, $l_v$, is restricted to be at least $0.1 \times$ the grid range to prevent excessively small values, and the variance parameter $\sigma^2_g$ is constrained to be less than $5^2$. Without these restrictions, the optimizer frequently converges to degenerate GP configurations that overfit the noise. A similar issue has been mentioned in other works \citep[e.g.,][]{lazaro2011variational}. 

\begin{algorithm}[!htbp]
	\caption{EB-EIM Design Strategy}\label{algorithm_empirical_MAP}
	\begin{algorithmic}[1] 
		\State \textbf{Input:} Initial design $D_0$, discretized set of design points $\mathcal{D}$, budget $B$
		\State Obtain point estimates $\bm{\hat{\theta}}_{MAP}$, and $\hat{\textbf{v}}_{0,MAP}$ via optimizing Eq~\eqref{MAP_likelihood}
		\Comment{Initialization}
		\For{$N = 0$ to $B-1$} \Comment{Sequential design iterations}
		\For{$\tilde{\mathbf{x}}_{N+1} \in \mathcal{D}$} \Comment{Lookahead acquisition step}
		\If{$\tilde{\mathbf{x}}_{N+1}$ is a new design point}
		\State Set $\hat{v}_{N+1}$ to be the mode of $p(v_{N+1} \mid \hat{\mathbf{v}}_{0:N, MAP}, \bm{\hat{\theta}}_{MAP})$
		\Else{ if $\tilde{\mathbf{x}}_{N+1}$ is a replicate}
		\State Set $\hat{v}_{N+1}$ to the existing MAP estimate of $v$ at that design point
		\EndIf
		\State Estimate $\mathbb{E}_{\tilde{y}_{N+1} \mid D_N} \left[ \text{BIMSPE} \big(D_N \cup \{(\tilde{\mathbf{x}}_{N+1}, \tilde{y}_{N+1})\}\big) \right]$ using \eqref{C1_MAP}
		
		\EndFor
		\State Select $\mathbf{x}_{N+1}$ using (6) in the main text and observe $y_{N+1}$ \Comment{Evaluation step}
		\State $D_{N+1} = D_N \cup \{(\mathbf{x}_{N+1}, y_{N+1})\}$   
		\State Obtain point estimates $\bm{\hat{\theta}}_{MAP}$, and $\hat{\textbf{v}}_{0:N+1,MAP}$ via optimizing Eq~\eqref{MAP_likelihood}
		\EndFor
		\State \Return Final estimates $(\bm{\hat{\theta}}_{MAP}, \mathbf{\hat{v}}_{0:B, MAP})$ 
	\end{algorithmic} 
\end{algorithm}

\FloatBarrier

\subsection{Fully Bayesian Inference with BIMSPE and Homoscedastic Noise (Homo-BIM)} \label{e:homo}

The Homo-BIM strategy differs from the FB-BIM strategy in its assumption of noise structure. Specifically, Homo-BIM assumes homoscedastic noise in the stochastic simulation, where the noise $\epsilon(\x)$ follows a normal distribution with zero mean and constant variance $\sigma^2_g$, i.e., $\epsilon(\x) \sim \mathcal{N}(0, \sigma^2_g)$. Meanwhile, a zero-mean GP prior is placed on the mean response surface $f$ with kernel $k_f(\x, \x')$, parameterized by $\sigma_f$ and $\textbf{l}_f$. Thus, the observation model $Y(\x)$ is fully specified by the parameter set $\bm{\theta}_{\text{homo}} = \{\sigma_f, \textbf{l}_f, \sigma_g\}$. In this case, the predictive distribution of $Y(\x_{*})$ given a vector of $n$ observations $\textbf{y}$ is written as:
\begin{equation}
	\E({Y}(\x_*)|\textbf{y}, \bm{\theta}_{homo})=\textbf{k}^T_*(\boldsymbol{K}_f+\sigma^2_g I)^{-1}\textbf{y} \label{pred_mean_homo}  
\end{equation}
\begin{equation}
	\V({Y}(\x_*)|\textbf{y}, \bm{\theta}_{homo})=\sigma_f^2-\textbf{k}^T_*(\boldsymbol{K}_f+\sigma^2_g I)^{-1}\textbf{k}_* \label{pred_var_homo}
\end{equation}
where $I$ is the identity matrix with rank $n$ and $\bm{k}_* =\big [k_f(\x_1, \x_*), \dots, k_f(\x_n, \x_*) \big ]^T$ is the vector of covariances between $\x_*$ and $\x_{1:n}$.

Now suppose we have properly weighted samples $\{\bm{\theta}_{homo}\}_{m=1}^{M}$ with normalized weights $\{\tilde{w}^{(m)}_N\}_{m=1}^{M}$ from $p(\bm{\theta}_{homo}|\textbf{y}_{0:N})$ at the $N$-th iteration. In the lookahead step, we draw $K$ samples $\tilde{y}_{N+1}^{(k)}$ from the normal distribution with mean and variance specified by \eqref{pred_mean_homo} and \eqref{pred_var_homo}. The estimates of the C1 and C2 terms in (7) in the main text under the homoscedastic assumption for candidate point $\tilde{\bfx}_{N+1}$ and one $\bfx_*$ are given by
\begin{equation}
	\widehat{C1}=\sum_{m=1}^{M} \tilde{w}^{(m)}_N  \V \left(f(\x_*) | \bm{\theta}^{(m)}_{homo}, D_N \right), \label{C1_hat_homo}
\end{equation}
\begin{equation}
	\widehat{C2}= \frac{1}{K}\sum_{k=1}^{K} \left[ \sum_{m=1}^{M}  \tilde{w}^{(m)}_N  \left(\tilde{\mu}^{(m,k)}_* \right)^2 -\bigg (\sum_{m=1}^{M} \tilde{w}^{(m)}_N \tilde{\mu}^{(m,k)}_* \bigg )^2\right],\label{C2_hat_homo}
\end{equation}
where $\tilde{\mu}^{(m,k)}_* = \E\left( f(\x_*) | \bm{\theta}^{(m)}_{homo}, D_N, \tilde{y}^{(k)}_{N+1} \right)$. The average of $\widehat{C1} + \widehat{C2}$ over all $\bfx_* \in \mathcal{D}$ completes our approximation of the expected BIMSPE selection criterion in (7) in the main text.

Similar to the FB-BIM strategy, we can apply the sequential importance sampling technique on Homo-BIM to improve computational efficiency. To handle the posterior update step with SIS, we write the decomposition of the target distribution as $p(\bm{\theta}_{homo}|\textbf{y}_{0:N+1})
\propto p(y_{N+1}|\bm{\theta}_{homo}, \textbf{y}_{0:N})p(\bm{\theta}_{homo}|\textbf{y}_{0:N})$,
where $p(\bm{\theta}_{homo}|\textbf{y}_{0:N})$ is treated as the proposal distribution and $p(y_{N+1}|\bm{\theta}_{homo}, \textbf{y}_{0:N})$ is the incremental importance weight, i.e., the likelihood of the observation $y_{N+1}$. Then $w^{(m)}_{N+1}$, namely the $m$-th un-normalized weight at iteration $N+1$, is computed as
$
w^{(m)}_{N+1} = \tilde{w}_{N}^{(m)}~ p(y_{N+1}|\bm{\theta}^{(m)}_{homo}, \textbf{y}_{0:N}),
$
and the updated normalized weights are
$
\Tilde{w}^{(m)}_{N+1}={w^{(m)}_{N+1}}/{\sum_{m=1}^{M}w^{(m)}_{N+1}}
$. We set $\tau=0.2$, the same as our default in the FB-BIM strategy. If the ESS of $\{\Tilde{w}^{(m)}_{N+1} \}_{m=1}^{M}$ falls below $\tau M$, we run MCMC to draw $M$ new samples from $p(\bm{\theta}_{homo} |\textbf{y}_{0:N+1})$ and reset the weights $\{\tilde{w}^{(m)}_{N+1}\}_{m=1}^{M}$ to $\frac{1}{M}$ before proceeding to the lookahead step of the next iteration. Otherwise we carry forward samples of $\{\bm{\theta}^{(m)}_{homo}\}_{m=1}^{M}$ with the updated weights. The complete implementation of Homo-BIM strategy with SIS is detailed in Algorithm \ref{algorithm_homo}. 

The prior settings for $\bm{\theta}_{homo}$ in the Homo-BIM strategy are the same as the default priors used in the FB-BIM strategy across all synthetic examples for the relevant parameters. Specifically, the homoscedastic noise variance parameter $\sigma^2_g$ corresponds to the average baseline variance $\exp(\mu_0)$ in the heteroscedastic model, so we set $\log(\sigma^2_g) \sim \text{Half-Normal}(0,5)$ to provide the equivalent prior.
Other implementation details, i.e., the configuration of the MCMC sampler, follow the specifications outlined in Section 2.5.1 of the main text.

\begin{algorithm}[!htbp]
	\caption{Homo-BIM with BIMSPE and SIS }\label{algorithm_homo}
	\begin{algorithmic}[1] 
		\State \textbf{Input:} Initial design $D_0$, discretized set of design points $\mathcal{D}$, budget $B$, number of samples $K$ and $M$, SIS threshold $\tau$
		\State Draw $M$ samples $\{\bm{\theta}^{(m)}_{homo}\}_{m=1}^{M}$ from $p(\bm{\theta}_{homo} \mid D_0)$ via MCMC  \Comment{Initialization}
		\State Set initial weights $\tilde{w}^{(m)}_0 = 1/M$ for $m=1, \dots, M$
		\For{$N = 0$ to $B-1$} \Comment{Sequential design iterations}
		\For{$\tilde{\mathbf{x}}_{N+1} \in \mathcal{D}$} \Comment{Lookahead acquisition step}
		\State Draw $K$ samples of $\tilde{y}_{N+1}$ from $p(\tilde{y}_{N+1} \mid D_N)$
		\State Estimate $\mathbb{E}_{\tilde{y}_{N+1} \mid D_N} \left[ \text{BIMSPE} \big(D_N \cup \{(\tilde{\mathbf{x}}_{N+1}, \tilde{y}_{N+1})\}\big) \right]$ using \eqref{C1_hat_homo} and \eqref{C2_hat_homo}
		
		\EndFor
		\State Select $\mathbf{x}_{N+1}$ using (6) in the main text and observe $y_{N+1}$ \Comment{Evaluation step}
		\State $D_{N+1} = D_N \cup \{(\mathbf{x}_{N+1}, y_{N+1})\}$   
		\State $w_{N+1}^{(m)} = \tilde{w}_{N}^{(m)} \cdot p(y_{N+1} \mid \bm{\theta}^{(m)}_{homo}, D_N)$ for each $m$  \Comment{Posterior update step}
		\State Normalize $\tilde{w}_{N+1}^{(m)} = w_{N+1}^{(m)} / \sum_{m=1}^{M} w_{N+1}^{(m)}$ and compute $\text{ESS} = \left( \sum_{m=1}^{M} (\tilde{w}_{N+1}^{(m)})^2 \right)^{-1}$
		\If{$\text{ESS} < \tau M$} 
		\State Run MCMC to draw fresh $\{\bm{\theta}^{(m)}_{homo}\}_{m=1}^{M}$  from $p(\bm{\theta}_{homo} \mid D_{N+1})$
		\State Reset weights $\tilde{w}^{(m)}_{N+1} = 1/M$
		\EndIf
		\EndFor
		\State \Return Final posterior samples $\{\bm{\theta}^{(m)}_{homo}\}_{m=1}^{M}$ and weights $\{\tilde{w}^{(m)}_{B}\}_{m=1}^{M}$
	\end{algorithmic} 
\end{algorithm}

\FloatBarrier

\subsection{hetGP \label{hetGP-method}}

The hetGP R package implements sequential design by optimizing IMSPE under the dual-GP joint modeling framework proposed in \cite{binois2018practical, binois2019replication}. This framework leverages replicated data to reduce computational cost and addresses data acquisition by balancing replication and exploration adaptively for various goals, such as for obtaining a globally accurate model \citep{binois2019replication}, for optimization, or for contour finding \citep{lyu2021evaluating}. 

To optimize the IMSPE, hetGP considers a horizon parameter $h \in \{0,1,2, \dots\}$ determining the number of design iterations to look ahead into the future, with $h$ = 0 representing a myopic IMSPE search. By looking over all decision paths (i.e., selecting new $\x$ or replicating existing points) over the next $h+1$ iterations, the path yielding the smallest IMSPE is identified. The first action along this optimal path determines the next design point, with $\x_{N+1}$ chosen as a new location if the explore-first path is optimal, and as a replicate otherwise.

To facilitate practical tuning of the horizon parameter, \citet{binois2019replication} proposed two online adjustment schemes. The \textit{target} scheme dynamically updates $h$ to maintain a desired ratio of unique design points $\rho = n/N$, thereby controlling surrogate modeling costs. In contrast, the \textit{adapt} scheme prioritizes IMSPE reduction irrespective of computational expense, with $h$ updates guided by a criterion inspired by \citet{ankenman2010stochastic}. In our experiments, we include hetGP with the \textit{adapt} scheme as well as fixed horizons $h = 0,\dots,4$ to enable a comprehensive comparison.

We implement this ``hetGP'' strategy following the guidelines provided by \cite{binois2021hetgp}. We use R version 4.5.1 and R package hetGP version 1.1.8. We use the \textit{find\_reps} function to preprocess the data, the \textit{mleHetGP} function to fit the surrogate model during the sequential process, and the \textit{IMSPE\_optim} function to calculate the IMSPE to select the next design point for evaluation. We use the default settings for most arguments and make a few implementation choices for a fair comparison between the hetGP and FB-BIM strategy as below.

For the \textit{mleHetGP} function that is used for model fitting and parameter estimation, we specify the \textit{covtype} argument as ``Gaussian'' to use the squared exponential function as the kernel function. To ensure that the strategy only returns models assuming heteroscedastic noise, we specify the argument \textit{checkHom}=FALSE in this function. We set the jitter \textit{eps}=$10^{-6}$ to match ours, which is used in the inversion of the covariance matrix for numerical stability. Note that we also implemented hetGP using its default \textit{checkHom}=TRUE setting, which allows the return of a homoscedastic model if it achieves a higher log-likelihood; however, this configuration yielded inferior performance in both predictive accuracy and robustness compared to the setup with \textit{checkHom}=FALSE and was therefore excluded from consideration. We de-couple the training of $l_f$ and $l_v$ by setting \textit{linkThetas}=``none''. To avoid degenerate GP fits caused by poor initialization, we specify both \textit{lower} and \textit{upper} bounds for the lengthscale parameters of $f$ and $v$. The lower bound is set to $0.01$ in all examples, while the upper bound is set to 2, 0.5, and 2 for the 1-, 2-, and 5-dimensional examples, respectively. In addition, we initialize the lengthscales of $f$ and $v$ at 0.05 and 0.2, respectively, to improve numerical stability during optimization.

For the \textit{IMSPE\_optim} function that is used to derive IMSPE and select the next design point in the sequential strategy, we pass in the discretized design points that are scaled to $[0,1]^d$ as candidate points using the \textit{Xcand} argument. When new data arrives, we can choose to update the existing model via the \textit{update} function or re-fit a new model via \textit{mleHetGP}. Instead of fully re-optimizing all hyperparameters, the \textit{update} function leverages previously estimated values. In particular, initializing the log-noise process from prior estimates can enhance numerical robustness. In our experiments, we considered both model update and refit, to select the final model with a higher log-likelihood value. However, since $B$ is relatively small, we observe that the incorporation of the \textit{update} procedure does not outperform refitting the model at each iteration. Consequently, we opt to always refit the model throughout our experiments.

Our implementation for one iteration of the hetGP strategy is detailed in Algorithm \ref{algorithm_hetGP} with R pseudo-code.

\begin{algorithm}[!htbp]
	\caption{Pseudo-code for hetGP}\label{algorithm_hetGP}
	\begin{algorithmic}[1] 
		\State \textbf{Input:} observations $D_N$, dimension $d$, upper bounds of lengthscales \textit{lengthscale\_max}
		
		\Procedure{hetGP}{$D_N$, $d$, \textit{lengthscale\_max}}
		\State Fit the model with \textit{mleHetGP} function
		\begin{verbatim}
			data_unique <- find_reps(X = as.matrix(sim_data[,1:d], ncol = d), 
			Z = sim_data[, d+1])
			model <- mleHetGP(X = data_unique, Z =  data_unique$Z, covtype = "Gaussian",
			settings = list(linkThetas="none", return.hom=FALSE, checkHom=FALSE), 
			eps=1e-6, lower = rep(0.01, d), upper=rep(lengthscale_max, d),
			init = list(theta = rep(0.05, d), theta_g = rep(0.2, d)))  
		\end{verbatim}
		\If{Adaptive horizon scheme}
		\State Derive the optimal horizon $h$
		\begin{verbatim}
			h <- horizon(model)
		\end{verbatim}
		\EndIf
		\State Select the next point via minimizing IMSPE with \textit{IMSPE\_optim} function
		\begin{verbatim}
			res <- IMSPE_optim(model, h = h, Xcand=matrix(x_grid, ncol=d))
		\end{verbatim}
		\EndProcedure
	\end{algorithmic}
\end{algorithm}

\FloatBarrier

\subsection{Variational Inference with BIMSPE (VI-BIM)} \label{LA-VI}

We also considered a variational inference (VI) approach to fitting the heteroscedastic dual GP model, since the key computational bottleneck for FB-BIM is that the posterior distribution is analytically intractable and requires expensive MCMC sampling. VI can mitigate this computational burden by approximating the intractable posterior with a tractable family and optimizing the approximation via the evidence lower bound (ELBO) \citep{Jordan1999,WainwrightJordan2008,Blei2017VI}. In our implementation, we used the \textit{GPflow} Python package following \citep{GPflow2017}, which fits two sparse variational GPs (SVGP) and returns hyperparameter estimates and Monte Carlo samples from the pseudo-posterior (i.e., variational predictive distribution) of the latent $v$-process. The returned VI-optimized $\bm{\theta}$ and $\textbf{v}$ samples can then be used to evaluate our expected BIMSPE criterion, thereby propagating uncertainty into sequential design without requiring an MCMC sampler. Thus, VI-BIM specifically differs from FB-BIM by replacing MCMC sampling with VI.

A recommended default is to use sparse variational GPs with inducing variables \citep{Titsias2009,Hensman2013,Hensman2015}; however, to ensure the best VI performance in our comparisons we placed inducing points at every observed location. Optimization is performed via a standard two-optimizer strategy for SVGPs \citep{Hensman2015}: (i) a natural-gradient step for the hyperparameters in the variational distributions, and (ii) an Adam step for the remaining trainable parameters. The natural-gradient optimizer uses $\gamma=0.1$ and Adam uses \textit{learning\_rate} $=10^{-2}$, with early stopping triggered by insufficient improvement in the training objective (\textit{patience} $=50$ and \textit{tolerance} $=10^{-3}$). The model is initialized using simple data-driven heuristics: the $f$-kernel lengthscale is initialized at $0.5\times\mathrm{median}(\|x_i-x_j\|)$ and variance at $\mathrm{Var}(\textbf{y}_{0:N})$, while the $v$-kernel is initialized with lengthscale $\mathrm{median}(\|x_i-x_j\|)$ and variance $1$; the log-variance mean is initialized at $\log(0.01\,\mathrm{Var}(\textbf{y}_{0:N})+10^{-6})$. 

Suppose we obtain the variational estimate $\hat{\bm{\theta}}_{VI}$ and $\{\textbf{v}^{(m)}_{0:N}\}^{M}_{m=1}$ from the variational distribution $q(v(\bfx))$ at the $N$-th iteration. In the lookahead step, we first draw $K$ samples of $\tilde{y}^{(k)}_{N+1}$ from its posterior distribution $p(\tilde{y}_{N+1}|D_N, \bm{\theta}=\hat{\bm{\theta}}_{VI})$. If $\tilde{\x}_{N+1}$ is a new point, then we sample one $\tilde{v}^{(m)}_{N+1}$ from its predictive distribution $p(v_{N+1}|{\textbf{v}}^{(m)}_{0:N}, \hat{\bm{\theta}}_{VI})$ for each ${\textbf{v}}^{(m)}_{0:N}$, $m = 1, \ldots, M$; if $\tilde{\x}_{N+1}$ is a replicate, we assign it the existing value from $\textbf{v}^{(m)}_{0:N}$. Subsequently, we estimate C1 and C2 in the expected BIMSPE based on the following:
\begin{equation}
	\widehat{C1}=\sum_{m=1}^{M} \tilde{w}^{(m)}_N  \V \left(f(\x_*) | \bm{\hat{\theta}}_{VI}, \textbf{v}_{0:N}^{(m)}, \tilde{v}^{(m)}_{N+1}, D_N \right), 
\end{equation}
\begin{equation}
	\widehat{C2}= \frac{1}{K}\sum_{k=1}^{K} \left[ \sum_{m=1}^{M}  \tilde{w}^{(m)}_N  \left(\tilde{\mu}^{(m,k)}_* \right)^2 -\bigg (\sum_{m=1}^{M} \tilde{w}^{(m)}_N \tilde{\mu}^{(m,k)}_* \bigg )^2\right],
\end{equation}
where $\tilde{\mu}^{(m,k)}_* = \E\left( f(\x_*) | \bm{\hat{\theta}}_{VI}, \textbf{v}_{0:N}^{(m)}, \tilde{v}_{N+1}^{(m)}, D_N, \tilde{y}^{(k)}_{N+1} \right)$. The average of $\widehat{C1} + \widehat{C2}$ over all $\bfx_* \in \mathcal{D}$ completes our approximation of the expected BIMSPE selection criterion in (7) of the main text.

Similar to the FB-BIM and Homo-BIM strategies, we can apply the sequential importance sampling technique on VI-BIM to improve computational efficiency. To handle the posterior update step with SIS, we write the decomposition of the target distribution as $p(\textbf{v}_{0:N+1} |\textbf{y}_{0:N+1}, \bm{\theta}_{VI})
\propto p(y_{N+1}|\textbf{v}_{0:N+1}, \bm{\theta}_{VI}, \textbf{y}_{0:N})q(\textbf{v}_{0:N+1}|\textbf{y}_{0:N}, \bm{\theta}_{VI})$,
where $q(\textbf{v}_{0:N+1}|\textbf{y}_{0:N}, \bm{\theta}_{VI})$ is the proposal variational distribution and $p(y_{N+1}|\textbf{v}_{0:N+1}, \bm{\theta}_{VI}, \textbf{y}_{0:N})$ is the importance weight. Then $w^{(m)}_{N+1}$, namely the $m$-th un-normalized weight at iteration $N+1$, is computed as
$
w^{(m)}_{N+1} = \tilde{w}_{N}^{(m)}~ p(y_{N+1}|\textbf{v}_{0:N+1}, \bm{\theta}_{VI}, \textbf{y}_{0:N}),
$
and the updated normalized weights are
$
\Tilde{w}^{(m)}_{N+1}={w^{(m)}_{N+1}}/{\sum_{m=1}^{M}w^{(m)}_{N+1}}
$. We set $\tau=0.2$, the same as our default in the FB-BIM strategy. If the ESS of $\{\Tilde{w}^{(m)}_{N+1} \}_{m=1}^{M}$ falls below $\tau M$, we run VI to draw $M$ new pseudo-samples of $\{\textbf{v}_{0:N+1}\}_{m=1}^{M}$ from $p(\textbf{v}_{0:N+1} |\textbf{y}_{0:N+1})$ and reset the weights $\{\tilde{w}^{(m)}_{N+1}\}_{m=1}^{M}$ to $\frac{1}{M}$ before proceeding to the lookahead step of the next iteration. Otherwise we carry forward samples of $\{\textbf{v}^{(m)}_{0:N}\}_{m=1}^{M}$ with the updated weights. 

\FloatBarrier

\section{Runtime of All Strategies \label{appendix: strategy_time}}

We evaluate the runtime of all strategies and summarize the results in Table~\ref{appendix: strategy_time_table}. The runtime is recorded after 100, 150, and 200 sequential iterations for the 1-, 2- and 5-dimensional examples, respectively. During benchmarking, expected BIMSPE computations in the relevant Bayesian strategies (i.e., FB-BIM, Homo-BIM, and VI-BIM) were parallelized across 24 CPU cores, while MCMC runs or model fitting were executed on a single core. In contrast, since selection criterion evaluations for EB-EIM and hetGP (Adapt) are extremely fast on a discretized design space, they will not tangibly benefit from parallelization; hence, both strategies were run on a single core, with the latter using the standard implementation in the hetGP R package. All experiments were conducted on a Linux workstation equipped with an Intel(R) Xeon(R) W7-2495X CPU (24 cores / 48 threads) and 250 GiB of RAM. Reported wall-clock runtimes are expressed in minutes and summarized as mean $\pm$ standard deviation across 10 macro-replications.

Among the 10 reported macro-replications of the FB-BIM strategy, Table~\ref{appendix: strategy_separate_time_table} reports the breakdown of runtime into lookahead acquisition and posterior update wall-clock times. These confirm that single-core MCMC sampling dominates FB-BIM's overall runtime compared to the parallelized lookahead step. Recall that MCMC is only run when the ESS falls below a threshold of $\tau M$ with $\tau = 0.2$; therefore, the reported posterior update time effectively represents the cumulative MCMC runtime over the sequential iterations in which MCMC is triggered, as SIS updates are extremely fast.

\begin{table}[ht]
	\centering
	\caption{Average runtime of all sequential strategies for the three examples. Reported wall-clock runtimes are expressed in minutes and summarized as mean $\pm$ standard deviation across 10 macro-replications.} \label{appendix: strategy_time_table}
	\label{tab:design_cost}
	\begin{tabular}{|c|p{3.5cm}|p{3.5cm}|p{3.5cm}|}
		\hline
		\textbf{Strategy} 
		& \textbf{1-D} 
		& \textbf{2-D}
		& \textbf{5-D}
		\\
		\hline
		FB-BIM 
		& 1.93 $\pm$ 0.66
		& 9.65 $\pm$ 1.7
		& 81.2 $\pm$ 8.91\\
		\hline
		EB-EIM 
		& 0.55  $\pm$ 0.01  
		& 1.7 $\pm$ 0.08
		& 6.03 $\pm$ 0.71\\
		\hline
		Homo-BIM 
		& 0.58  $\pm$ 0.1 
		& 3.46 $\pm$ 0.19
		& 19.7 $\pm$ 0.83\\
		\hline
		VI-BIM 
		& 2.46 $\pm$ 0.12  
		& 8.97 $\pm$ 0.43
		& 39.4 $\pm$ 2.12\\
		\hline
		hetGP (Adapt) 
		& 0.54$\pm$ 0.0 
		& 1.68 $\pm$ 0.0
		& 6.61 $\pm$ 0.0\\
		\hline
	\end{tabular}
\end{table}

\begin{table}[ht]
	\centering
	\caption{Separate lookahead criterion evaluation and model fitting runtimes for the FB-BIM strategy. Reported wall-clock runtimes are expressed in minutes and summarized as mean $\pm$ standard deviation across 10 macro-replications.} \label{appendix: strategy_separate_time_table}
	\begin{tabular}{|c|p{3.5cm}|p{3.5cm}|p{3.5cm}|}
		\hline
		\textbf{Step} 
		& \textbf{1-D} 
		& \textbf{2-D}
		& \textbf{5-D}
		\\
		\hline
		Lookahead
		& 0.14  $\pm$ 0.0 
		& 2.54 $\pm$ 0.02
		& 19.8 $\pm$ 0.15\\
		\hline
		Posterior update 
		& 1.78 $\pm$ 0.65
		& 7.09 $\pm$ 1.7
		& 61.3 $\pm$ 8.14\\
		\hline
	\end{tabular}
\end{table}

\FloatBarrier

\section{Final Design Visualization}\label{appendix: final_designs}

\begin{figure}
	\centering
	\includegraphics[width=1\linewidth]{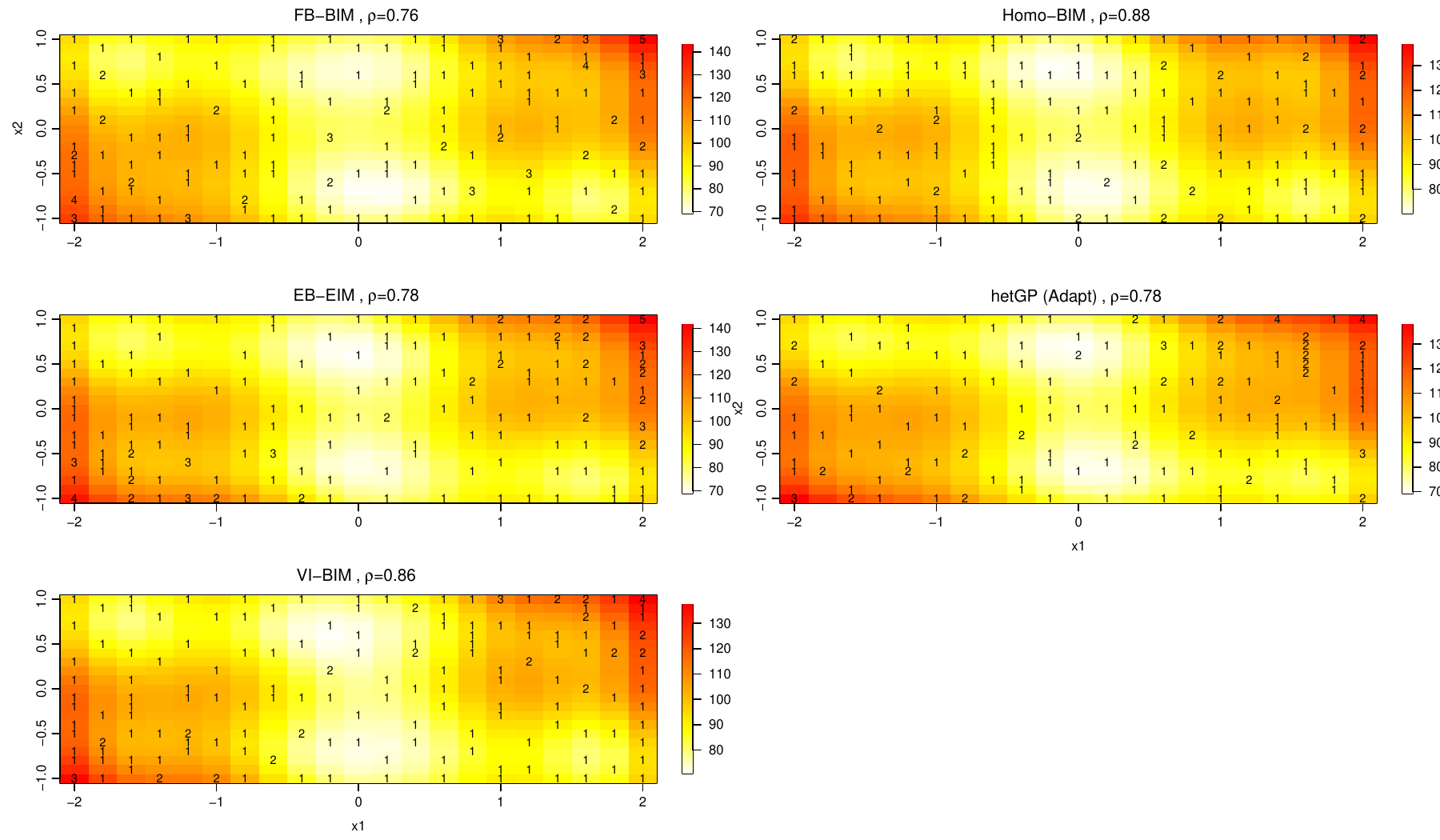}
	\caption{Final designs on the 2-D example generated by the different strategies for one macro-replication. The strategy used is indicated at the top of the plots along with the ratio of unique design points to total number of observations, denoted as $\rho$. The numbers overlaid indicate observed design locations and numbers of replicates.}
	\label{fig:final-design-2d}
\end{figure}

Figure~\ref{fig:final-design-2d} depicts the final designs under the different strategies for one macro-replication of the 2-D example. The fitted mean response surfaces appear broadly similar across strategies, while the varying values of $\rho$ suggest differences in each strategy’s preference for replication versus exploration.

\FloatBarrier
\bigskip

\section{Comparing hetGP with Different Horizons}\label{appendix: hetGP_details}

\begin{figure}[!b]
	\captionsetup[subfigure]{labelformat=empty}
	\begin{subfigure}[b]{1\linewidth}
		\centering
		\includegraphics[width=0.9\textwidth]{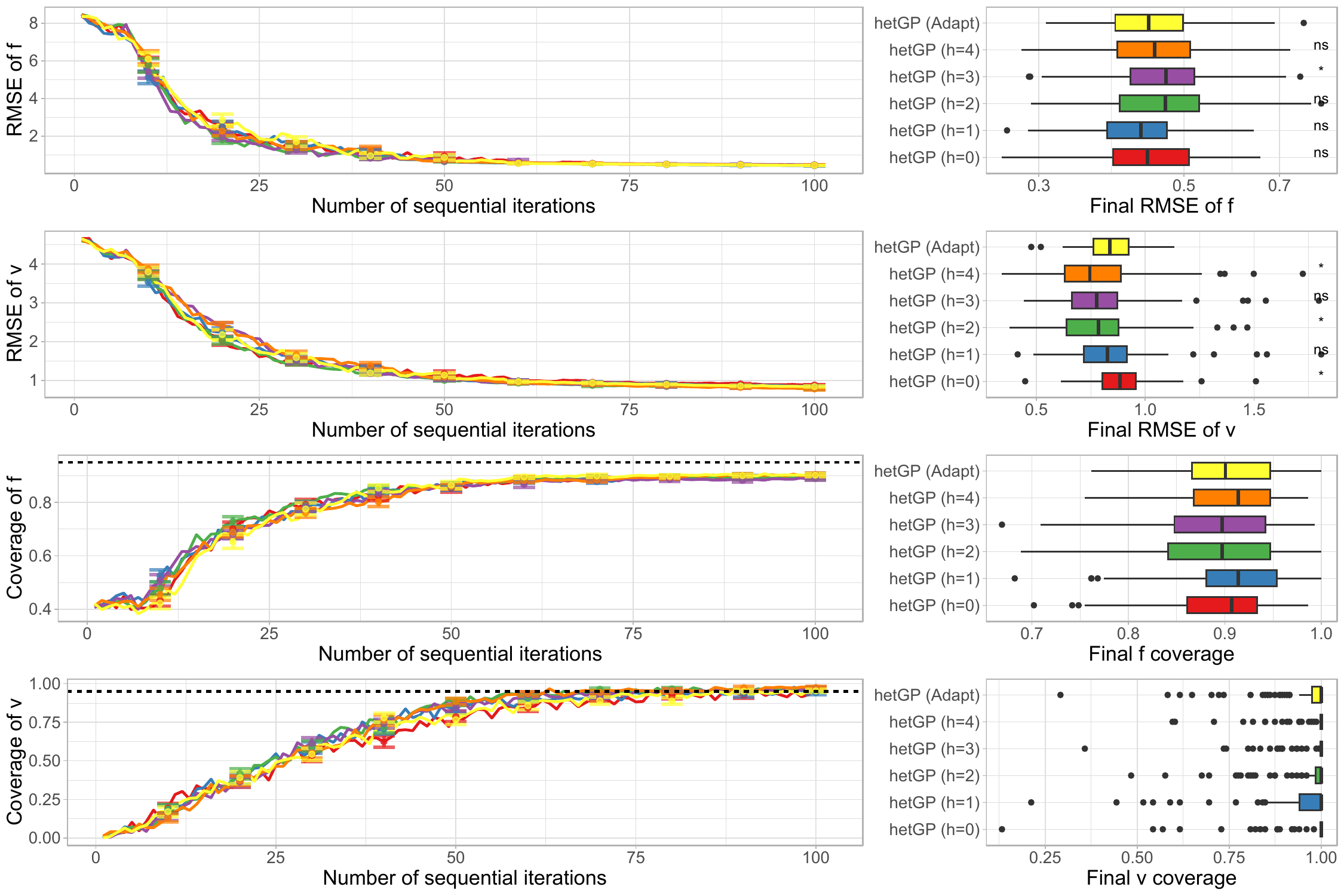}
		\caption{(a) 1-D example}
	\end{subfigure} 
	\hfill
\end{figure}
\begin{figure}[ht]\ContinuedFloat
	\begin{subfigure}[b]{1\linewidth}
		\centering
		\includegraphics[width=0.9\textwidth]{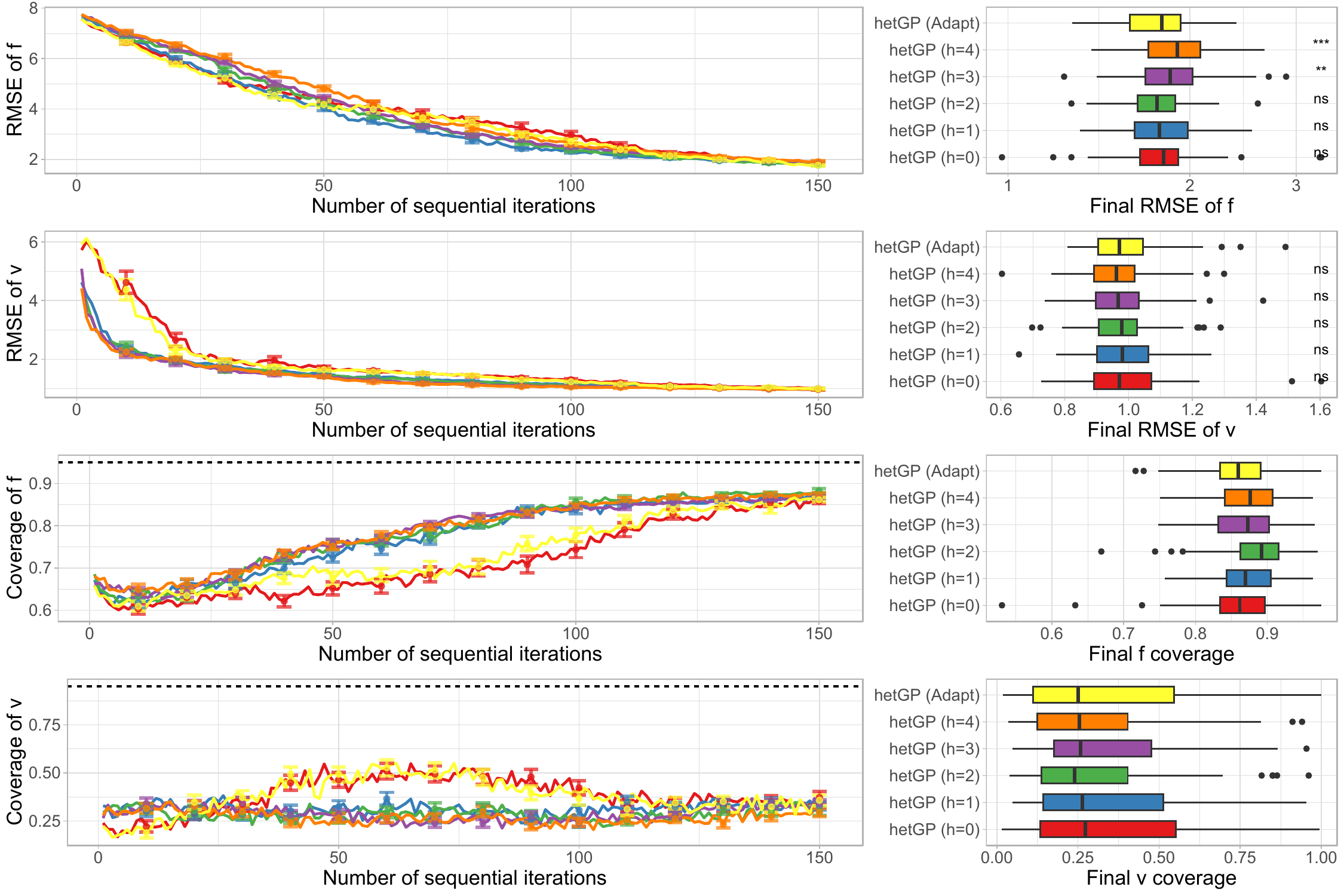}
		\caption{2-D example}
	\end{subfigure}
\end{figure}
\hfill
\begin{figure}[ht]\ContinuedFloat
	\begin{subfigure}[b]{1\linewidth}
		\centering
		\includegraphics[width=0.9\textwidth]{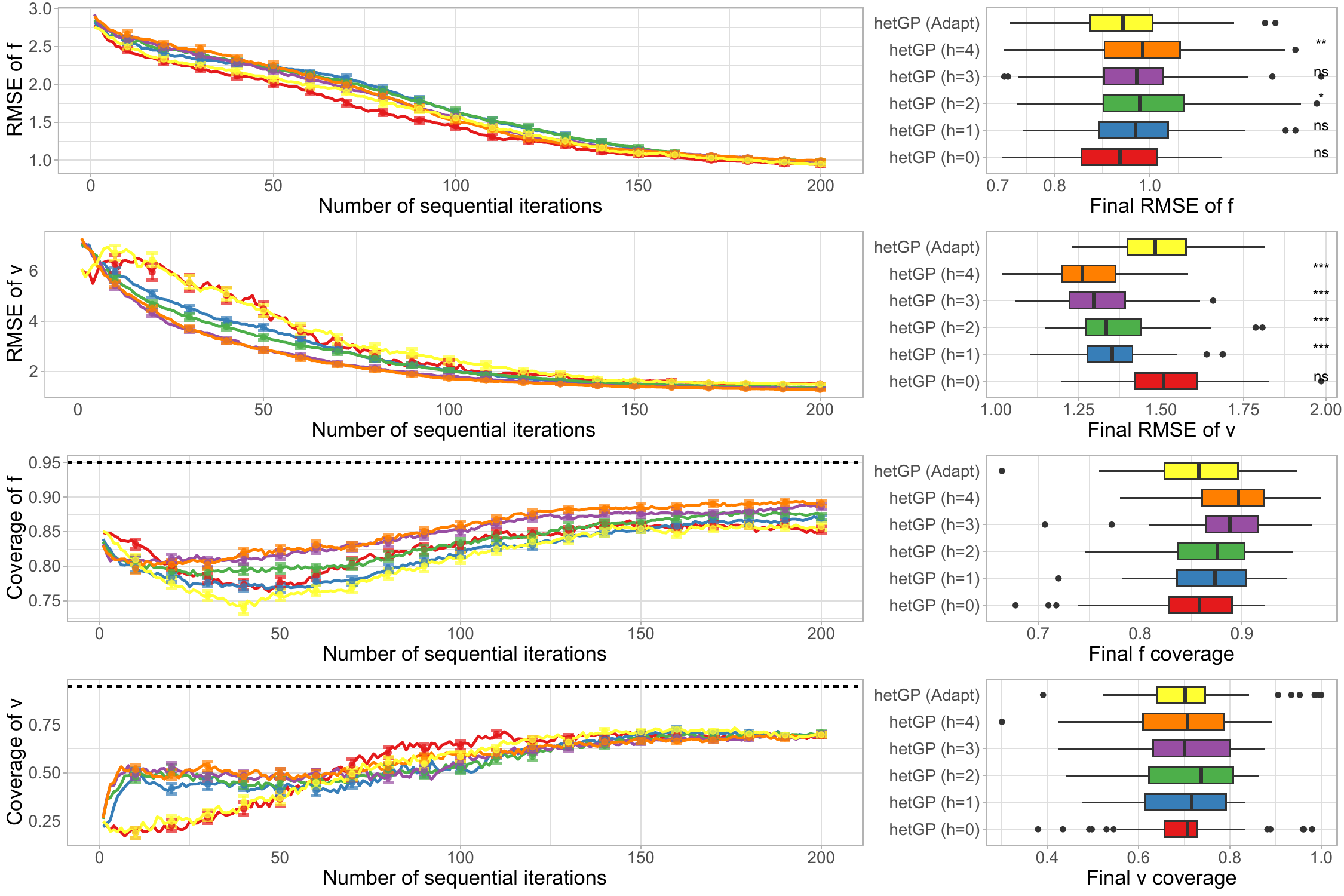}
		\caption{5-D example}
	\end{subfigure}
	\caption{Line plots of prediction performance (left) and final iteration boxplots (right) for hetGP with different horizons $h$, denoted by red ($h=0$), blue ($h=1$), green ($h=2$), purple ($h=3$), orange ($h=4$), and yellow (adaptive scheme), respectively, for each example. The left column shows the trajectories of four metrics over iterations: RMSE for the mean response, RMSE for the log-noise variance, and empirical coverage probabilities for the mean response and log-noise variance. The black dashed line in the coverage plots indicates the 95\% nominal level. Error bars (mean $\pm$ standard deviation) are displayed every 10 iterations. The right column presents boxplots of the corresponding metric values at the final iteration. A two-sample t-test is conducted on the results at the final iteration to compare the mean RMSE metrics between the reference strategy used in the main text (hetGP with the adaptive scheme) and each alternative horizon. Significance levels are indicated beside the boxplots: ``ns'' ($p>0.05$), ``*'' ($p\le 0.05$), ``**'' ($p\le 0.01$), ``***'' ($p\le 0.001$).}
\end{figure}

\end{document}